\documentclass[final,3p,times]{elsarticle}

\usepackage{amssymb}
\usepackage{amsmath}

\usepackage{tabularx}
\usepackage{makecell}
\usepackage{epsfig}
\usepackage{titlesec}
\usepackage{graphicx}
\usepackage{subcaption}
\usepackage{longtable}
\usepackage{rotating}
\usepackage{float}
\usepackage{hyperref}
\usepackage{url} 
\usepackage{booktabs}
\usepackage{titlesec}
\usepackage{enumitem}
\usepackage{soul}
\usepackage{xcolor}
\usepackage{titlesec}
\usepackage{multirow}

\begin{document}

\begin{frontmatter}

\title{TUS-REC2024: A Challenge to Reconstruct 3D Freehand Ultrasound Without External Tracker}

\author[ucl]{\texorpdfstring{Qi Li\corref{cor1}}{Qi Li}}
\cortext[cor1]{Corresponding author.}
\ead{qi.li.21@ucl.ac.uk}
\author[ucl,queen_marry]{Shaheer U. Saeed}
\author[ucl]{Yuliang Huang}
\author[MUSIC_Lab_1,MUSIC_Lab_2,MUSIC_Lab_3]{Mingyuan Luo}
\author[MUSIC_Lab_1,MUSIC_Lab_2,MUSIC_Lab_3]{Zhongnuo Yan}
\author[MUSIC_Lab_1,MUSIC_Lab_2,MUSIC_Lab_3,MUSIC_Lab_4]{Jiongquan Chen}
\author[MUSIC_Lab_1,MUSIC_Lab_2,MUSIC_Lab_3]{Xin Yang}
\author[MUSIC_Lab_5,MUSIC_Lab_2,MUSIC_Lab_3]{Dong Ni}
\author[ISRU_1]{Nektarios Winter}
\author[ISRU_2]{Phuc Nguyen}
\author[ISRU_3]{Lucas Steinberger}
\author[ISRU_1]{Caelan Haney}
\author[zjr_1]{Yuan Zhao}
\author[zjr_1]{Mingjie Jiang}
\author[zjr_2]{Bowen Ren}
\author[AMI_Lab_1]{SiYeoul Lee}
\author[AMI_Lab_1]{Seonho Kim}
\author[AMI_Lab_1]{MinKyung Seo}
\author[AMI_Lab_2,AMI_Lab_3]{MinWoo Kim}
\author[Madison_Elastography_Lab_1,Madison_Elastography_Lab_2]{Yimeng Dou}
\author[Madison_Elastography_Lab_1,Madison_Elastography_Lab_2]{Zhiwei Zhang}
\author[Madison_Elastography_Lab_3,Madison_Elastography_Lab_4]{Yin Li}
\author[Madison_Elastography_Lab_1,Madison_Elastography_Lab_2]{Tomy Varghese}

\author[ucl]{Dean C. Barratt}
\author[ucl]{Matthew J. Clarkson}
\author[kcl]{Tom Vercauteren}
\author[ucl]{Yipeng Hu}

\affiliation[ucl]{
            organization={UCL Hawkes Institute, Department of Medical Physics and Biomedical Engineering, University College London},
            city={London},
            postcode={WC1E 6BT}, 
            country={U.K.}}

\affiliation[queen_marry]{
            organization={Centre for Bioengineering, Digital Environment Research Institute, School of Engineering and Materials Science, Queen Mary University of London},
            city={London},
            postcode={E1 4NS}, 
            country={U.K.}}
            
\affiliation[MUSIC_Lab_1]{
            organization={National-Regional Key Technology Engineering Laboratory for Medical Ultrasound, School of Biomedical Engineering, Health Science Center, Shenzhen University},
            city={Shenzhen},
            postcode={518000}, 
            country={China}} 

\affiliation[MUSIC_Lab_5]{
            organization={School of Artificial Intelligence, Shenzhen University},
            city={Shenzhen},
            postcode={518060}, 
            country={China}}
    
\affiliation[MUSIC_Lab_2]{
            organization={Medical UltraSound Image Computing (MUSIC) Lab, Shenzhen University},
            city={Shenzhen},
            postcode={518000}, 
            country={China}} 
\affiliation[MUSIC_Lab_3]{
            organization={Marshall Laboratory of Biomedical Engineering, Shenzhen University},
            city={Shenzhen},
            postcode={518000}, 
            country={China}} 
\affiliation[MUSIC_Lab_4]{
            organization={Shenzhen RayShape Medical Technology Inc.},
            city={Shenzhen},
            postcode={518000}, 
            country={China}}

\affiliation[ISRU_1]{
            organization={DKFZ (German Cancer Research Center) Heidelberg},
            country={Germany}
            }
\affiliation[ISRU_2]{
            organization={University of Cincinnati, Cincinnati, OH, 45219},
            country={USA}
            }
\affiliation[ISRU_3]{
            organization={Tufts University},
            country={USA}
            }
            
\affiliation[zjr_1]{
            organization={Hong Kong Centre for Cerebro-cardiovascular Health Engineering, Rm 1115-1119, Building 19W, Hong Kong Science Park},
            city={Hong Kong SAR},
            country={China}
            }
\affiliation[zjr_2]{
            organization={City University of Hong Kong, Kowloon},
            city={Hong Kong SAR},
            country={China}
            }

\affiliation[AMI_Lab_1]{
            organization={Department of Information Convergence Engineering, Pusan National University},
            city={Yangsan},
            country={Korea}
            }
\affiliation[AMI_Lab_2]{
            organization={School of Biomedical Convergence Engineering, Pusan National University},
            city={Yangsan},
            country={Korea}
            }
\affiliation[AMI_Lab_3]{
            organization={Center for Artificial Intelligence Research, Pusan National University},
            city={Busan},
            country={Korea}
            }

\affiliation[Madison_Elastography_Lab_1]{
            organization={Department of Medical Physics, University of Wisconsin (UW) School of Medicine and Public Health},
            city={Madison},
            postcode={WI 53705}, 
            country={USA}
            }

\affiliation[Madison_Elastography_Lab_2]{
            organization={Department of Electrical and Computer Engineering, UW–Madison},
            city={Madison},
            postcode={WI 53706}, 
            country={USA}
            }
\affiliation[Madison_Elastography_Lab_3]{
            organization={Department of Biostatistics and Medical Informatics, UW School of Medicine and Public Health},
            city={Madison},
            postcode={WI 53726}, 
            country={USA}
            }
\affiliation[Madison_Elastography_Lab_4]{
            organization={Department of Computer Sciences, UW–Madison},
            city={Madison},
            postcode={WI 53706}, 
            country={USA}
            }

\affiliation[kcl]{
            organization={School of Biomedical Engineering \& Imaging Sciences, King’s College London},
            city={London},
            postcode={WC2R 2LS}, 
            country={U.K.}}

\begin{abstract}
Trackerless freehand ultrasound reconstruction aims to reconstruct 3D volumes from sequences of 2D ultrasound images without relying on external tracking systems. By eliminating the need for optical or electromagnetic trackers, this approach offers a low-cost, portable, and widely deployable alternative to more expensive volumetric ultrasound imaging systems, particularly valuable in resource-constrained clinical settings.
However, predicting long-distance transformations and handling complex probe trajectories remain challenging. 
The TUS-REC2024 Challenge establishes the first benchmark for trackerless 3D freehand ultrasound reconstruction by providing a large publicly available dataset, along with a baseline model and a rigorous evaluation framework.
By the submission deadline, the Challenge had attracted 43 registered teams, of which 6 teams submitted 21 valid dockerized solutions. The submitted methods span a wide range of approaches, including the state space model, the recurrent model, the registration-driven volume refinement, the attention mechanism, and the physics-informed model.
This paper provides a comprehensive background introduction and literature review in the field, presents an overview of the challenge design and dataset, and offers a comparative analysis of submitted methods across multiple evaluation metrics. These analyses highlight both the progress and the current limitations of state-of-the-art approaches in this domain and provide insights for future research directions.
All data and code are publicly available to facilitate ongoing development and reproducibility. 
As a live and evolving benchmark, it is designed to be continuously iterated and improved. The Challenge was held at MICCAI 2024 and is organised again at MICCAI 2025, reflecting its sustained commitment to advancing this field.

\end{abstract}


\end{frontmatter}


\section{Introduction}
\label{Introduction}

Ultrasound imaging remains a cost-effective, non-invasive modality with real-time capabilities, making it a valuable tool across a wide range of clinical applications \cite{solberg2007freehand}. However, it only provides incomplete 3D information because the locations of ultrasound frames are unknown.
This poses challenges for applications requiring accurate volumetric information, such as biometric quantification, image registration, and 3D visualisation \cite{adriaans2024trackerless}. While expert clinicians can often infer 3D structure mentally or through standardised acquisition protocols (e.g., standard planes), the absence of inter-frame positional data limits reproducibility and the integration of ultrasound images into advanced image analysis workflows.

Ongoing works seek to address this limitation by using 3D ultrasound probes to enable 3D reconstruction. 
3D ultrasound probes are capable of acquiring volumetric data directly, using dedicated mechanical probes or 2D array transducers \cite{huang2017review}. While these probes provide valuable 3D imaging capabilities and offer flexible scanning trajectories, their higher cost, limited user experience in clinical practice, and physical constraints (e.g., interference in 2D array transducers) may restrict their use in some clinical settings, such as low-resource environments, point-of-care scenarios, or mobile and emergency units where portability and affordability are critical \cite{park2023cost}.

In comparison, freehand 2D ultrasound imaging has advantage of widespread availability and long-standing integration into clinical workflows. It has been used for decades across a broad range of applications, and clinicians are highly familiar with both the use and interpretation of it. 
Building on this established foundation, tracker-based freehand ultrasound reconstruction techniques have been introduced to enable the generation of 3D anatomical representations. These methods aim to enhance conventional 2D ultrasound by incorporating spatial information from external tracking systems, such as optical \cite{wiles2004accuracy} or electromagnetic (EM) \cite{franz2014electromagnetic} trackers. 
This enables conventional 2D ultrasound probes to be used for 3D imaging, providing a more flexible and accessible solution in clinical and research applications where dedicated and bulky 3D ultrasound systems may be impractical.
However, optical and EM tracking systems present other challenges in clinical environments. Optical tracking requires an unobstructed line of sight between the tracker and the camera \cite{he2015inertial}. Although approaches such as using multiple cameras have been proposed to mitigate this limitation, challenges related to system complexity and calibration remain \cite{moller2016enhanced,leizea2023calibration}. EM tracking is sensitive to nearby metal objects and electromagnetic interference, which can affect accuracy \cite{cavaliere2024enhancing}.

Trackerless freehand ultrasound reconstruction refers to generating 3D volumetric representations from sequential 2D ultrasound frames in a handheld freehand scan, without using external tracking systems. This kind of method computes the relative spatial transformations among frames using images themselves. Common approaches include non-learning-based methods such as  speckle decorrelation \cite{chang20033,chen1997determination,gee_2006_SD_3} and learning-based motion estimation such as convolutional neural network (CNN) \cite{prevost20183d,guo2020sensorless} and recurrent neural network (RNN) \cite{luo2023recon,li2023long}. Additionally, trackerless freehand ultrasound reconstruction may further enhance existing 3D ultrasound systems, rather than serving solely as alternatives.

This application involves practical challenges such as handling both 2D and 3D imaging data, incorporating tracking information, and managing multiple spatial coordinate systems, all of which contribute to a significant barrier for newcomers to this field and may impede broader progress and adoption. In addition, trackless reconstruction encounters other challenges: 1) the difficulty of predicting accurate poses when the ultrasound sequence length is large; and 2) the high variability across different datasets, which complicates the validation and fair comparison of methods. While benchmarking may be essential to address this variability, progress has been limited by the scarcity of publicly available datasets, which are critical for both performance evaluation and the development of learning-based approaches.

Furthermore, comparison of methods in the existing literature is often conducted on relatively small, private datasets, using a variety of evaluation metrics to assess performance \cite{adriaans2024trackerless}. This variability complicates the comparison of different methods' strengths and weaknesses and may lead to biased conclusions, due to dataset characteristics, evaluation metric choices, and inherent differences in the methods' underlying assumptions. For example, learning-based approaches may assume that training and testing data come from similar distributions, while classical methods may rely on consistently available speckle patterns.

To bridge these gaps, we organised the TUS-REC2024 (Trackerless 3D Freehand Ultrasound Reconstruction) Challenge. This Challenge is designed to foster both algorithmic innovation and practical clinical applicability by promoting reproducibility, benchmarking, and methodological transparency. TUS-REC2024 Challenge provides an \textit{in vivo} ultrasound dataset, consisting of scans from both the left and right forearms of 85 volunteers (2,040 scans, 1,025,448 frames in total), acquired using a time-synchronised optical tracking system. We aim to conduct a comprehensive comparison among methods, evaluating their strengths and weaknesses on a common dataset, using a consistent set of carefully-defined performance metrics. This approach will ensure a more objective and transparent assessment of the methods' relative efficacy, and more importantly, to drive the development of new techniques for trackerless freehand ultrasound reconstruction.

This Challenge has three key contributions. Firstly, it establishes a rigorous and standardised benchmarking framework for trackerless freehand ultrasound reconstruction, advancing the development of novel algorithms and promoting objective performance evaluation through withheld test data and unified assessment metrics. Secondly, it provides the necessary infrastructure to support this benchmarking effort, including a large publicly available dataset in the field, and detailed preliminary materials and accompanying code that describe the end-to-end pipeline for trackerless freehand ultrasound reconstruction. 
Thirdly, beyond the outcomes of the Challenge itself, this summary paper delivers additional insights, including a comprehensive literature review, a detailed comparative analysis of the participating algorithms and a discussion of the design choices and performance trade-offs for future method development.

The rest of the article is organised as below. Sections~\ref{Preliminaries} introduces the basic knowledge of trackerless freehand ultrasound reconstruction, aiming to provide researchers with clear technical background and consistent terminologies in this field. Sections~\ref{Related_work} summarises the state-of-the-art methods in trackerless freehand ultrasound reconstruction, comparing traditional and deep learning techniques across key methodological aspects.  Section~\ref{Challenge_design} reports details of the Challenge, including dataset curation, evaluation metrics and so on. 
Section~\ref{Challenge_Outcome} describes the participation statistics and the methodologies submitted by participating teams, accompanied by performance analysis of each method.
Section~\ref{Discussions}  discusses the limitations of the Challenge and outlines potential directions for future work. 
Finally, Section~\ref{Conclusion} concludes the study by summarising the outcomes of the Challenge and highlighting its key contributions, benefits, and future directions.

\section{Preliminaries}
\label{Preliminaries}
The goal of trackerless freehand ultrasound reconstruction is to estimate the transformation between pairs of ultrasound frames within a scan without relying on any external tracking device, thereby enabling the reconstruction of 2D ultrasound images into 3D space. Table~\ref{terminologies} summarises the terminologies commonly used in this field and the notations used across this paper.

\subsection{Coordinate Systems and Spatial Transformations}
\label{tracking_system}

For learning-based methods, a tracking system is typically used to directly capture the pose of each ultrasound frame, providing labels for training and ground truth for evaluation. 
The most commonly utilised tracking modalities are optical tracking systems and EM tracking systems.
The optical tracking system consists of rigid tracking tools attached to the ultrasound probe and cameras that capture position of the tracking tool \cite{treece2003high}. The tool typically includes at least three passive or active markers, which enable the determination of the probe's six-degree-of-freedom (6-DoF) pose. After spatial calibration, as detailed in Section~\ref{calibration}, the system can obtain the pose of each ultrasound frame itself. The tracking data are timestamped, and subsequently transferred and stored using an interface such as the open-source PLUS platform \cite{lasso2014plus}. While ultrasound machine and tracking device typically have their own API for data management, softwares such as PLUS provide unified interface and consistent data formats, making practical integration more convenient, though not strictly necessary.
The EM tracking system \cite{kindratenko2000survey} comprises three main components: the transmitter, the system control unit, and the tracked receiver. When the probe is moved within the magnetic field produced by the transmitter, the receiver mounted on the probe detects induced electrical currents, for calculating the spatial location relative to a predefined reference. 

The remainder of this section describes the three coordinate systems and their spatial transformations involved in freehand ultrasound reconstruction, using an optical tracking system as a representative example.

\begin{figure}[htbp]
    \centering
    \begin{subfigure}[b]{0.49\textwidth}
        \includegraphics[width=\textwidth]{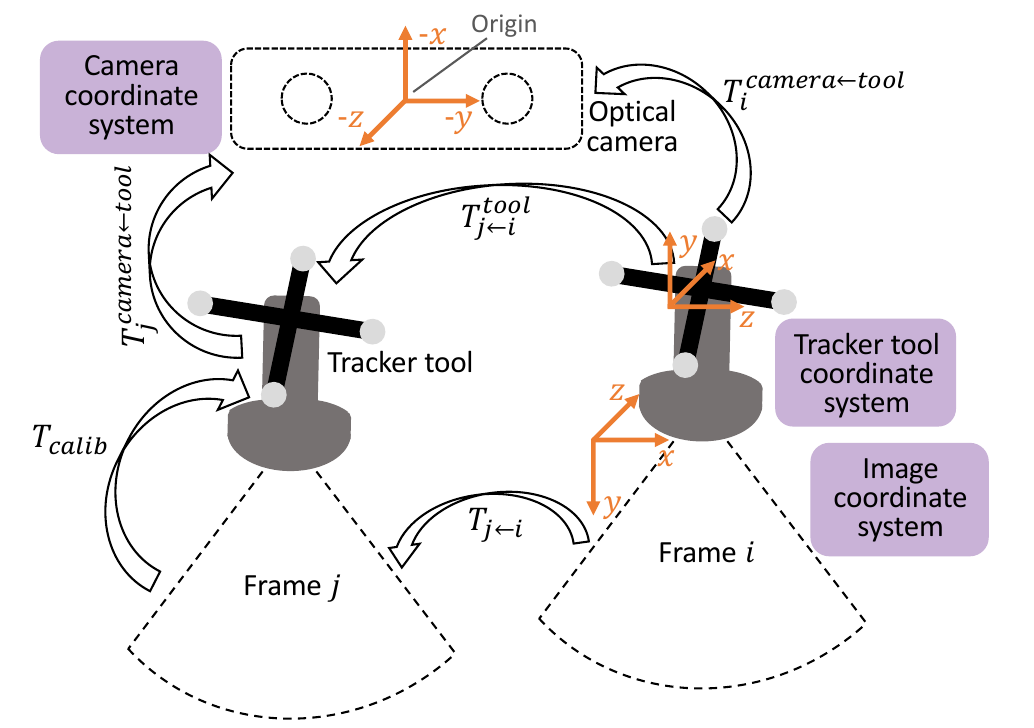}
        \caption{}
        \label{coordinate_systems}
    \end{subfigure}
    \hfill
    \begin{subfigure}[b]{0.49\textwidth}
        \includegraphics[width=\textwidth]{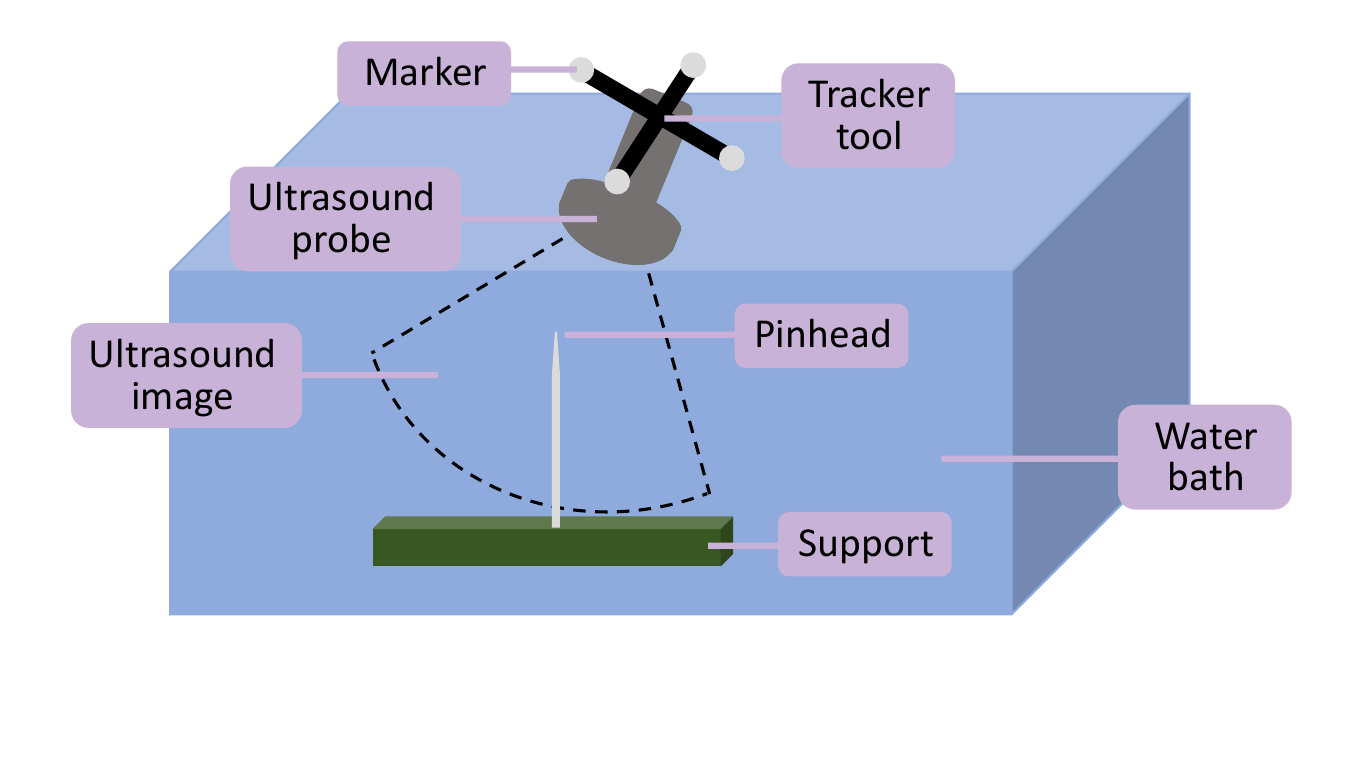}
        \caption{}
        \label{fig_calibration}
    \end{subfigure}
    \caption{(a): Schematic illustration of three coordinate systems: the image coordinate system, the tracker tool coordinate system, and the camera (or world) coordinate system. (b) Schematic illustration of the calibration setup for freehand ultrasound calibration, where an ultrasound probe with an attached tracker tool images a pinhead submerged in a water bath. The pinhead acts as a calibration target, allowing computation of the spatial transformation between the ultrasound image coordinate system and the tracker tool coordinate system.}
    \label{coordinate_systems_calibration}
\end{figure}

As shown in Fig.~\ref{coordinate_systems}, there are three coordinate systems: the image coordinate system, the tracker tool coordinate system, and the camera (or world) coordinate system, as defined in Table~\ref{terminologies}. 
The transformation recorded by the optical tracker is from tracker tool coordinate system to camera coordinate system, which represents the location of the tracking tool in camera coordinate system.
However, this tracker-reported transformation does not directly provide transformation between the coordinate system of the ultrasound image itself and the other two coordinate systems. 
Consequently, a transformation, commonly referred to as the calibration matrix, is necessary to map the ultrasound image coordinate system to the tracker tool coordinate system. It usually incorporates both scaling matrix that converts image coordinate unit from pixels to millimeters, as well as rigid transformation between the image coordinate system (in millimeters) and the tracker tool coordinate system (in millimeters). This transformation is crucial for locating each pixel in the 2D ultrasound image in the 3D space for reconstruction purpose.

\begin{table}[htbp]
\caption{Terminologies in freehand ultrasound reconstruction and notations in this paper.}\label{terminologies}%
\renewcommand{\arraystretch}{1.4} 
\newcolumntype{L}[1]{>{\raggedright\arraybackslash}p{#1}}

\begin{tabularx}{\textwidth}{L{2cm}L{3.5cm}L{3cm}L{1cm}L{4.8cm}}
\toprule
Terminology & Definition &  \multicolumn{1}{p{3cm}}{\raggedright\arraybackslash Example Origin Position} &   Unit & Example Axis Directions \\ 
\midrule
Image coordinate system & A 2D coordinate system defining pixel positions in an image & Top-left corner & pixel & 
\parbox[t]{5cm}{X axis: along the image width, increasing from left to right;\\
Y axis: along the image height, increasing from top to bottom;\\
Z axis: perpendicular to the image plane, increasing into the image.}

 \\

Tracker tool coordinate system &  A 3D coordinate system defined by three or four sphere markers which are attached to a rigid body with a unique geometry &  Origin of the marker attached to the object of interest (phantom, cadaver, patient, etc.) & mm &  As defined by the tracking system / marker manufacturer \\

Camera (or world) coordinate system & A 3D coordinate system defined by the tracking system manufacturer & Origin of the tracking system (midpoint between the two camera lenses) & mm & \parbox[t]{5cm}{X axis: increasing downward from the center between the two lenses;\\
Y axis: increasing toward the camera's right;\\
Z axis: inward, toward the back of the device.}
\vspace{0.1em}
  \\

\bottomrule
\end{tabularx}

\begin{tabularx}{\textwidth}{L{2cm}L{12.3cm}}
\toprule
Notation & Definition \\
\midrule
$T$ & The transformation between two coordinate systems, which changes the coordinate of the same point represented in one coordinate system to another.\\
$T_{scale}$ & The scaling matrix to change the unit of the image coordinate system from pixels to millimeters.\\
$T_{rigid}$ & The transformation from the image coordinate system (in millimeters) to the tracker tool coordinate system (in millimeters).\\
 $T_i^{camera \leftarrow tool}$ & The transformation from the tracker tool coordinate system (in millimeters) of frame $i$ to the camera coordinate system (in millimeters). \\

$T_{j\leftarrow i}^{tool}$ & \multicolumn{1}{p{14cm}}{The transformation from the tracker tool coordinate system (in millimeters) of frame $i$ to that of frame $j$.}     \\ 
$T_{j\leftarrow i}$ & \multicolumn{1}{p{14cm}}{The transformation from image coordinate system (in millimeters) of frame $i$ to that of frame $j$.}     \\

\bottomrule
\end{tabularx}

\end{table}

Let $T_i^{camera \leftarrow tool}$ denote the transformation from the tracker tool coordinate system of frame $i$ to the camera coordinate system, as recorded by the optical tracking system during scanning. Then, the rigid transformation from image coordinate system (in millimeters) of frame $i$ to image coordinate system (in millimeters) of frame $j$, $T_{j\leftarrow i}$, is given by:

\begin{equation}\label{image_mm_to_image_mm}
T_{j\leftarrow i}= T_{rigid}^{-1} \cdot T_{j\leftarrow i}^{tool} \cdot T_{rigid}
\end{equation}
where $T_{j\leftarrow i}^{tool}$ denotes the transformation from the $i^{th}$ tracker tool to the $j^{th}$ tracker tool. $T_{rigid}=\left[ \begin{array}{cc} \mathbf{R_{3\times3}} & \mathbf{t_{3\times1}}  \\ \mathbf{0} & 1 \end{array} \right]$ represents the transformation from the image coordinate system (in millimeters) to the tracker tool coordinate system, where $\mathbf{R}$ is a $3\times3$ rotation matrix and $\mathbf{t}$ is a $3\times1$ translation vector. The transformation $T_{rigid}$ is obtained through the calibration process, as described in Section.~\ref{calibration}, while $T_{j\leftarrow i}^{tool}$ can be computed using Eq.~\eqref{transformation}.

\begin{equation}\label{transformation}
T_{j\leftarrow i}^{tool}=(T_{j}^{camera \leftarrow tool})^{-1}\cdot T_{i}^{camera \leftarrow tool}
\end{equation}

Reconstructing the 3D ultrasound volume requires determining the position of each frame.
Let the first frame serve as the reference frame. If the transformations from each frame to the reference frame are known, the coordinates of all pixels within the scan can be computed using Eq.~\eqref{coordinates}.

\begin{equation}\label{coordinates}
P_{(x,y,z)} = T_{1\leftarrow i} \cdot T_{scale} \cdot p_{(u,v)}
\end{equation}
where $p_{(u,v)}=(u,v,0,1)^T$ and $P_{(x,y,z)}=(x,y,z,1)^T$ represent the homogeneous coordinates of the same point in the image coordinate system of the $i^{th}$ frame (in pixels) and the image coordinate system of the first frame (in millimeters), respectively. $T_{scale}=diag(s_x,s_y,1,1)$ is the diagonal scaling matrix that converts from pixels to millimeters, where  $s_x$ and $s_y$ represent the scaling factors along the $x$ and $y$ axes respectively.

The transformation from the $i^{th}$ frame to the first frame, $T_{1 \leftarrow i}$, can be computed by recursively multiplying the previously estimated relative transformations, as shown in Eq.~\eqref{chain-multiplying}.

\begin{equation}\label{chain-multiplying}
T_{1\leftarrow i}= T_{1\leftarrow 2} \cdot T_{2\leftarrow 3}  \cdots  T_{i-1\leftarrow i} 
\end{equation}
However, if any transformation in Eq.~\eqref{chain-multiplying} has a large error, it may propagate and affect the subsequent trajectory.

\subsection{Calibration}
\label{calibration}

The calibration process in freehand ultrasound reconstruction involves both spatial and temporal components. Temporal calibration synchronises timestamps from the ultrasound machine and the optical tracking system, ensuring that the timestamps of each frame are aligned with those of pose measurements.
This calibration can be performed using the PLUS Toolkit~\cite{lasso2014plus}, as well as other established methods described in the literature~\cite{treece2003high,barratt2001optimisation,barratt2001accuracy,nakamoto2003temporal}. For example, synchronisation can be achieved by identifying the optimal time offset that maximises the correlation between the probe’s motion measured by the tracking system and positions of the bottom of a water bath observed in the image stream.
Spatial calibration is required to determine the transformation between the ultrasound image coordinate system (in pixels) and the tracker tool coordinate system.

In this study, a pinhead based method was employed for spatial calibration (as shown in Fig.~\ref{fig_calibration}). The calibration was performed in a water medium to ensure optimal ultrasound imaging quality. A pinhead sitting under water served as the calibration phantom and was repeatedly imaged in the ultrasound images while the optical tracker simultaneously recorded the corresponding transformation matrices. During data acquisition, the ultrasound probe, equipped with tracking markers, was moved at various angles and distances (typically tens of positions) relative to the pinhead. The pinhead would therefore appear at different locations within the ultrasound image planes. 

The goal of the spatial calibration is to estimate the transformation matrix that can transform 2D points in different ultrasound frames to the same point in the fixed camera coordinate system, as well as to estimate the unknown but fixed location of the pinhead. Let $\{p_{(u_i,v_i)}|i=1,...,n\}$ denote the set of 2D coordinates of the pinhead in the ultrasound image coordinate system, and $P$ the corresponding 3D coordinates in the camera coordinate system. Since the pinhead remains stationary throughout the acquisition, each 2D image point, when transformed into 3D space using the estimated calibration matrix, should correspond to the same 3D location $P$, as shown in Eq.~\eqref{eq:calibration}.

\begin{eqnarray}\label{eq:calibration}
P &=& T_{1}^{camera \leftarrow tool} \cdot T_{rigid} \cdot T_{scale} \cdot p_{(u_1,v_1)}\nonumber\\
P &=& T_{2}^{camera \leftarrow tool} \cdot T_{rigid} \cdot T_{scale} \cdot p_{(u_2,v_2)}\nonumber\\
&\vdots& \nonumber \\
P &=& T_{n}^{camera \leftarrow tool} \cdot T_{rigid} \cdot T_{scale} \cdot p_{(u_n,v_n)}\nonumber\\
\end{eqnarray}
where $\{T_{i}^{camera \leftarrow tool}|i=1,...,n\}$ denotes the transformation matrices corresponding to each 2D image location of the pinhead, from tracker tool coordinate system to camera coordinate system, recorded from optical tracker. 
The complete calibration matrix is expressed as $T_{calib}=T_{rigid}\cdot T_{scale}$. This composition ensures that pixels are first scaled into physical unit, and then mapped into the tracker tool coordinate system using a rigid transformation. 

Specifically, in each ultrasound frame, the 2D pixel location of the pinhead $p_{(u_i,v_i)}$ is manually identified, while its corresponding physical position in the camera coordinate system, denoted as $P$, remains constant but unknown throughout the acquisition. Consequently, in Eq.~\eqref{eq:calibration}, the parameters to be estimated include the scaling factors ($s_x$ and $s_y$), the 6-DoF (three rotation angles and three translation components) comprising the rigid transformation $T_{rigid}$, and $P$. These parameters are jointly estimated using a nonlinear least-squares optimisation algorithm~\cite{mercier2005review}, in which the objective is to minimise the distance between the transformed 3D locations and the estimated fixed 3D location of the pinhead. The optimisation formulation is given as:

\begin{eqnarray}\label{optimisation}
\mathcal{L}&=&\min_{T_{rigid},\,T_{scale},\,P}\,\sum_{i=1}^n\,dist\,(T_{i}^{camera \leftarrow tool} \cdot T_{rigid} \cdot T_{scale} \cdot p_{(u_i,v_i)},P)\nonumber \\
&=&\min_{\mathbf{R},\,\mathbf{t},\,s_x,\,s_y,\,x,y,z}\,\sum_{i=1}^n\,dist\,(T_{i}^{camera \leftarrow tool} \cdot \left[ \begin{array}{cc} \mathbf{R_{3\times3}} & \mathbf{t_{3\times1}}  \\ \mathbf{0} & 1 \end{array} \right] \cdot \left[ \begin{array}{cccc} s_x & 0&0&0  \\ 0 & s_y&0&0\\0&0&1&0\\0&0&0&1 \end{array} \right] \cdot \left[ \begin{array}{cccc} u_i  \\ v_i\\0\\1 \end{array} \right] ,\left[ \begin{array}{cccc} x  \\ y\\z\\1 \end{array} \right])
\end{eqnarray}
where $dist\,(\cdot)$ denotes the Euclidean distance computed between corresponding pairs of transformed 3D coordinates and $P$.

\subsection{Transformation Estimation}
\label{transformation_estimation}

The learning-based trackerless freehand ultrasound reconstruction task can be formulated as a pose regression problem, where the goal is to estimate transformations directly from ultrasound frames. The network architecture can vary in terms of its input configuration (e.g., frame pairs or sequences), output (and its supervisory label) representation (rigid or non-rigid; 6-DoF, $4\times 4$ transformation matrix, or points coordinates), and loss function design (e.g., in Euclidean space or parameter space).

\textit{Input configuration.} 
Given an ultrasound scan, the network can process a sequence of ultrasound frames to estimate the relative transformations between them. A common approach is to use two adjacent frames as input, which can be seen as a special case of the general sequence input.
The 3D ultrasound volume can then be reconstructed by composing these estimated relative transformations in order, as described in Eq.~\eqref{chain-multiplying}. 
As the estimation accuracy is influenced by the inter-frame distance, there are also some studies that estimate the transformation between non-adjacent frames. A limitation of this approach is that it may not provide transformation estimations for all frames in the sequence.

\textit{Supervisory label / Output representation.} Since the transformation from the tracker tool coordinate system to the camera coordinate system, $T_{i}^{camera \leftarrow tool}$, is defined relative to the camera pose, it depends on the external configuration of the camera. This means that scanning the same object from different camera poses produces different transformations (from the tracker tool to the camera coordinate system). As a result, the same set of images can be associated with different supervisory labels, which introduces ambiguity and hinders model generalisation. Therefore, transformations that are invariant to the camera’s pose are generally preferred in this application.
The supervisory label can be designed as the transformation between two tracker tool coordinate systems corresponding to two ultrasound frames, denoted as $T_{j\leftarrow i}^{tool}$. Alternatively, the calibration matrix can be incorporated into labels to express the transformation in millimeters, $T_{j\leftarrow i}$, or in pixels, $T_{scale}^{-1} \cdot T_{j\leftarrow i} \cdot T_{scale}$. It is important to note that all the three supervisory labels, $T_{j\leftarrow i}^{tool}$, $T_{j\leftarrow i}$, and $T_{scale}^{-1} \cdot T_{j\leftarrow i} \cdot T_{scale}$, are rigid transformations. It is crucial to highlight the relationship between the supervisory labels and evaluation metrics discussed in Section~\ref{definition}, where the supervisory labels serve as ground truth for model training, and the evaluation metrics are employed to assess model performance. 

Regardless of the form of labels being used, they can be represented as 6-DoF vector, consisting of three rotation and three translation parameters, or 7-DoF vector, comprising four parameters for quaternions and three translations. 
Additionally, point coordinates derived from Eq.~\eqref{coordinates} can also be used as supervisory labels. 

\textit{Loss functions.}  Depending on the form of supervision label and network output, loss function can vary but in essence should characterise the difference between the output prediction and label. In addition, it is worth noting that the representation of the network output can be similar to, but does not need to match, the format of the labels used during training. For example, if the ground truth is provided as a $4\times4$ transformation matrix and the network predicts a 6-DoF vector, the loss can be computed based on the difference between the predicted and ground truth 6-DoF parameters, where the latter are derived from the transformation matrix. Alternatively, the loss can be defined as the point-wise Euclidean distance between points transformed by the ground truth matrix and those transformed by the prediction.

\subsection{Three-dimensional Reconstruction}
\label{3D_reconstruction}

To reconstruct the volume for the entire scan, the relative transformation between each frame and a reference frame must be known. This can be achieved either by directly estimating the relative transformations with respect to the reference frame, or by estimating the transformations between pairs of consecutive frames (or non-adjacent frame pairs) and composing them sequentially. 
The reconstruction can be considered complete once all frame positions are estimated in a common reference coordinate system. 
Although interpolating scattered pixel intensities onto a regular voxel grid is useful in some applications, it is not essential for other clinical applications and falls outside the scope of this Challenge.

\section{Related Work}
\label{Related_work}

Trackerless freehand ultrasound reconstruction has evolved rapidly over the past two decades, in particular, driven by advances in computer vision, machine learning, and hardware. This section is organised in two parts. The first part provides an overview of both non-deep learning and deep learning-based approaches, analysing representative studies to track the methodological evolution of the field. The second part offers a comparative analysis of deep learning-based methods, examining them across key dimensions, including input, network architecture, output parameterisation, supervision strategies, datasets, and evaluation metrics. More details can also be found in Tables ~\ref{related_work_table1} and \ref{related_work_table2} in~\ref{related_work_app} and the Supplementary Material\footnote{\url{https://github-pages.ucl.ac.uk/tus-rec-challenge/TUS-REC2024/img/supplementary.xlsx}}.

\subsection{Overview}

\subsubsection{Non-Learning-Based Methods}

Before the era of deep learning, trackerless freehand ultrasound reconstruction was primarily addressed using classical signal-processing and correlation-based techniques \cite{chen1997determination,prager2003sensorless}. Most approaches exploited speckle decorrelation to infer elevational motion \cite{chang20033,gee_2006_SD_3,housden2008rotational,liang2012feature,6615950}. Validation was typically performed on phantoms \cite{chen1997determination,prager2003sensorless,gee_2006_SD_3,housden2006subsample,housden2007sensorless,housden2008rotational,liang2012feature,tetrel2016learning,harindranath2024imu,dai2024advancing}, simulations \cite{housden2008rotational,liang2012feature,dai2024advancing}, or small \textit{ex vivo} datasets \cite{gee_2006_SD_3,housden2006sensorless,housden2007sensorless,6615950}. Ground truth was usually provided by motor stages \cite{6615950,tetrel2016learning}, dial gauges \cite{gee_2006_SD_3,housden2006subsample}, or external optical trackers \cite{prager2003sensorless,housden2006sensorless,housden2007sensorless,6615950,tetrel2016learning}. Evaluation was based on accumulated offset errors \cite{housden2006sensorless,housden2006subsample,housden2007sensorless,housden2008rotational,6615950}, displacement or angle estimation \cite{gee_2006_SD_3,6615950}.

Chen et al. \cite{chen1997determination} proposed an approach to estimate scan-plane motion in ultrasound by analysing the rate of change in the correlation of echo signal intensities across B-mode images.
Prager et al. \cite{prager2003sensorless} proposed a sensor-free method for estimating probe trajectory in freehand 3D ultrasound using linear regression of echo-envelope intensity signals, derived from a probabilistic speckle analysis, where the regression gradient directly relates to probe motion.
Chang et al. \cite{chang20033} first derived a coarse motion estimate from speckle decorrelation, then refined the pose via image registration between the 3D ultrasound datasets and an orthogonal reference image to achieve more accurate alignment. 
Gee et al. \cite{gee_2006_SD_3} proposed a heuristic technique to quantify the amount of coherency at each point in the B-scans, enabling an adapted elevational decorrelation scheme.
Housden et al. \cite{housden2006sensorless,housden2007sensorless} presented reconstruction methods that handled unconstrained freehand sweeps such as irregular frame spacing, non-monotonic out-of-plane motion, and significant in-plane motion, allowing estimation of more complex motion patterns.
The follow-up work \cite{housden2006subsample} introduced a novel interpolation strategy for trackerless freehand 3D ultrasound and demonstrated its robustness to simultaneous lateral and elevational probe motion.
They later analysed the effect of probe rotation on the decorrelation curve and proposed a method to correct these curves using measurements from the orientation sensor for improved reconstruction \cite{housden2008rotational}.
Liang et al. \cite{liang2012feature} analysed two types of speckle pattern variations, the geometric transformation and the intensity change of speckle patterns, and demonstrated that a coupled filtering method can compensate for both types to provide accurate strain estimation under large tissue deformation or rotation.
Most methods above were based on fully developed speckle (FDS) which rarely existed in real tissues. \cite{6615950} developed methods that could use statistics of non-FDS to estimate probe motion. They combined a closed-form derivation with a linear regression model of the ultrasound beam profile, and applies second-order statistics of the Rician-Inverse Gaussian model to improve reliability and flexibility in speckle tracking.
Tetrel et al. \cite{tetrel2016learning} reduced error accumulation by modeling frames and motion measurements as a graph, generating random trajectories via constrained shortest paths, and refining estimates with edge weights predicted by a Gaussian process regressor.
Ito et al. \cite{ito2017probe} presented a probe-camera system for 3D ultrasound reconstruction, using phantom-based calibration and structure from motion (SfM) for probe localisation and camera motion estimation.
Balakrishnan et al. \cite{balakrishnan2019novel} proposed a similarity metric that correlated the parametric representations of image texture between consecutive ultrasound images, modeling texture dynamics with a parametrical auto-regressive model, and estimated out-of-plane motion using a trained fine Gaussian SVM regression model.
Harindranath et al. \cite{harindranath2024imu} presented an affordable IMU-assisted manual 3D ultrasound scanner, combining a consumer-grade IMU with Kalman filter-based orientation estimation and a scanline-based reconstruction method.
More recently, Dai et al. \cite{dai2024advancing} developed a novel coupling pad with 3 N-shaped lines to provide 3D spatial information without external tracking devices and introduced a coarse-to-fine optimisation method that refined sequential 2D ultrasound image poses via distance-topology discrepancy reduction.

Taken together, these classical methods established the feasibility of trackerless freehand ultrasound reconstruction and demonstrated that speckle statistics encode useful elevational motion cues. However, their accuracy in predicting out-of-plane motion is still limited. Moreover, their reliance on speckle decorrelation made them highly sensitive to noise and tissue variability. These constraints ultimately motivated the transition toward learning-based and hybrid approaches that now dominate the field.

\subsubsection{Learning-Based Methods}

Early CNN-based approaches demonstrated that learned image representations could outperform traditional speckle-based techniques in estimating inter-frame motion \cite{prevost2017deep,prevost20183d}. 
Subsequent works expanded on this by incorporating spatial and temporal modeling through Long Short-Term Memory (LSTM) networks, enabling more stable trajectory estimation over longer scan sequences \cite{luo2023recon,li2023long}.

Guo et al. \cite{guo2020sensorless} proposed a deep contextual network that applied 3D convolutions to ultrasound video segments with a self‑attention module to emphasise speckle‑rich regions and a case‑wise correlation loss to stabilise training.
Afterwards, they expanded the work by introducing contrastive learning strategy \cite{guo2022ultrasound}.
Miura et al. \cite{miura2020localizing} presented a CNN with feature extraction and motion estimation components, along with a consistency loss. 
In one subsequent work, they introduced two loss functions to guarantee the consistency of forward and backward motion of the probe \cite{miura2021probe}.
Afterwards, they integrated CNN and RNN to exploit long‑term temporal context, predicting both relative and absolute probe poses directly from image sequences \cite{miura2021pose}.
Yeung et al. \cite{yeung2021learning} utilised a CNN to predict the position of 2D ultrasound fetal brain scans in 3D atlas space with self-supervised learning strategy along with an attention module. 
Building on that, they proposed a framework for localising 2D ultrasound images in a 3D anatomical atlas, trained on co-aligned 3D volumes and fine-tuned on freehand scans using an unsupervised cycle-consistency constraint \cite{yeung2022adaptive}.
Luo et al. \cite{luo2021self} proposed a recurrent convolutional LSTM based online learning framework via a differentiable reconstruction algorithm, a self‑supervised learning method that exploits contextual cues, and adversarial training for anatomical shape prior learning, improving robustness to complex scan sequences. They further extended this work by introducing path-level supervision and a motion-weighted training loss \cite{luo2023recon}.
Leblanc et al. \cite{leblanc2022stretched} combined segmentation based on Mask RCNN, in-plane registration, and CNN-predicted out-of-plane translation to generate a 3D stretched reconstruction of the femoral artery.
Di Vece et al. \cite{di2022deep} used a regression CNN to estimate the ultrasound plane poses in obstetric imaging. 
Chen et al. \cite{chen2023freehand} employed a 3D CNN-LSTM to estimate pose, using as input both original ultrasound frames and frames generated through Bezier interpolation and speckle decorrelation.
To investigate the influence of past and future frames in trackerless freehand ultrasound reconstruction, Li et al. \cite{li2023trackerless} proposed a multi-task learning algorithm which utilised a large number of auxiliary transformation-predicting tasks. Then, they explored the impact of long-term dependencies on reconstruction performance, demonstrating how sequence length, anatomical content, and scanning protocol influence reconstruction quality, providing insights for optimising training data collection, scanning procedures, and network design \cite{li2023long}. They further introduced a new multi-task learning framework, which leveraged anatomical and protocol information as privileged inputs and optimised the branching location of these auxiliary tasks via a differentiable architecture \cite{li2023privileged}.
Most recently, Ramesh et al. \cite{ramesh2024geometric} proposed an uncertainty-aware deep learning model for 3D plane localisation in 2D fetal brain images, using a multi-head network that predicted different geometric transformations and uncertainty.

In addition, those learning approaches that predicted the 6-DoF parameters treated rotation as simple Euclidean vectors and used $L2$-norm as the loss function, ignoring the non-linear manifold structure of Lie group $SE(3)$.
To address this problem, \cite{hou2018computing} presented a Riemannian formulation for pose estimation, training CNNs on the $SE(3)$ manifold with a left-invariant Riemannian metric and using geodesic distance as the loss to couple translation and rotation.

Transformer- and Mamba \cite{gu2023mamba}-based architectures have recently been introduced to better model long-range dependencies and spatial coherence. 
For example, Ning et al. \cite{ning2022spatial} leveraged transformer for performing regression tasks on the sequence, within a joint local and global information encoding approach.
Yan et al. \cite{yan2024fine} utilised a multi-directional state space model (SSM) for extracting multi-scale spatio-temporal information and fused the auxiliary information from multiple IMUs to enhance spatio-temporal perception, along with an online alignment strategy to further improve reconstruction performance.
Sun et al. \cite{sun2025ultrasom} developed a Mamba-based spatio-temporal attention module, integrated with optical flow, to capture global spatio-temporal correlations.

Recent approaches integrated optical flow and spatio-temporal attention \cite{guo2020sensorless,xie2021image} to better capture dense inter-frame motion and global spatio-temporal consistency \cite{sun2025ultrasom}.
Xie et al. \cite{xie2021image} fused image and optical flow features via a pyramid warping layer with channel and spatial attention.
Mikaeili et al. \cite{mikaeili2022trajectory} combined a densely connected network \cite{huang2017densely} and FlowNet \cite{dosovitskiy2015flownet} for pose estimation.
Based on a Siamese architecture, El hadramy et al. \cite{el2023trackerless} leveraged RNN and optical flow for pose estimation.
To better leverage the contextual cues between frames, speckle decorrelation was incorporated as a physics-based constraint \cite{dou2023sensorless}.
Furthermore, they utilised a two-stream model to separately extract spatial and temporal information, in addition to the incorporation of speckle decorrelation \cite{10684746}.
Most recently, Lee et al. \cite{lee2025enhancing} introduced a motion-based learning network with a global-local self-attention module, incorporating correlation features, global-local attention, and a motion-aware supervision strategy to enhance motion estimation.

Several works have investigated sensor signals and auxiliary tracking modalities to enhance pose estimation accuracy. IMUs have been integrated with learning-based systems to provide motion information or correct drift \cite{prevost20183d}.
Luo et al. \cite{luo2022deep} introduced IMU for estimating elevational displacements outside the plane with an online self-supervised strategy. 
They further proposed an online self-consistency network with multiple IMUs for improving reconstruction performance, along with a modal-level self-supervised strategy \cite{luo2023multi}.
They then expanded \cite{luo2022deep} by adding scan-level, path-level, and patch-level consistency \cite{luo2025monetv2}.

Recent research in trackerless freehand 3D ultrasound reconstruction has pursued multiple directions to improve motion estimation accuracy. One of the promising direction is the fusion of deep learning-based trajectory estimation with volumetric consistency optimisation. Several methods combine learning-based inter-frame pose estimation with multi-view or global model refinement to improve alignment and robustness across long scan sequences. For example, Wein et al. \cite{wein2020three} proposed a pipeline that reconstructed 3D volumes from two freehand ultrasound sweeps (transverse and sagittal) using deep‑learning trajectory estimation and image‑based 3D model optimisation.
Li et al. \cite{li2024nonrigid} proposed a co‑optimisation framework that jointly estimated rigid and non‑rigid transformations, with a fast scattered‑data interpolation approach. 
In addition, point cloud-based registration technique has been used in \cite{grossbrohmer20253d}. 

A notable recent development involves the application of implicit neural representations to ultrasound imaging. These methods aim to compress the volumetric information and encode it as parameters of a model, offering potential gains in memory efficiency and spatial resolution compared with explicit representation.
Yeung et al. \cite{yeung2021implicitvol,yeung2024sensorless} introduced a model to represent the 3D volume implicitly as a continuous function, while jointly refining scan locations.
Wysocki et al. \cite{wysocki2024ultra} proposed a physics-enhanced implicit neural representation for ultrasound imaging that leveraged ray-tracing-based neural rendering, enabling more faithful synthesis of geometrically accurate B-mode images from overlapping sweeps.
Gaits et al. \cite{gaits2024ultrasound} formulated the ultrasound volume reconstruction problem as the optimisation of a 3D function, parameterised by a deep neural network.
Dou et al. \cite{dou2024pitfalls} addressed pixel mismatches in freehand ultrasound Neural radiance field (NeRF) caused by transducer pressure variations, introducing a modified positional encoding that improved 3D representation learning. In a later work, they proposed a probabilistic and differentiable method that used a continuous Bernoulli distribution to model scatterer distributions and backscattered pixel intensities, enabling more realistic 3D ultrasound reconstruction with NeRF \cite{dou2024continuous}.
Eid et al. \cite{eid2025rapidvol} used tensor-rank decomposition to speed up slice-to-volume ultrasound reconstruction for fetal brain images.

Additionally, an evaluation of low-cost tracking alternatives has shown no statistically significant difference between high and low-end optical trackers, further supporting the feasibility of cost-effective freehand 3D ultrasound setups~\cite{leger2021evaluation}.
Another active area focuses on addressing domain shifts introduced by different ultrasound transducers. Domain adaptation strategies have been proposed to improve cross-device generalisation, particularly through the learning of transducer-invariant features \cite{guo2021transducer}.

\subsection{Analysis of Learning-Based Methods}

\subsubsection{Algorithm Inputs}

Most approaches operate on B-mode images, typically consisting of short frame sequences of 3-10 images \cite{guo2020sensorless,guo2021transducer,guo2022ultrasound,mikaeili2022trajectory,chen2023freehand,el2023trackerless,10684746,lee2025enhancing,sun2025ultrasom}. Short sequences provide useful local spatiotemporal context, but they alone cannot guarantee globally consistent reconstructions, leading to drift over long sweeps. A special case is using pairwise frames as input \cite{prevost2017deep,prevost20183d,miura2020localizing,miura2021probe,xie2021image,leblanc2022stretched,dou2023sensorless}, which reduces computational demand but makes the system more vulnerable to noise, as the model lacks additional spatial and temporal context to stabilise motion estimation. Other studies utilise longer frame sequences \cite{miura2021pose,li2023trackerless,li2023privileged,li2023long,li2024nonrigid}, allowing the network to exploit richer spatiotemporal context that can stabilise motion estimation over longer sweeps. Several studies even attempt to process an entire scan or very long subsequences in a single forward pass \cite{luo2021self,luo2022deep,luo2023recon,luo2023multi,yan2024fine,luo2025monetv2}, providing full spatial and temporal context across the sweep. This strategy, however, substantially increases computational demands.
In addition to image data, a few works incorporate auxiliary modalities, for instance IMU measurements \cite{prevost20183d,ning2022spatial,luo2022deep,luo2023multi,yan2024fine,grossbrohmer20253d,luo2025monetv2}, which supply translational and angular acceleration measures beyond raw B-mode intensity.

From a methodological perspective, different input sequence lengths entail distinct trade-offs. 
Frame-pair inputs save computational cost but lack global regularisation. 
Short subsequences (e.g., 3-10 frames) improve local robustness by leveraging limited spatial and temporal context. Longer clips or full-scan inputs provide richer spatiotemporal cues and can mitigate local ambiguities, yet impose additional memory burdens.
In addition, current models almost universally operate on preprocessed B-mode images, whereas richer physics-level information such as raw radio-frequency data \cite{zhang2021ultrasound} or probe-specific beam geometry \cite{goudarzi2022deep} could improve robustness to ultrasound’s inherent variability across scanners, probes, and patients. Integrating such physics-informed features remains an underexplored but potential direction for trackerless freehand ultrasound reconstruction.

\subsubsection{Secondary Information}

The majority of works use only raw B-mode frames without additional derived features. Among those that incorporate secondary information, optical flow is the most frequently employed \cite{prevost2017deep,prevost20183d,wein2020three,miura2020localizing,miura2021probe,miura2021pose,xie2021image,luo2023recon,el2023trackerless,sun2025ultrasom}, providing explicit motion cues but introducing potential domain-shift issues, since most existing optical flow models are trained on natural images and may not transfer reliably to ultrasound images. In the context of slice-to-volume reconstruction with implicit neural representations, positional encoding has been employed to map slice coordinates into a higher-dimensional feature space \cite{yeung2021implicitvol,ning2022spatial,yeung2024sensorless,gaits2024ultrasound,dou2024pitfalls,eid2025rapidvol,sun2025ultrasom}. Bezier interpolation has been applied among input frames to achieve frame rate consistent with that of external tracking devices \cite{chen2023freehand}. Edge-based features such as canny edge maps highlight structural boundaries that can stabilise motion estimation \cite{luo2023recon}. Geometry-based signals have also been integrated such as bone surface point clouds and gradient keypoints \cite{grossbrohmer20253d}.

These strategies are based on the assumption that auxiliary cues can inject domain knowledge and structural priors that are difficult to infer from intensity data alone. While raw-frame inputs remain the most practical and widely adopted, the selective use of physics-informed or geometry-aware secondary information has shown potential to improve stability and accuracy, particularly in anatomies with strong structural boundaries. Future work may benefit from systematically combining such cues with learned representations.

\subsubsection{Network Architectures and Adaptive Reconstruction}

CNN backbones dominate the field \cite{prevost2017deep,prevost20183d,wein2020three,miura2020localizing,miura2021probe,guo2021transducer,yeung2021learning,xie2021image,di2022deep,mikaeili2022trajectory,yeung2022adaptive,li2023trackerless,li2023privileged,li2023long,dou2023sensorless,10684746,ramesh2024geometric}, which may be due to their stability and efficiency. A number of studies augment CNNs with temporal modules to exploit sequential dependencies, often modeled with recurrent architectures such as ConvLSTMs \cite{luo2021self,miura2021pose,luo2022deep,chen2023freehand,luo2023recon,li2023trackerless,luo2023multi,el2023trackerless,li2023long,luo2025monetv2,lee2025enhancing}.
Some methods apply 3D convolutions to stacked frame sequences, effectively treating them as spatiotemporal volumes \cite{guo2020sensorless,guo2022ultrasound}.
More recent studies adopt transformer-based architectures to capture long-range dependencies globally \cite{ning2022spatial}.
Several works also incorporate attention mechanisms to enhance feature representation within their networks \cite{yeung2021learning,miura2021pose,xie2021image,guo2022ultrasound,lee2025enhancing,sun2025ultrasom}.

Beyond transformers, SSM such as Mamba \cite{gu2023mamba} have recently emerged as an alternative for sequence modeling. These architectures offer linear-time complexity with respect to sequence length, enabling efficient representation of long-range temporal dependencies. Very recently, two studies have adapted Mamba to trackerless freehand ultrasound reconstruction. One study employs Mamba to capture fine-grained temporal dependencies while integrating multi-modal alignment \cite{yan2024fine}. Another integrates Mamba as a spatiotemporal attention module within its reconstruction framework, enabling the capture of global motion dependencies beyond conventional optical-flow features \cite{sun2025ultrasom}.
Although these works remain early explorations, they highlight the potential of SSMs to balance scalability with robust temporal modeling in ultrasound reconstruction.

A smaller number of studies explore test time computing strategies \cite{luo2021self,luo2022deep,luo2023recon,luo2023multi,yan2024fine,luo2025monetv2}, in which the model parameters are updated during inference as new frames are processed. These approaches apply self-supervised consistency objectives directly at test time to refine predictions and improve reconstruction accuracy. While this introduces additional computational burden, it demonstrates the potential of online optimisation to enhance performance on unseen data, and thus represents a promising direction toward clinically robust deployment.

Several methods integrate deep learning with registration refinement, where predicted transformations are further optimised using image-based similarity metrics. \cite{wein2020three} adopts a classical refinement strategy, combining the efficiency of learning-based motion estimation with the precision of registration. More recently, registration refinement has been adapted to simultaneously estimate rigid transformations and nonrigid deformations, thereby improving the global consistency of the reconstructed volume \cite{li2024nonrigid}. Finally, some works explore reference-based point-cloud registration, where point clouds from freehand sweeps are aligned against a reference model to achieve global consistency \cite{grossbrohmer20253d}.

An emerging line of work explores implicit neural representations \cite{yeung2021implicitvol,wysocki2024ultra,yeung2024sensorless,gaits2024ultrasound,dou2024pitfalls,dou2024continuous,eid2025rapidvol}. By learning a continuous volumetric field that jointly encodes all frames, these approaches allow reconstructing high quality images from sparse data. However, they require substantial computational resources due to large amount of neural network inference for volumetric sampling. 

Overall, CNNs with temporal extensions remain the practical default, but the literature is trending toward hybrid designs that integrate attention mechanism, state space modeling, implicit representations, or registration, either as refinement or as a primary strategy, aiming to balance robustness, accuracy, and deployability.

\subsubsection{Network Outputs}

Most surveyed methods regress rigid 6-DoF transformation parameters between adjacent frames. For networks that take two frames as input, the output is a single relative transformation \cite{prevost2017deep,prevost20183d,wein2020three,miura2020localizing,miura2021probe,xie2021image}, whereas sequence-based models typically estimate a series of consecutive inter-frame transformations \cite{luo2021self,miura2021pose,ning2022spatial,luo2022deep,mikaeili2022trajectory,luo2023recon,luo2023multi,yan2024fine,luo2025monetv2,lee2025enhancing,sun2025ultrasom}.
Related extensions are explored in \cite{li2023trackerless,li2023privileged,li2023long}, where the network not only regresses transformations between adjacent frames but also predicts interval transformations spanning multiple frames. These objectives supply denser supervision across varying temporal gaps and encourage the model to learn more robust motion representations, improving training stability and generalisation.

As an alternative, a few studies attempt to directly predict an absolute trajectory from current frame to the reference frame \cite{miura2021pose,li2024nonrigid}, but this formulation is inherently challenging, as predictions for long sequences tend to have large errors.
Another line of work provides a different design in which a sequence of frames is provided as input but only a single transformation is predicted, with ground truth labels defined as the mean of the relative transformations between the neighboring frames \cite{guo2020sensorless,guo2021transducer,guo2022ultrasound}.
In addition, \cite{li2024nonrigid} jointly estimates global rigid transformations and a dense displacement field, offering greater flexibility than purely rigid regression.

Another formulation is introduced in \cite{yeung2021learning,yeung2022adaptive}, where the output is defined as the 3D Cartesian coordinates of key points representing each imaging plane.
A more comprehensive formulation is proposed in \cite{ramesh2024geometric}, which outputs multiple pose parameterisations simultaneously (quaternions, axis-angles, Euler angles, rotation matrices, along with translation displacements and scaling factors, and direct landmark positions), along with associated variances. 
More recently, some works also predict volumetric fields using implicit representations \cite{yeung2021implicitvol,wysocki2024ultra,yeung2024sensorless,gaits2024ultrasound,dou2024pitfalls,dou2024continuous,eid2025rapidvol}. 

Taken together, these studies present a wide spectrum of output formulations for trackerless freehand ultrasound reconstruction. Rigid inter-frame regression remains most common and is valued for its simplicity. Nonrigid displacement fields provide greater flexibility at the cost of dimensionality and supervision complexity. Landmark-based parameterisations and multi-representation schemes offer alternative ways to encode motion and quantify uncertainty. Implicit volumetric fields provide additional information of the global anatomy. This diversity underscores an ongoing trade-off between efficiency, robustness, and clinical applicability, with no single output design yet achieving a definitive balance.

\subsubsection{Loss Functions -- Fidelity}

The predominant supervision strategy in trackerless freehand ultrasound reconstruction is mean squared error (MSE) or $L2$ loss, applied directly to rigid transformation parameters \cite{prevost2017deep,prevost20183d,guo2020sensorless,wein2020three,miura2020localizing,miura2021probe,guo2021transducer,miura2021pose,ning2022spatial,di2022deep,guo2022ultrasound,chen2023freehand,el2023trackerless,dou2023sensorless,10684746,sun2025ultrasom}. Alternatively, the loss may be defined on spatial coordinates, obtained either directly from the network output \cite{yeung2021learning,yeung2022adaptive} or derived from the predicted transformations \cite{li2023trackerless,li2023privileged,li2023long,li2024nonrigid}. 
A number of studies employ $L1$ or mean absolute error (MAE) losses \cite{luo2021self,xie2021image,luo2022deep,luo2023recon,luo2023multi,yan2024fine,luo2025monetv2,lee2025enhancing} on transformations or point coordinates.

Another widely used formulation is case-wise correlation loss \cite{guo2020sensorless,luo2021self,guo2022ultrasound,luo2022deep,luo2023recon,luo2023multi,10684746,yan2024fine,luo2025monetv2,lee2025enhancing,sun2025ultrasom}, which is based on the Pearson correlation coefficient between the predicted and ground-truth trajectories. This can enforce consistency of motion estimates across frames within a sequence and thus encourages globally coherent reconstructions. 

Several works employ reprojection-based objectives \cite{miura2020localizing}. Extensions of this idea incorporate forward consistency losses \cite{miura2021probe,miura2021pose,li2023trackerless,el2023trackerless,yeung2022adaptive}, which penalise the Euclidean distance between points projected forward through successive transformations and their ground truth positions, and backward consistency losses \cite{miura2021probe,miura2021pose}, which further reproject points back to the original frame to enforce bidirectional alignment. Together, these objectives encourage temporal coherence and mitigate drift.

A subset of works evaluate the fidelity of reconstructed anatomy based on pixel intensities \cite{gaits2024ultrasound,li2024nonrigid,dou2024pitfalls,dou2024continuous}. Examples include image similarity metrics for instance structural similarity index measure (SSIM) \cite{yeung2021implicitvol,wysocki2024ultra,yeung2024sensorless,eid2025rapidvol}. These objectives promote anatomically faithful reconstructions, though their use is constrained by the computational burden for example volumetric rendering.

Recent work has also explored additional objectives at inference time, refining predictions by enforcing consistency with auxiliary signals such as IMU measurements, reconstructed slices, or reference scans. One line of work incorporates Pearson correlation or MAE losses between estimated motion parameters and IMU-derived data such as acceleration or Euler angle \cite{luo2022deep,luo2023multi,yan2024fine,luo2025monetv2}. Another strategy applies self-supervised objectives that compare input frames to slices extracted from the reconstructed volume at the estimated positions \cite{luo2021self,luo2023recon}. A third example is the path-level constraints, in which a prior training scan with ground-truth transformations most similar to the current test sequence is selected, and the network is optimised by minimising MAE and Pearson correlation losses between the training scan’s estimated and ground-truth transformations \cite{luo2023recon}. These inference-time objectives improve performance but increase computational cost and may be sensitive to sensor noise, sequence selection, or slice reconstruction quality.

To summarise, most methods use $L1$/$L2$ losses for simplicity, but often lack regularisation to enforce temporal coherence.
Complementary strategies, such as case-wise correlation and reprojection-based consistency losses, address this gap by enforcing temporal coherence and reducing drift. Volumetric losses connect supervision more directly to clinically meaningful outcomes by promoting anatomy-preserving reconstructions. Finally, inference-time consistency losses represent an emerging paradigm that leverages auxiliary signals, such as IMU data or reconstructed slices, to refine models online. These strategies highlight a shift towards hybrid and adaptive supervision, where $L1$/$L2$ losses are complemented by temporal, volumetric, and multimodal constraints.

\subsubsection{Loss Functions -- Regularisation}

Several works enforce motion consistency within a sequence, for example by constraining that composed transformations remain consistent with the corresponding direct predictions \cite{li2023trackerless,luo2025monetv2}.
\cite{luo2023multi,luo2025monetv2} sample subsequences from the original sequence and estimate the inter-frame transformations within each subsequence, either directly using the subsequence as input, or indirectly by first predicting transformations over the entire sequence and then extracting those corresponding to the subsequence. A consistency constraint loss is introduced to enforce agreement between the two sets of predictions.

Some works explicitly link motion with anatomical content. For example, motion-weighted regularisation \cite{luo2023recon} enforces a positive correlation between probe motion speed and appearance variance. 
Additionally, \cite{luo2025monetv2} penalises discrepancies between normalised image content differences and the normalised estimated distance.
Others enforce smoothness by using a bending-energy term \cite{li2024nonrigid} or penalise large displacements using $L1$ loss \cite{grossbrohmer20253d}. 

Several studies employ representation-based objectives that operate in feature space, including discrepancy \cite{guo2021transducer}, margin ranking loss and its variants \cite{guo2022ultrasound,10684746,lee2025enhancing}. 
Discrepancy loss encourages the network to learn domain-robust feature embeddings by penalising feature discrepancies from different domains. Margin ranking loss links representation similarity to transformation consistency, by enforcing a minimum separation margin between positive and negative pairs in latent space.

Adversarial objectives are also introduced, including global shape priors that encourage reconstructions to resemble distributions of real volumes \cite{luo2021self,luo2023recon}.

A recent work incorporates IMU-based regularisation, using auxiliary sensor signals to stabilise motion estimation. \cite{luo2023multi} enforces agreement across predictions obtained with signals from different IMUs for the same scan. Rather than relying on ground truth, these objectives encourage the network to produce similar estimates from different sensor inputs. Such multi-IMU consistency improves robustness by reducing sensor-specific noise, but introduces additional hardware requirements and calibration challenges.

Compared to fidelity terms, regularisation losses in trackerless freehand ultrasound reconstruction are markedly more diverse, reflecting attempts to address drift and instability through complementary constraints. Temporal consistency objectives offer lightweight yet effective mechanisms to mitigate long-distance prediction error. Physics-based deformation priors and adversarial shape constraints point towards more anatomically plausible reconstructions. 
Representation-based objectives encourage discriminative, domain-robust feature embeddings, offering promising avenues for cross-domain generalisation.
Multi-modal constraints, particularly those leveraging IMU signals, highlight the potential of sensor fusion but are limited by clinical practicality. Overall, a major research opportunity lies in combining lightweight consistency terms with physics-aware or multimodal priors to improve robustness and generalisability without prohibitive annotation cost.

\subsubsection{Datasets and Clinical Applications}

The datasets span a broad range of anatomies, including peripheral limbs (forearm or arm \cite{prevost2017deep,prevost20183d,miura2020localizing,miura2021probe,miura2021pose,xie2021image,ning2022spatial,luo2022deep,chen2023freehand,li2023trackerless,li2023privileged,luo2023multi,li2023long,10684746,yan2024fine,li2024nonrigid,grossbrohmer20253d,luo2025monetv2,lee2025enhancing,sun2025ultrasom} and lower leg \cite{prevost2017deep,prevost20183d}), prostate \cite{guo2020sensorless,guo2021transducer,guo2022ultrasound,dou2023sensorless}, thyroid \cite{wein2020three,luo2025monetv2}, liver \cite{el2023trackerless}, carotid \cite{prevost20183d,luo2022deep,luo2023multi,yan2024fine,luo2025monetv2}, fetal imaging \cite{yeung2021learning,luo2021self,yeung2021implicitvol,di2022deep,yeung2022adaptive,luo2023recon,yeung2024sensorless,ramesh2024geometric,eid2025rapidvol}, spine \cite{luo2023recon} and hip \cite{luo2021self,luo2023recon}.
Among these, the arm and forearm datasets are by far the most widely used, owing to their accessibility, repeatability, and suitability for controlled evaluation. In addition, numerous studies rely on phantom data (e.g., BluePhantom ultrasound biopsy phantom \cite{prevost2017deep,prevost20183d}, breast \cite{miura2020localizing,miura2021probe,miura2021pose}, hypogastric \cite{miura2020localizing,miura2021probe,miura2021pose}, abdominal \cite{10684746,dou2024pitfalls} and lumbar spine phantoms \cite{wysocki2024ultra}) as well as simulated datasets \cite{wysocki2024ultra,gaits2024ultrasound,dou2024continuous}.

The size of datasets varies substantially across studies. At one extreme, some studies utilise only a handful of phantom or \textit{ex vivo} scans, whereas others report relatively large collections comprising hundreds of sweeps or hundreds of thousands of frames. For instance, \cite{el2023trackerless} includes six scans from an \textit{ex vivo} swine liver, while the prostate dataset in \cite{dou2023sensorless}  has more than 1,900 transrectal ultrasound scans. In several datasets, 2D slices are generated from 3D ultrasound volumes, enabling large-scale training. Examples include fetal brain datasets \cite{yeung2021learning}, where slice-sampling yields more than 190,000 frames, and 3D hip volumes \cite{luo2021self,luo2023recon}, where complex scans are simulated by combining various scanning trajectories such as loop, fast-and-slow, and sector trajectories.

The number of human participants also varies widely across studies. Many datasets are relatively small, involving only a handful of volunteers (2-6) \cite{miura2020localizing,miura2021pose} or modest groups of 10-20 participants \cite{prevost2017deep,prevost20183d,luo2021self,leblanc2022stretched,li2023trackerless,li2023privileged,li2023long,10684746,lee2025enhancing,sun2025ultrasom}. Some medium-scale collections with 40-80 subjects have been reported in carotid \cite{luo2022deep} and forearm datasets \cite{li2024nonrigid}, often comprising several hundred sweeps with diverse probe trajectories. Larger clinical datasets remain rare but notable: prostate datasets in \cite{guo2020sensorless,guo2022ultrasound} encompass more than 600 patients with transrectal ultrasound video sequences, while fetal datasets in \cite{luo2021self} and \cite{luo2023recon} involve 78 and 128 pregnant volunteers, respectively.
This strong imbalance, with the majority of studies relying on fewer than 30 subjects, highlights the ongoing challenge of collecting sufficiently large and diverse cohorts for robust model development and evaluation.

Scanning protocols also differ considerably. Some studies restrict acquisition to well-defined trajectories (e.g., transversal and sagittal sweeps for thyroid imaging \cite{wein2020three}), whereas others deliberately vary probe trajectories across linear, curved, loop, or S-shaped sweeps to test robustness \cite{luo2022deep,li2023trackerless,li2023privileged,luo2023multi,li2023long,10684746,yan2024fine,li2024nonrigid,grossbrohmer20253d,luo2025monetv2,lee2025enhancing,sun2025ultrasom}. Travel distances range from short sweeps of approximately 53.71 mm in carotid dataset \cite{luo2022deep} to more than 300 mm in arm datasets \cite{luo2023multi,yan2024fine,luo2025monetv2}. 

The current landscape of datasets for freehand ultrasound reconstruction is characterised by significant imbalance. On the one hand, anatomically diverse datasets exist, ranging from peripheral limbs to prostate, thyroid, fetal brain, and hip, supplemented by phantom and simulated data. On the other hand, dataset scale and subject numbers remain highly uneven. While large collections exist for prostate and fetal imaging, the majority of studies rely on small datasets with fewer than 30 participants. This disparity constrains the ability to compare methods fairly and limits model generalisability. 
A further challenge is that most datasets are private, restricting reproducibility and hindering the development of widely accepted benchmarks. Together, these limitations underscore the need for larger, more diverse, and openly available datasets, ideally collected across multiple centres, to enable robust evaluation and accelerate clinical translation.

\subsubsection{Ground Truth Acquisition}

Ground truth for training and evaluation is most commonly obtained using external tracking systems, with optical trackers \cite{prevost2017deep,prevost20183d,wein2020three,miura2020localizing,miura2021probe,miura2021pose,xie2021image,ning2022spatial,leblanc2022stretched,li2023trackerless,li2023privileged,li2023long,10684746,li2024nonrigid,grossbrohmer20253d,lee2025enhancing,sun2025ultrasom} and EM trackers \cite{guo2020sensorless,guo2021transducer,guo2022ultrasound,luo2022deep,luo2023recon,luo2023multi,el2023trackerless,gaits2024ultrasound,10684746,yan2024fine,dou2024pitfalls,luo2025monetv2,lee2025enhancing} representing the predominant approaches. A few employ robotic tracking \cite{wysocki2024ultra} or IMU \cite{mikaeili2022trajectory} tracking. Beyond hardware-based tracking, alternative strategies include sampling 2D slices from pre-aligned 3D ultrasound volumes \cite{yeung2021learning,yeung2021implicitvol,di2022deep,yeung2022adaptive,yeung2024sensorless,ramesh2024geometric,eid2025rapidvol}, particularly in fetal brain datasets, and generating synthetic sequences from 3D volumes by simulating diverse probe trajectories \cite{luo2021self,luo2023recon}.

\subsubsection{Training and Testing Protocol}

The majority of studies adopt a straightforward strategy in which training, validation, and testing data are drawn from the same dataset \cite{guo2020sensorless,wein2020three,luo2021self,yeung2021implicitvol,ning2022spatial,guo2022ultrasound,luo2022deep,mikaeili2022trajectory,chen2023freehand,luo2023recon,li2023trackerless,li2023privileged,luo2023multi,el2023trackerless,li2023long,dou2023sensorless,yeung2024sensorless,gaits2024ultrasound,10684746,ramesh2024geometric,yan2024fine,li2024nonrigid,dou2024pitfalls,dou2024continuous,eid2025rapidvol,luo2025monetv2,lee2025enhancing,sun2025ultrasom}. A smaller subset of works employ cross-validation strategies such as 2-fold \cite{prevost2017deep,prevost20183d}, 5-fold \cite{leblanc2022stretched} or 10-fold \cite{xie2021image,guo2022ultrasound} schemes. One study also explores variation within a single dataset by stratifying the dataset according to acquisition protocols, for example splitting it according to probe trajectory types (e.g., straight vs. C-shaped sweeps) \cite{li2023long}. 
A smaller number of studies explicitly separate source and target domains (e.g., transrectal vs. transabdominal scans) to assess generalisation \cite{guo2021transducer}. While cross-domain and cross-device validation has been explored in a few studies (e.g., \cite{prevost2017deep,prevost20183d,yeung2021learning,yeung2022adaptive,luo2023recon,luo2025monetv2,lee2025enhancing}), such efforts remain limited, and the majority of works continue to rely on single-domain training and evaluation, underscoring an important gap in assessing clinical generalisability.

\subsubsection{Evaluation Metrics}

The most frequently reported evaluation measures are trajectory-based, with drift-related metrics serving as the predominant benchmark. Drift is typically defined as the distance between the estimated and ground truth positions at the final frame of a sweep \cite{prevost2017deep,prevost20183d,guo2020sensorless,miura2021probe,guo2021transducer,miura2021pose,xie2021image,ning2022spatial,leblanc2022stretched,guo2022ultrasound,chen2023freehand,li2023trackerless,li2023privileged,li2023long,dou2023sensorless,10684746,sun2025ultrasom}. Variants include final drift rate \cite{luo2021self,guo2022ultrasound,luo2022deep,luo2023recon,luo2023multi,el2023trackerless,yan2024fine,luo2025monetv2,lee2025enhancing,sun2025ultrasom}, average drift rate \cite{luo2021self,luo2022deep,luo2023recon,luo2023multi,el2023trackerless,yan2024fine,luo2025monetv2,sun2025ultrasom}, maximum drift \cite{luo2021self,luo2022deep,luo2023recon,luo2023multi,yan2024fine,luo2025monetv2,sun2025ultrasom}, and sum of drift \cite{luo2021self,luo2022deep,luo2023recon,luo2023multi,yan2024fine,luo2025monetv2,sun2025ultrasom}. In addition, several works report parameter-wise error \cite{prevost2017deep,prevost20183d,miura2020localizing,miura2021probe,miura2021pose,mikaeili2022trajectory,chen2023freehand}, measuring the discrepancy between predicted and ground truth rigid transformation parameters (e.g., translation in millimetres, rotation in degrees).
Other formulations have also been proposed. Frame error \cite{guo2022ultrasound,li2023trackerless,li2023privileged,li2023long,10684746,li2024nonrigid,lee2025enhancing} assesses the mismatch between the predicted and ground truth positions of consecutive frames, providing a local consistency measurement. Accumulated tracking error \cite{guo2020sensorless,xie2021image,ning2022spatial,guo2022ultrasound,li2023trackerless,li2023privileged,li2023long,10684746,li2024nonrigid,lee2025enhancing,sun2025ultrasom} extends this idea by computing the average Euclidean distance between predicted and ground-truth reconstructed pixel locations across the entire sequence. 
Together, these metrics are widely adopted as they directly reflect the stability of sequential pose predictions and offer finer-grained perspectives on error propagation and temporal coherence. However, they primarily capture geometric consistency and do not fully assess anatomical fidelity.

A smaller subset of works report target registration error (TRE) \cite{grossbrohmer20253d}, which offers clinically interpretable validation by linking reconstruction accuracy to independently defined and usually manually annotated anatomical correspondences. However, it is limited by the need for manual annotations or reliable anatomical landmark availability. Several studies evaluate reconstruction quality through volumetric or segmentation-based measures. Examples include the Dice coefficient that is used to quantify overlap between segmented structures \cite{wein2020three,leblanc2022stretched,guo2022ultrasound} or reconstructed volumes \cite{li2023trackerless,li2023privileged,li2023long}. A subset of works employ image-based similarity metrics, such as SSIM \cite{yeung2021implicitvol,wysocki2024ultra,yeung2024sensorless,ramesh2024geometric,dou2024pitfalls,dou2024continuous,eid2025rapidvol} and cross-correlation \cite{yeung2021learning,yeung2021implicitvol,ramesh2024geometric}, typically applied by comparing input slices to those resampled from reconstructed volumes.

Overall, evaluation in trackerless freehand ultrasound reconstruction remains dominated by trajectory-based metrics, with drift established as the most widely adopted benchmark. While these measures are effective for quantifying sequential errors and provide insight into error propagation, they offer only a partial view of reconstruction performance, derailing from anatomical and clinical relevance. Complementary measures such as TRE, Dice scores, or image similarity indices provide richer assessments of structural fidelity, but their use is relatively sparse due to the challenges of annotation and computational cost. Notably, only two studies explicitly report inference time \cite{10684746,grossbrohmer20253d}, which is also important in real-world applications.
\subsubsection{Scales of Expected Errors}

Reported accuracy varies substantially across datasets, anatomies, and acquisition conditions. Translation errors span a broad range: in some phantom datasets, mean values are 1-10 mm \cite{prevost2017deep,prevost20183d,mikaeili2022trajectory,chen2023freehand}, whereas in more challenging scenarios such as forearm or prostate scans, they rise to 10-30 mm \cite{prevost2017deep,miura2021probe,miura2021pose}. Rotation errors also vary widely, from 1-10° \cite{prevost2017deep,prevost20183d,miura2020localizing,chen2023freehand} in controlled sweeps to 10-20° \cite{miura2021probe,miura2021pose} in more variable acquisitions.

Drift-related metrics exhibit broad variability. Final drift is often reported 1-20 mm in small phantoms or constrained sweeps \cite{prevost2017deep,prevost20183d,guo2020sensorless,guo2021transducer,xie2021image,ning2022spatial,leblanc2022stretched,guo2022ultrasound,chen2023freehand,li2023privileged,li2023long,dou2023sensorless,10684746,sun2025ultrasom}, but becomes 30-96 mm on larger datasets \cite{prevost2017deep,miura2021probe,miura2021pose,ning2022spatial,li2023trackerless}. Drift rate, when normalised by scan length, ranges from 5\% to 15\% \cite{luo2021self,guo2022ultrasound,luo2022deep,luo2023recon,luo2023multi,yan2024fine,luo2025monetv2,lee2025enhancing,sun2025ultrasom} in typical cases but can surpass 20\% \cite{el2023trackerless} in unconstrained freehand trajectories. 

Frame error values are frequently below 2 mm \cite{guo2022ultrasound,li2023trackerless,li2023privileged,li2023long,10684746,lee2025enhancing}, indicating good local consistency, while accumulated tracking error can reach 3-30 mm \cite{xie2021image,ning2022spatial,guo2022ultrasound,li2023trackerless,li2023privileged,li2023long,10684746,li2024nonrigid,lee2025enhancing,sun2025ultrasom} depending on scan length and motion complexity. Dice scores often fall between 0.60-0.90 \cite{wein2020three,li2023trackerless,li2023privileged,li2023long}, while SSIM values typically range from 0.50-0.75 \cite{yeung2021implicitvol,wysocki2024ultra,ramesh2024geometric,dou2024pitfalls,dou2024continuous} when comparing resampled slices to input images.

Taken together, the metrics show wide variation across datasets and clinical applications, with errors ranging from a few millimeters to tens of millimeters.
The limited reporting of frame-wise, accumulated, and anatomical metrics further constrains comparability across studies. Broader adoption of complementary measures, including volumetric fidelity and perceptual similarity indices, would provide a more balanced understanding of reconstruction accuracy and clinical relevance.

\subsubsection{Publication Trends}

An analysis of publication venues and years provides additional context for the evolution of trackerless freehand ultrasound reconstruction research. The field is heavily concentrated in computer vision and medical imaging conferences, with the \emph{International Conference on Medical Image Computing and Computer Assisted Intervention (MICCAI)} and its associated workshops constituting the largest proportion of works (13 papers), followed by the \emph{IEEE International Symposium on Biomedical Imaging} conference with 5 publications. High-impact medical imaging journals such as \emph{Medical Image Analysis} (4 papers), \emph{International Journal of Computer Assisted Radiology and Surgery} (3 papers), \emph{IEEE Transactions on Medical Imaging} (1 paper) also contribute significantly. Contributions are also distributed across broader engineering and biomedical venues (e.g., \emph{IEEE Transactions on Biomedical Engineering}, \emph{Annual International Conference of the IEEE Engineering in Medicine and Biology Society (EMBC)}), and recent deep learning-based approaches have also been published in AI-centric venues (e.g., \emph{Medical Imaging with Deep Learning (MIDL)}).

\begin{figure}[t]
\centering
\includegraphics[width=.8\textwidth]{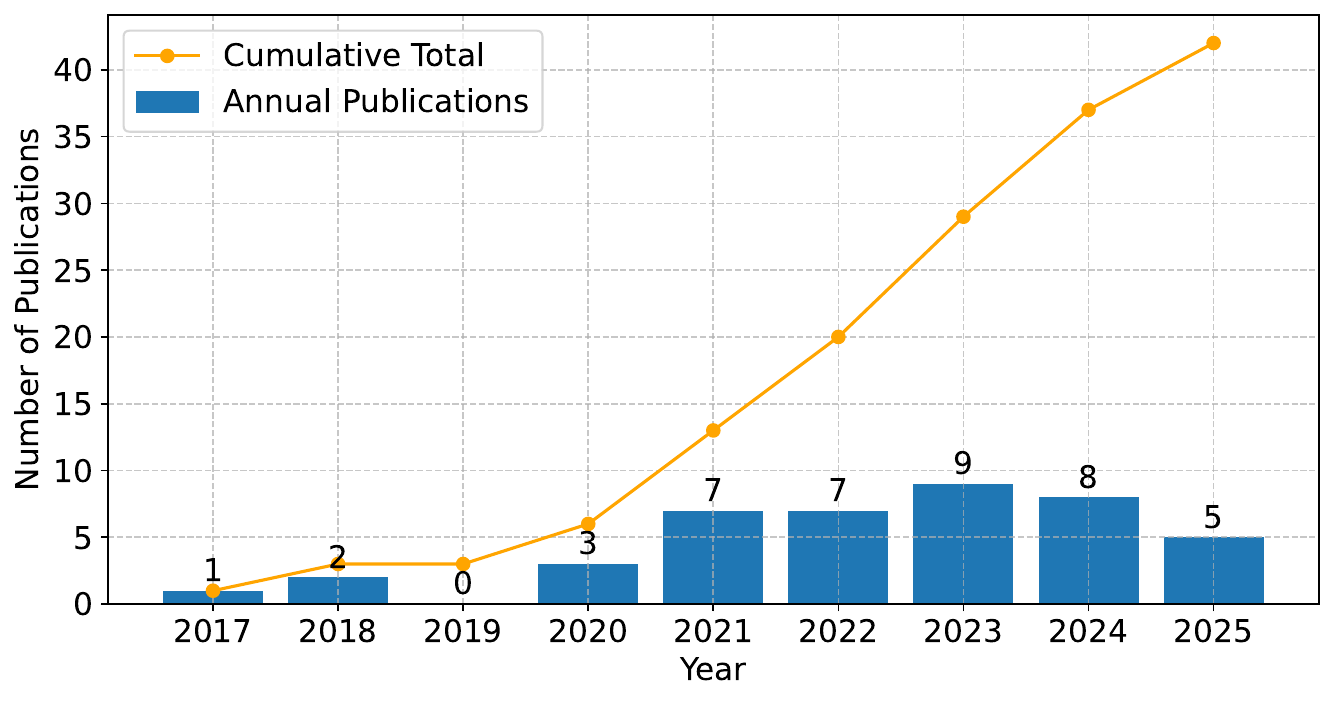}
\caption{Annual and cumulative publications on deep learning-based trackerless freehand ultrasound reconstruction.}\label{publications}
\end{figure}

The tendency of publications on deep learning based trackerless freehand ultrasound reconstruction highlights a distinct acceleration in research activity over recent years. As shown in Fig.~\ref{publications}, early work was sparse, but the adoption of deep learning around 2021 led to a rapid expansion of the field. More than half of all surveyed studies have been published in the past three years, indicating increasing interest in this topic.

\subsubsection{Summary}

To summarise, most of the surveyed literature share a common methodological template: most works rely on short B-mode frame sequences, CNN backbones with modest temporal modeling, regression of rigid 6-DoF transformations supervised by MSE, and evaluation on relatively small, often private datasets using drift-based metrics. While this design achieves reasonable accuracy in controlled settings, it remains limited by drift, poor generalisability, and restricted clinical interpretability. Our analysis identifies several under-explored but promising directions, including the use of physics-aware or multi-modal inputs, hybrid architectures that integrate attention mechanism or SSM, multi-task output designs combining pose and volumetric predictions, and supervision strategies that regularise regression with temporal, structural, or inference-time consistency. Dataset diversity and evaluation metrics also remain key bottlenecks: most studies rely on arm datasets or small single-centre cohorts, with sparse adoption of cross-domain validation and limited attention to inference-time efficiency. Addressing these challenges will require not only algorithmic innovation but also coordinated efforts in dataset curation, benchmarking, and clinically meaningful evaluation, ultimately enabling a transition from experimental prototypes toward robust, generalisable, and clinically deployable trackerless freehand ultrasound reconstruction systems.

\section{Challenge Design}
\label{Challenge_design}
The TUS-REC2024 Challenge\footnote{\url{https://github-pages.ucl.ac.uk/tus-rec-challenge/TUS-REC2024/}}\footnote{\url{https://doi.org/10.5281/zenodo.10991500}} is designed following the BIAS \cite{maier2020bias} Reporting Guideline for enhanced quality and transparency of biomedical research. This Challenge is associated with 5th International Workshop of Advances in Simplifying Medical UltraSound (ASMUS) at MICCAI 2024. The training and validation datasets are publicly available under CC BY-NC-SA license. The Challenge is an open-ended challenge, and submissions are welcome even after the official deadline. The test set remains held out to ensure fair benchmarking of reconstruction performance.

\subsection{Task Description}
\label{Task_description}   

Aiming at estimating the location for each ultrasound frame in 3D space, this Challenge is tasked to predict two different sets of transformation-representing dense displacement field (DDF), at global and local levels, respectively.
The global DDFs denote the displacement between the current frame and the first frame, and the local DDFs represent the displacement between the current frame and the previous frame. 
There are no restrictions on the internal design of the algorithm, for example, whether it is learning-based, whether it processes data at the frame, sequence, or scan level, and whether it assumes rigid, affine, or non-rigid transformations.

Participating teams are provided with ultrasound sequence and corresponding transformations. Each team’s model should take an ultrasound scan as input and output four sets of displacement vectors, representing the transformations to a reference frame (i.e., the first frame or the previous frame in the sequence). During evaluation, the submitted dockerised models will be used to generate these displacement fields, from which accuracy scores will be computed to assess reconstruction performance at both local and global levels.

\subsection{Dataset}
\label{Dataset}

\subsubsection{Data Collection}
\label{data_collection} 
\begin{figure}[t]
\centering
\includegraphics[width=\textwidth]{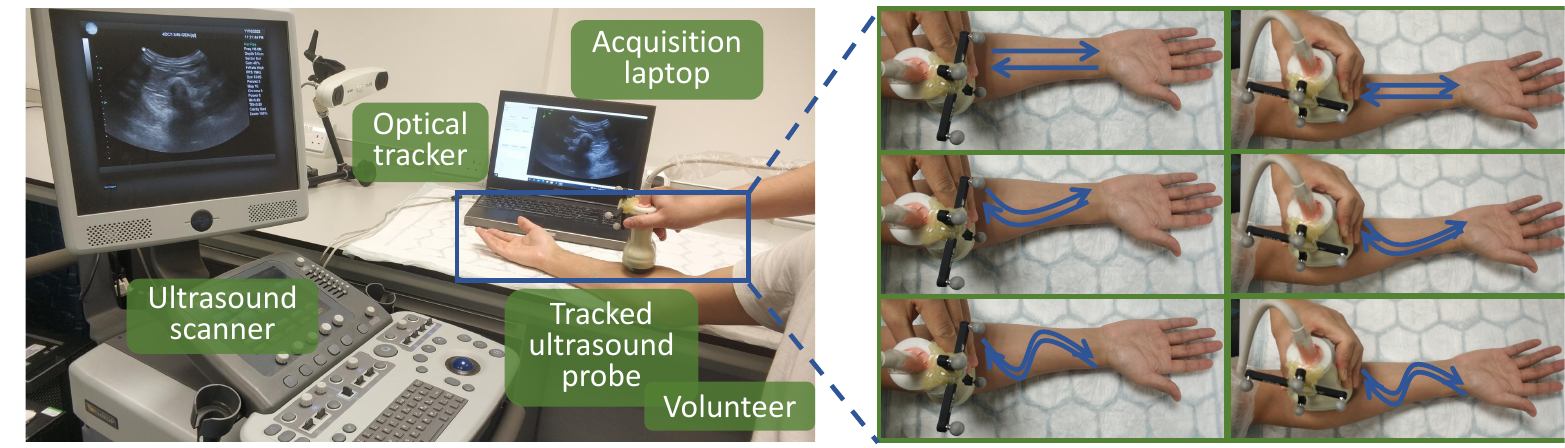}
\caption{Experimental setup for freehand ultrasound data acquisition. The setup consists of a tracked ultrasound probe, an ultrasound scanner, an optical tracker, and an acquisition laptop. The optical tracker monitors the probe's transformation during scanning, while a volunteer is scanned using predefined probe trajectories.}\label{set_up}
\end{figure}

The dataset\footnote{\url{https://doi.org/10.5281/zenodo.11178508}}\footnote{\url{https://doi.org/10.5281/zenodo.11180794}}\footnote{\url{https://doi.org/10.5281/zenodo.11355499}}\footnote{\url{https://doi.org/10.5281/zenodo.12752245}} used in this Challenge was collected from both the left and right forearms of 85 volunteers at University College London (UCL), United Kingdom. This study was performed in accordance with the ethical standards in the 1964 Declaration of Helsinki and its later amendments or comparable ethical standards. Approval was granted by the Ethics Committee of local institution on 20th Jan. 2023 [24055/001]. The subject cohort was diverse in terms of race, gender, and age. Fig.~\ref{set_up} illustrates the equipment setup used during data acquisition. There were no specific exclusion criteria, except for individuals with allergies or other skin conditions that could be aggravated by the ultrasound gel. All scanned forearms were confirmed to be in healthy condition. 

2D ultrasound images were acquired using an Ultrasonix machine (BK, Europe) equipped with a curvilinear probe (4DC7-3/40). The ultrasound frames were captured at a rate of $20$ frames per second, with a dimension of $480 \times 640$ pixels and a pixel size of around 0.2 $mm$, without applying speckle reduction as the speckle pattern may be useful for the task. Imaging was performed at a frequency of $6$ MHz, with a dynamic range of $83$ dB, an overall gain of $48\%$, and a depth of $9$ cm. Both the left and right forearms of volunteers were scanned. For each forearm, the ultrasound probe was moved along three trajectories, \textit{Straight line shape}, \textit{C shape}, and \textit{S shape}, in both \textit{distal-to-proximal} and \textit{proximal-to-distal} directions. These scans were performed with the ultrasound image plane \textit{perpendicular} of or \textit{parallel} to the scanning direction. 

In this study, an optical tracking system was chosen due to its greater accuracy and operational convenience compared to EM tracking systems. Specifically, the NDI Polaris Vicra (Northern Digital Inc., Canada) was used. During acquisition of ultrasound images, the position data recorded by the optical tracker was captured by the PLUS Toolkit~\cite{lasso2014plus}, represented as a homogeneous transformation matrix from the tracker tool to the camera coordinate system.

As described previously in Section~\ref{calibration}, the calibration matrix was obtained using the pinhead-based method~\cite{hu2017freehand}, and the timestamps of the ultrasound frames and the transformations were aligned.

\subsubsection{Sources of Errors}
\label{error_source} 

The primary source of error arises from the precision limitation of the optical tracker. All labels were obtained using an optical tracker with the 3D root-mean-square (RMS) volumetric accuracy acceptance criterion being less than or
equal to 0.25 mm and the 3D RMS repeatability acceptance criterion being less than or equal to 0.20 mm. Slight forearm movements may occur during scanning, which is also expected in clinical environments where this technique would be deployed. These motion-induced errors are assumed to be random across different cases. 
Additional sources of error include inaccuracies in the calibration process (both spatial and temporal), pressure-induced skin deformation, as well as the intra-observer variability during ultrasound data acquisition, affecting probe positioning and image quality. 

\subsubsection{Data Pre-processing}
\label{data_pre_processing} 

For each scan, ultrasound frames with invalid transformation matrices, which were typically caused by occluded line of sight, were excluded. The remaining raw images, along with their corresponding transformation matrices, were temporally ordered and stored as key-value pairs in a \textit{.h5} file. 

\subsubsection{Data Split}
\label{data_split} 

Statistical power analysis was performed to determine the appropriate test sample size and minimise the likelihood of Type I and Type II errors in hypothesis testing. The effect size was calculated using Cohen’s D value \cite{cohen2013statistical}, where the system error of the optical tracker ($0.25$ mm) was considered the meaningful difference between group means, and the standard deviation ($0.46$ mm) was derived from the results reported by~\cite{li2023trackerless}. A statistical power analysis for a pairwise t-test, assuming a significance level of $0.05$ and a statistical power of $0.9$, indicated a required test sample size of $31$. To ensure adequate power, we rounded up to $32$ samples ($768$ scans in total). This setup limits the probability of a Type II error to $10\%$ and a Type I error to $5\%$.

The dataset was randomly divided into training, validation, and test sets, comprising $50$, $3$, and $32$ subjects, respectively. This corresponds to $1200$, $72$, and $768$ scans, amounting to $606597$, $34746$, and $384105$ frames. Ultrasound scans from the same subject will be assigned to the same set which avoids the information leak. Detailed information is described in Table~\ref{data}. 
Specifically, the structure of the validation dataset is the same as that of the test set to ensure compatibility with the pre-defined folder hierarchy and naming conventions. This design allows submitted Docker images to run seamlessly on the test set.

No specific constraints are imposed on the use of the training and validation datasets. For example, participants are free to use all data from both sets for model training, or they may split the training set into training, validation, and test subsets for parameter tuning. Additionally, the use of both public and private data is permitted, but participants must disclose any external data sources they utilise.

\begin{table}[H]
\begin{center}
\caption{Overview of the freehand ultrasound dataset used in the TUS-REC2024 Challenge. The table summarises the number of subjects, scans, and frames across the training, validation, and test sets, categorised by scan trajectory shapes (\textit{Straight line shape}, \textit{C shape}, \textit{S shape}), scanning directions (\textit{Parallel} vs. \textit{Perpendicular}, \textit{Distal-to-proximal} vs. \textit{Proximal-to-distal}), and scanned arms (\textit{Left arm} vs. \textit{Right arm}).}\label{data}
\small
\centering
\begin{tabular*}{.9\textwidth}{@{\extracolsep{\fill}}cccccccc@{\extracolsep{\fill}}}
\addlinespace[5pt]
\hline
\addlinespace[5pt]
 Protocol& Statistics & Train & Validation & Test   \\
 \addlinespace[5pt]
 \hline
 \addlinespace[5pt]
\multirow{3}{*}{All}& \# Subjects & 50 & 3 & 32 \\
& \# Scans & 1200 & 72 & 768 \\
& \# Frames &  606597  &  34746   &  384105  \\
\addlinespace[5pt]
\hline
\addlinespace[5pt]
\multirow{3}{*}{\textit{Straight line shape}} & \# Subjects &50 & 3 & 32  \\
& \# Scans &400  &24 & 256  \\
& \# Frames & 192117  &10515  & 119421  \\

\addlinespace[5pt]

\multirow{3}{*}{\textit{C shape}} & \# Subjects & 50 & 3 & 32   \\
&\# Scans & 400 &24 &256   \\
&\# Frames & 202654  & 11655 & 128721   \\
\addlinespace[5pt]

\multirow{3}{*}{\textit{S shape}}& \# Subjects & 50 & 3 & 32  \\
& \# Scans & 400 &24& 256  \\
& \# Frames & 211826  & 12576 & 135963  \\
\addlinespace[5pt]
\hline
\addlinespace[5pt]

\multirow{3}{*}{\textit{Parallel} scanning} & \# Subjects &50 & 3 & 32  \\
& \# Scans & 600  & 36 & 384  \\
& \# Frames & 298722  & 17228 & 188399  \\

\addlinespace[5pt]
\multirow{3}{*}{\textit{Perpendicular} scanning} &\# Subjects &50 & 3 & 32  \\
&\# Scans & 600  & 36 & 384  \\
&\# Frames & 307875  & 17518 & 195706 \\
\addlinespace[5pt]
\hline
\addlinespace[5pt]
\multirow{3}{*}{\textit{Left arm}}&\# Subjects & 50 & 3 & 32 \\
&\# Scans & 600  & 36 & 384   \\
& \# Frames & 301155  & 17081 & 192118   \\
\addlinespace[5pt]
\multirow{3}{*}{\textit{Right arm}}&\# Subjects & 50 & 3 & 32 \\
&\# Scans & 600  & 36 & 384    \\ 
&\# Frames & 305442  & 17665 & 191987    \\ 
\addlinespace[5pt]
\hline
\addlinespace[5pt]
\multirow{3}{*}{\textit{Distal-to-proximal} scanning} &\# Subjects & 50 & 3 & 32 \\
&\# Scans & 600 & 36 & 384 \\
&\# Frames & 298803 &16908  &181844  \\
\addlinespace[5pt]
\multirow{3}{*}{\textit{Proximal-to-distal} scanning} &\# Subjects & 50 & 3 & 32 \\
&\# Scans &  600 & 36 & 384  \\
& \# Frames & 307794  &17838  &202261   \\
\addlinespace[5pt]
\hline
\addlinespace[5pt]
\end{tabular*}
\end{center}
\end{table}

\subsection{Evaluation Metrics}
\label{Evaluation_metrics}   

\subsubsection{Metrics Definition}
\label{definition}   

We use DDFs to evaluate the reconstruction performance, borrowing the widely recognised term used in non-rigid image registration for clarity and intuition. For each scan, participating methods are required to generate two types of DDFs representing frame-to-frame transformations, hereinafter referred to as predictions, at both global and local levels: 1) global displacement vectors are used to reconstruct all frames (excluding the first) relative to the first frame of the scan, which serves as the global reference frame;
2) local displacement vectors are used to reconstruct each frame (excluding the first) relative to its immediately previous frame, which serves as the local reference frame.

The performance of each submitted method will be assessed for every scan using two metrics: landmark reconstruction error and pixel reconstruction error: 1) landmark reconstruction error is defined as the average Euclidean distance between the ground-truth-reconstructed frame and the prediction-reconstructed frame, computed over a predefined set of landmarks; 2) pixel reconstruction error is similarly defined as the average Euclidean distance between the ground-truth and predicted reconstructions, calculated over all pixels in every frame except the first. The scale-invariant feature transform (SIFT) \cite{lowe2004distinctive} algorithm was applied to detect landmarks. For each scan, 20 landmarks with the highest response values were selected. 

Accordingly, each method should produce the following four sets of vectors: 
\begin{itemize}[itemsep=-4pt, topsep=1pt]
\setlength{\labelsep}{5pt} 
  \renewcommand{\labelitemi}{\scalebox{0.6}{\textbullet}} 
    \item Global-Pixel (GP) vectors – one per pixel (excluding the first frame) for global-level pixel reconstruction;
    \item Global-Landmark (GL) vectors – one per landmark for global-level landmark reconstruction;
    \item Local-Pixel (LP) vectors – one per pixel (excluding the first frame) for local-level pixel reconstruction;
    \item Local-Landmark (LL) vectors – one per landmark for local-level landmark reconstruction.
\end{itemize}

Based on these outputs, four evaluation metrics will be computed:
\begin{itemize}[itemsep=-4pt, topsep=1pt]
\setlength{\labelsep}{5pt} 
  \renewcommand{\labelitemi}{\scalebox{0.6}{\textbullet}} 
    \item Global Pixel Reconstruction Error (GPE) – the pixel reconstruction error calculated using GP vectors;
    \item Global Landmark Reconstruction Error (GLE) – the landmark reconstruction error calculated using GL vectors;
    \item Local Pixel Reconstruction Error (LPE) – the pixel reconstruction error calculated using LP vectors;
    \item Local Landmark Reconstruction Error (LLE) – the landmark reconstruction error calculated using LL vectors.

\end{itemize}

Runtime will be included as an additional evaluation metric. It is defined as the consumed time of predicting the positions for all frames but the first frame in a scan, averaged across all scans in the test set.

\subsubsection{Rationale of Evaluation Metrics}
\label{Rationale} 
\textit{Use of Euclidean distance-based error metrics vs. transformation parameter-based errors.}
Directly evaluating the accuracy in transformation parameter space can be biased, as the weighting of rotational and translational components can vary significantly depending on experimental setups, imaging configurations, reference coordinate systems, and definitions of rotational axes. These factors are also often application-dependent. Therefore, this Challenge adopts Euclidean distance-based metrics, which offer a less biased assessment of the discrepancy between ground truth and predicted positions in physical space. 

\textit{Use of displacement-based transformation representations vs. rigid / affine matrices.}
Although ground-truth transformations are provided in the form of rigid transformation, we argue, based on practical experience in developing similar numerical algorithms, that requiring submissions to output homogeneous transformation matrices is not only unnecessary, but sometimes misleadingly encourages a more numerically challenging solution due to issues such as gimbal lock in using rotation matrix, local minima in numerical optimisation. In contrast, displacement-based representations allow flexibility for a quantitatively more accurate reconstruction, with a near-rigid transformation, which may be clinically sufficient~\cite{li2024nonrigid}. Nevertheless, there are no restrictions on the internal methodology: participants may choose to internally estimate a rigid transformation matrix and convert it into the four required displacement vector sets for submission.

\textit{Justification for local and global reconstruction error metrics.}
Local and global reconstruction errors capture complementary aspects of algorithm performance. Global reconstruction (relative to the first frame) can reveal accumulated drift over time, while local reconstruction (relative to the immediately previous frame) assesses frame-level reconstruction. These metrics are therefore indicative of both short- and long-term accuracy. Although other monotonic metrics such as final drift and Dice overlap are also commonly used~\cite{li2023long}, they are excluded here to streamline evaluation. In practice, one might choose to reconstruct a sequence of ultrasound frames (as opposed to the entire scan or two adjacent frames, which are represented by local and global errors, respectively), using a pre-optimised sequence length that is the most suitable to specific downstream application. Since this Challenge is designed without targeting a specific clinical use case, both local and global reconstruction errors are included to span the spectrum of reconstruction performance and provide a comprehensive assessment of algorithmic accuracy.

\subsection{Ranking Scheme}
\label{Ranking_scheme}   

The ranking follows the ``aggregate then rank” strategy~\cite{maier2018rankings}. For each test scan, the four reconstruction error metrics will be normalised to the range $[0, 1]$ using the formulas below.

\begin{align}
    GPE^*&=(GPE_{max}-GPE)/(GPE_{max}-GPE_{min}) \nonumber\\
    GLE^*&=(GLE_{max}-GLE)/(GLE_{max}-GLE_{min}) \nonumber\\
    LPE^*&=(LPE_{max}-LPE)/(LPE_{max}-LPE_{min}) \nonumber\\
    LLE^*&=(LLE_{max}-LLE)/(LLE_{max}-LLE_{min})
\end{align}
where the superscript $^*$ denotes the normalised reconstruction error, and the subscript ${min}$ and ${max}$ denote the minimum and maximum errors among all participating submissions for each corresponding metric.
For each scan, the final score is computed as a weighted average of the four normalised metrics:
\begin{equation}
final\, score = 0.25\times GPE^* + 0.25\times GLE^* + 0.25\times LPE^* + 0.25\times LLE^*
\end{equation}
In this Challenge, equal weighting is used, with motivation explained later.

Each team's overall score was calculated as the average final score across all test scans. This score, ranging from 0 to 1, determines the final ranking of all submitted algorithms. Scores were reported to three decimal places, with higher values indicating better performance.

For further insight, we also reported four other categories of scores, for reference and research interest without formal ranking: global reconstruction score = $0.5\times GPE^* + 0.5\times GLE^*$, local reconstruction score = $0.5\times LPE^* + 0.5\times LLE^*$, landmark reconstruction score = $0.5\times GLE^* + 0.5\times LLE^*$ and pixel reconstruction score = $0.5\times GPE^* + 0.5\times LPE^*$.

All evaluation metrics are normalised to a common scale to prevent metrics with inherently larger magnitudes from disproportionately influencing the overall score. The two levels of measurement (global and local) and the two types of displacement vectors (pixel-based and landmark-based) are considered equally important in achieving desirable reconstruction performance. Consequently, equal weighting is applied to each metric to establish a fair and balanced benchmark for the Challenge. A minimum score of 0 was assigned to any case where the submitted code failed to execute or the evaluation metrics could not be computed successfully.
In the event of tied overall scores, ranking was determined based on runtime. A smaller runtime was awarded a higher rank. To encourage usability in the clinical applications, a maximum runtime limit of 2 minutes per scan was enforced for all Challenge submissions. Additionally, the raw (unnormalised) values of all defined evaluation metrics were made publicly available for transparency and further analysis.

\subsection{Validation and Submission}
\label{Submission} 

A small validation set was provided mainly for sanity checking on previously unseen data. An example Docker template\footnote{\url{https://github.com/QiLi111/tus-rec-challenge_baseline/tree/main/submission}} for evaluation on the validation dataset was provided, along with implementation of the corresponding evaluation metrics. This facilitates the preparation of valid Docker images by participants and improves overall transparency in the evaluation process.
Docker images from participating teams were submitted via an online form, which included a brief method description and step-by-step instructions for downloading and executing the Docker image. All submitted methods must operate in a fully automatic manner.
Participants are permitted to modify and resubmit their Docker image if it fails to run on the test set due to issues such as incorrect input/output formatting or mismatched file types and data structures. 
Each team was allowed to make up to five submissions, provided that each submission represents substantively different approaches rather than minor variations in hyperparameters of other submissions. The best-performing result in terms of final scores among these was considered as the team's final result.
All submitted Docker images were independently tested by two members of the Challenge organisition team using the hidden test dataset. Evaluations were conducted on two separate platforms with identical hardware configurations: Ubuntu 18.04.6 LTS, Intel(R) Xeon(R) Gold 5215 CPU @ 2.50GHz (10 cores), NVIDIA Quadro GV100 GPU (32GB VRAM), and 128GB RAM.
            
\subsection{Awards}
\label{Awards} 

Results from all participants were publicly displayed on the official leaderboard\footnote{\url{https://github-pages.ucl.ac.uk/tus-rec-challenge/TUS-REC2024/leaderboard.html}}, except in cases where submissions encountered errors during the evaluation process. Additional certificates of recognition were awarded to the first-place team and the runner-up. All teams with successfully evaluated submissions received certificates of participation. Members of the organizers’ institutes may participate but not eligible for awards and not listed in leaderboard.

\subsection{Timeline}
\label{Time_line} 

The TUS-REC2024 Challenge is an open-call event designed to encourage broad community participation. Although this edition was structured as a one-time event tied to MICCAI 2024, its infrastructure and open-submission framework support potential future iterations, enabling continued engagement beyond the initial evaluation cycle.

\begin{figure}[t]
\centering
\includegraphics[width=\textwidth]{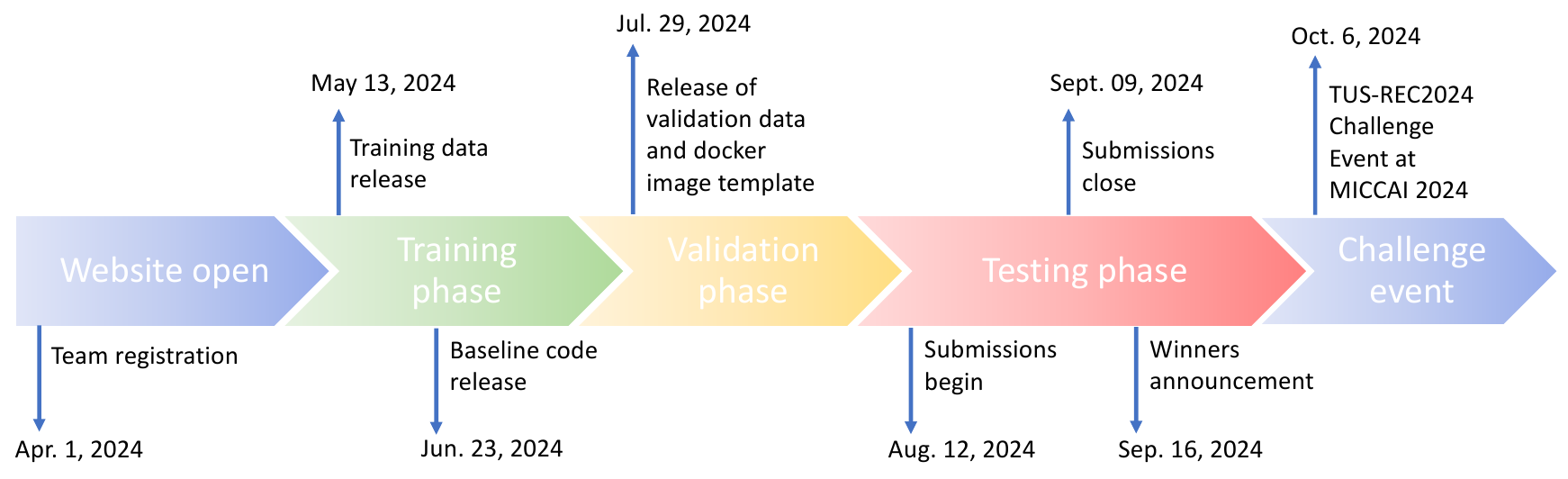}
\caption{Timeline of the TUS-REC2024 Challenge. Key milestones include the release of training data, baseline code, and validation resources, followed by the submission phase and final Challenge event at MICCAI 2024.}\label{timeline}
\end{figure}

The official timeline is aligned with MICCAI 2024, as detailed in Fig.~\ref{timeline}. The Challenge began with the launch of website and team registration on April 1, 2024, followed by the release of training data on May 13, and release of baseline code on June 23. On July 29, validation data and the Docker image template were released, offering participants a clear instruction for submission. This template, along with an evaluation script that incorporates the Challenge metrics, aimed to ensure transparency and reproducibility in assessment and was designed to align with the BIAS Reporting Guideline.
The submission window officially opened on August 12 and closed on September 9. The announcement of the winning teams took place on September 16, and the TUS-REC2024 Challenge event was held on October 6, 2024, during MICCAI 2024. The participating teams can publish their results separately after publication of the joint challenge paper.

\begin{figure}[t]
\centering
\includegraphics[width=\textwidth]{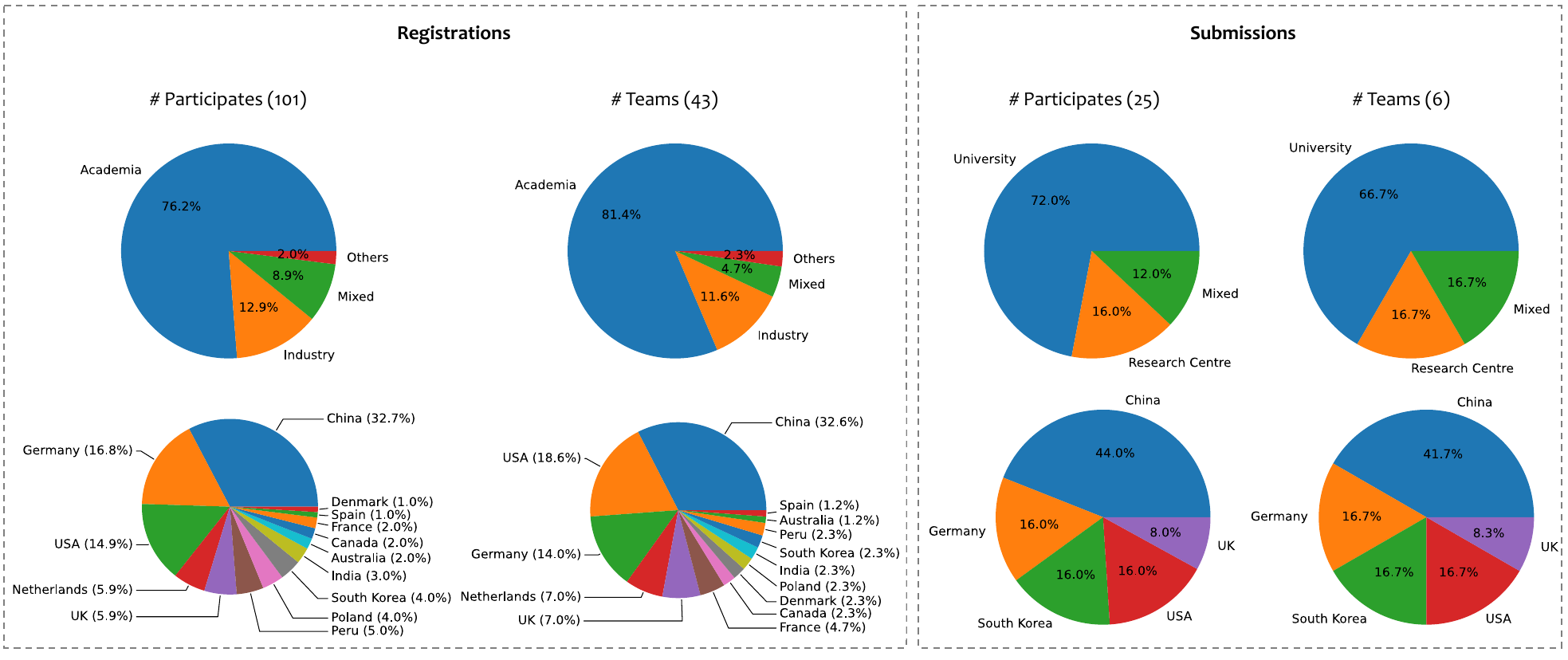}
\caption{Participant and team statistics summarising engagement across TUS-REC2024 Challenge, including 101 registered participants from 43 teams, and participation by 25 individuals grouped into 6 teams.
}\label{statistics}
\end{figure}

\section{Challenge Outcome}
\label{Challenge_Outcome} 

\subsection{Participation Statistics}
\label{Participation_statistics} 

Fig.~\ref{statistics} presents the participant statistics for TUS-REC2024 Challenge. By the submission deadline, a total of 101 individuals registered, representing 43 teams comprising members from both academia and industry. Participants came from 14 countries across 5 continents, reflecting the international interest and global reach of the Challenge. Despite strong initial engagement, 6 teams submitted their Docker images by the submission deadline, involving a total of 25 participants. In total, 21 valid docker images were received. The number of submissions varied across teams, with several teams submitting multiple Docker images for performance optimisation. This decline from registration to submission may reflect the technical complexity of the task or limited preparation time. Notably, the majority of registered and submitting teams were affiliated with academic institutions, particularly universities. Overall, the statistics highlight both the broad appeal of the Challenge, and the practical hurdles faced by participants in progressing from initial registration to successful submission, such as time constraints and difficulties in model development.

\subsection{Methods}
\label{Methods}

\subsubsection{\texorpdfstring{Methodologies of Baseline and Participating Teams\protect\footnote{This is a summary of TUS-REC2024 Challenge, rather than proposing these methods. The authors may publish their own technical papers enabling reproducibility of their methods.}}{Overview of methodologies of baseline and participating teams}}

\label{Methodologies}

This section presents the approaches of the top five participating teams, alongside the baseline approach provided by the organisers. Table~\ref{teams} summarises key information about these teams, including their model abbreviations, methodological highlights, team names, and institutional affiliations. Table~\ref{imp} summarises the implementation details of the baseline and top five participating methods. It includes model architectures, backbones, training configurations, loss functions, and other relevant technical aspects that highlight the diversity of approaches adopted in the Challenge.

\begin{table}
\begin{center}
\caption{Overview of the top five participating teams in TUS-REC2024 Challenge, including model abbreviations, methodological descriptions, team names, and institutional affiliations.}\label{teams}
\small
\renewcommand{\arraystretch}{1.3} 
\begin{tabular*}{\textwidth}{@{\extracolsep{\fill}}lllllll@{\extracolsep{\fill}}}
\toprule
Rank & \multicolumn{1}{p{2cm}}{\raggedright Model Abbreviation} & Method & Team Name & Affiliation(s)  \\ 
\midrule
1 & FiMoNet  & Enhanced Fine-grained Motion Network & MUSIC Lab & \multicolumn{1}{p{5cm}}{\raggedright Shenzhen University; Shenzhen RayShape Medical Technology Inc.}\\

2 & RecuVol  & \multicolumn{1}{p{5cm}}{\raggedright Recurrent CNN-LSTM Trackerless Freehand 3D Ultrasound Reconstruction}  & ISRU@DKFZ & \multicolumn{1}{p{5cm}}{\raggedright DKFZ (German Cancer Research Center) Heidelberg; University of Cincinnati; Tufts University}    \\ 

3 & FlowNet  & \multicolumn{1}{p{5cm}}{\raggedright Three-dimensional Ultrasound Reconstruction using CNN Learned by Flow Field Transformation} & zjr & \multicolumn{1}{p{4cm}}{\raggedright Hong Kong Centre for Cerebro-cardiovascular Health Engineering; City University of Hong Kong}\\

4 & MoGLo-Net  & \multicolumn{1}{p{5cm}}{\raggedright Motion-based Learning Networks with Global-Local Attention for Ultrasound Scan Motion Estimation} & AMI-Lab & Pusan National University\\ 

5 & PLPPI  & \multicolumn{1}{p{5cm}}{\raggedright Physics Guided Learning-based Prediction of Pose Information} & \multicolumn{1}{p{2cm}}{\raggedright UW-Madison Elastography Lab} & \multicolumn{1}{p{5cm}}{\raggedright University of Wisconsin-Madison}\\

\bottomrule
\end{tabular*}
\end{center}
\end{table}

\makeatletter
\renewcommand\paragraph{\@startsection{paragraph}{4}{\z@}
  {-1.25ex\@plus -1ex \@minus -.2ex}
  {0.75ex \@plus 0.1ex}
  {\normalfont\normalsize\itshape}}
\makeatother    
\paragraph{Baseline Algorithm\protect\footnote{\url{https://github.com/QiLi111/tus-rec-challenge_baseline}}}\label{Baseline}

The baseline method utilises the EfficientNet-B1 architecture \cite{tan2019efficientnet}, taking as input a pair of adjacent ultrasound frames. The network predicts a 6-DoF transformation, representing the transformation from the image coordinate system (in mm) of one frame to that of the other. 
The training loss is formulated as the mean squared error (MSE) between point coordinates transformed by the ground truth and the predicted transformations:

\begin{equation}
\mathcal{L}=D\,(T^{gt}_{j\leftarrow i} \cdot T_{scale} \cdot \mathbf{p}_{corner},\,T_{j\leftarrow i} \cdot T_{scale} \cdot \mathbf{p}_{corner}),\,\,i=j+1
\end{equation}
where $D(,\cdot,)$ denotes the MSE computed over the $x$, $y$ and $z$ coordinates of corresponding points.  $T_{j\leftarrow i}$ is obtained by converting the predicted 6-DoF into a homogeneous transformation matrix. $T^{gt}_{j\leftarrow i}$ represents the ground truth transformation matrix, as defined in Eq.~\eqref{image_mm_to_image_mm}.
$\mathbf{p}_{corner}$ denotes the pixel coordinates of the four corner points in the image coordinate system.

During inference, the 6-DoF transformations between adjacent frames are estimated by sequentially inputting frame pairs into the network. The local DDFs are computed by $DDF_{local}^{(i)} = T_{local}^{(i)} \cdot T_{scale} \cdot \mathbf{p} - T_{scale} \cdot \mathbf{p}$, where $\mathbf{p}$ denotes the coordinates of all pixels within an image in the image coordinate system (in pixels). $T_{local}^{(i)}=T_{i-1\leftarrow i}$ is the transformation matrix from frame $i$ to frame $i-1$, converted from the 6-DoF.
The global transformation from any frame $i$ to the first frame, $T_{global}^{(i)}$, is derived by composing predicted local transformations through $T_{global}^{(i)}=T_{local}^{(2)}\cdot T_{local}^{(3)} \cdots T_{local}^{(i)}$. The global DDFs are computed using $DDF_{global}^{(i)}=T_{global}^{(i)} \cdot T_{scale} \cdot \mathbf{p}- T_{scale} \cdot \mathbf{p}$. 
The local and global DDFs at predefined landmark locations can be obtained either by indexing the corresponding positions from $DDF_{local}$ and $DDF_{global}$, respectively, or by calculating them using the formula above, replacing all pixel coordinates $\mathbf{p}$ with the landmark locations. 
    
\paragraph{FiMoNet\protect\footnote{\url{https://github.com/Lmy0217/FiMoNet}}}\label{FiMoNet}

Fine-grained spatio-temporal learning is essential for trackerless freehand 3D ultrasound reconstruction. FiMoNet adapts Mamba~\cite{gu2023mamba} to address the complexities of long-range dependencies introduced by diverse probe motions as well as the large number of patches involved in spatio-temporal modeling. Mamba utilises the state space model’s capacity to manage long-range dependencies, providing an effective solution for this task.

Ensemble learning is used to combine two models:
\begin{itemize}[itemsep=-4pt, topsep=1pt]
\setlength{\labelsep}{5pt} 
  \renewcommand{\labelitemi}{\scalebox{0.6}{\textbullet}} 
    \item Model 1 consists of ResNet18 and ReMamba~\cite{yan2024fine}. Following the method in \cite{yan2024fine}, convolutional blocks from ResNet18 and ReMamba blocks are applied to extract fine-grained image features at multiple scales. A fully connected layer is then employed to regress the 6-DoF transformation parameters. 
    \item Model 2 integrates ResNet18 with a multi-layer Mamba block. Inspired by \cite{luo2022deep}, a cascaded architecture is designed. Specifically, the final fully connected layer of ResNet18 is removed and replaced with a multi-layer Mamba block, followed by a fully connected layer to produce the output.
\end{itemize}

For the 6-DoF transformations between adjacent frames, estimated by the network as $\theta$ and the ground truth 6-DoF transformation $\theta^{gt}$, both the $L1$ loss and Pearson correlation loss are used:
\begin{equation}
\mathcal{L}=\parallel{\theta^{gt}}-\theta\parallel_1+\left(1-\frac{Cov(\theta^{gt},\theta)}{\sigma(\theta^{gt})\sigma(\theta)}\right)
\end{equation}
where $Cov(\theta^{gt},\theta)$ represents the covariance between ground truth and predicted 6-DoF parameters. $\sigma(\cdot)$ denotes the standard deviation. During inference, the model takes the entire scan as input and outputs the 6-DoF transformation between all adjacent frames. These local transformations are then converted to global 6-DoF transformations, which are used to generate the global DDFs.

\makeatletter
\renewcommand\paragraph{\@startsection{paragraph}{4}{\z@}
  {-1.25ex\@plus -1ex \@minus -.2ex}
  {0.75ex \@plus 0.1ex}
  {\normalfont\normalsize\itshape}}
\makeatother
\makeatletter
\renewcommand\paragraph{\@startsection{paragraph}{4}{\z@}
  {-1.25ex\@plus -1ex \@minus -.2ex}
  {0.75ex \@plus 0.1ex}
  {\normalfont\normalsize\itshape}}
\makeatother
\paragraph{RecuVol\protect\footnote{\url{https://github.com/ISRU-DKFZ/RecuVol}}}\label{RecuVol}

This approach utilises an EfficientNet-based CNN (pre-trained on ImageNet) to extract features from pairs of consecutive frames. These features are processed sequentially by a LSTM network to model temporal dependencies. The network predicts 3D translation and rotation parameters for each frame pair. Training is performed by minimising the MSE loss on these parameters, enabling the model to learn robust frame-to-frame alignments. TrivialAugment \cite{muller2021trivialaugment} is used for data augmentation, and sequences of 16 frames are processed at a time, with adjacent frames concatenated prior to input into the CNN.

RecuVol applied a 5-fold cross-validation strategy for training. However, one fold displayed instability and was consequently excluded. The remaining four folds were ensembled by computing the median of the predicted 6-DoF transformation parameters, yielding a single final prediction. To further enhance the performance, a second 4-fold ensemble was trained on data downsampled by a factor of 1.25. The final submission was composed of both 4-fold ensembles (original and downsampled), resulting in a total of eight models.

During inference, the model estimates the rigid transformation parameters between each consecutive pair of frames within a scan. By sequentially concatenating these pairwise transformations starting from the first frame, the method computes the global pose of each frame relative to the first frame. Using all frames' global transformations, the 3D volume is reconstructed. The DDF is derived by back-mapping voxel coordinates from a reference 3D grid to their original frame positions. The model only explicitly predicts local transformations, while global transformations are obtained by sequentially accumulating these local estimates. Both local and global transformations are rigid and derived from the model's frame-to-frame predictions.

\makeatletter
\renewcommand\paragraph{\@startsection{paragraph}{4}{\z@}
  {-1.25ex\@plus -1ex \@minus -.2ex}
  {0.75ex \@plus 0.1ex}
  {\normalfont\normalsize\itshape}}
\makeatother
\paragraph{FlowNet}\label{FlowNet}

The network is based on EfficientNet-B6, taking $n=10$ consecutive ultrasound frames $S \in \mathbb{R}^{n\times h\times w}$ as input. It outputs a set of transformation parameters $Y \in \mathbb{R}^{(n-1)\times6}$, where each 6-DoF vector represents the rigid transformation from the last ultrasound frame $S^{(n)}$ to a preceding frame $S^{(i)},\,i\in [1,n-1]$. $Y$ is used to compute a flow field $F$, enabling the warping of $n-1$ frames to generate $S^{warp}\in \mathbb{R}^{(n-1)\times h\times w}$. $Y$ can also be converted to matrices $T_Y$, where each $T_Y^{(i)}$ denotes the transformation matrix from $S^{(n)}$ to $S^{(i)}$. 
The transformation matrix between any two frames $S^{(i)}$ and $S^{(j)}$ can be obtained by calculating $T_Y^{(j)} \cdot (T_Y^{(i)})^{-1}$, forming the dense transformation matrix set $T^{ds}$. The resulting dense point coordinates $P^{ds}$ are then used to calculate the overall loss:
\begin{equation}
\mathcal{L}=\text{MSE}\left(P_{gt}^{ds},P^{ds}\right)+0.5\times \text{MSE}\left(T_{gt}^{ds},T^{ds}\right)+0.5\times \text{MSE}\left(S,S^{warp}\right)
\end{equation}
where $P_{gt}^{ds}$ and $T_{gt}^{ds}$ denote the ground truth points coordinates and transformations, respectively.

Given the full scan $\mathcal{S}\in \mathbb{R}^{N\times h\times w}$, with local and global transformations $T_{local},T_{global}\in \mathbb{R}^{(N-1)\times4\times4}$, sequences of $n$ frames are sequentially processed using a stride of $n-1$, such that the last frame of one sequence is the first of the next. Three models are selected: the final epoch model, the model from 100 epochs earlier, and the one with the lowest validation distance. Predictions from the three models are averaged to obtain the final $Y$ and $T_Y$. 
For the first sequence, local transformation could be calculated by $T_{local}^{(i)}=T_Y^{(i-1)}\cdot (T_Y^{(i)})^{-1}$, and global transformation is calculated by $T_{global}^{(i)}=T_Y^{(1)}\cdot (T_Y^{(i)})^{-1}$. 
After computing local transformations for all frames in the first sequence, subsequent sequences are processed sequentially using the same method to obtain local transformations. The global transformation for the $k^{th}$ frame in $l^{th}$ sequence $S^{((l-1)\times (n-1)+k)}$ is computed as $T_{global}^{((l-1)\times (n-1)+k)}=T^{(1)}_{Y,s_1}\cdot T^{(1)}_{Y,s_2} \cdots T^{(1)}_{Y,s_l} \cdot (T_{Y,s_l}^{(k)})^{-1}$, where $T^{(k)}_{Y,s_l}$ denotes the transformation matrix from $n^{th}$ frame to the $k^{th}$ frame in the $l^{th}$ sequence. This yields $T_{local}$ and $T_{global}$ for the full scan. The scan is then reversed, and the same procedure is applied to obtain $T_{local}^{reverse}$ and $T_{global}^{reverse}$. Final predictions are obtained by averaging the forward and reversed results: $T^{avg}_{local} = (T_{local}+T_{local}^{reverse})/2$, $T^{avg}_{global}=(T_{global}+T_{global}^{reverse})/2$.

To further improve local transformation prediction, the scan is offset by excluding the first $m=1,2,3,4$ frames, yielding sub-scans $\mathcal{S}_m\in \mathbb{R}^{(N-m)\times h\times w}$. Corresponding local transformations $T^{avg}_{local,m}$ are computed for each offset. The final local transformation is obtained by averaging all predictions: $T^{final}_{local} = (T^{avg}_{local}+ T^{avg}_{local,1}+T^{avg}_{local,2}+T^{avg}_{local,3}+T^{avg}_{local,4})/5$.

\begin{figure}
\centering
\includegraphics[width=0.8\textwidth]{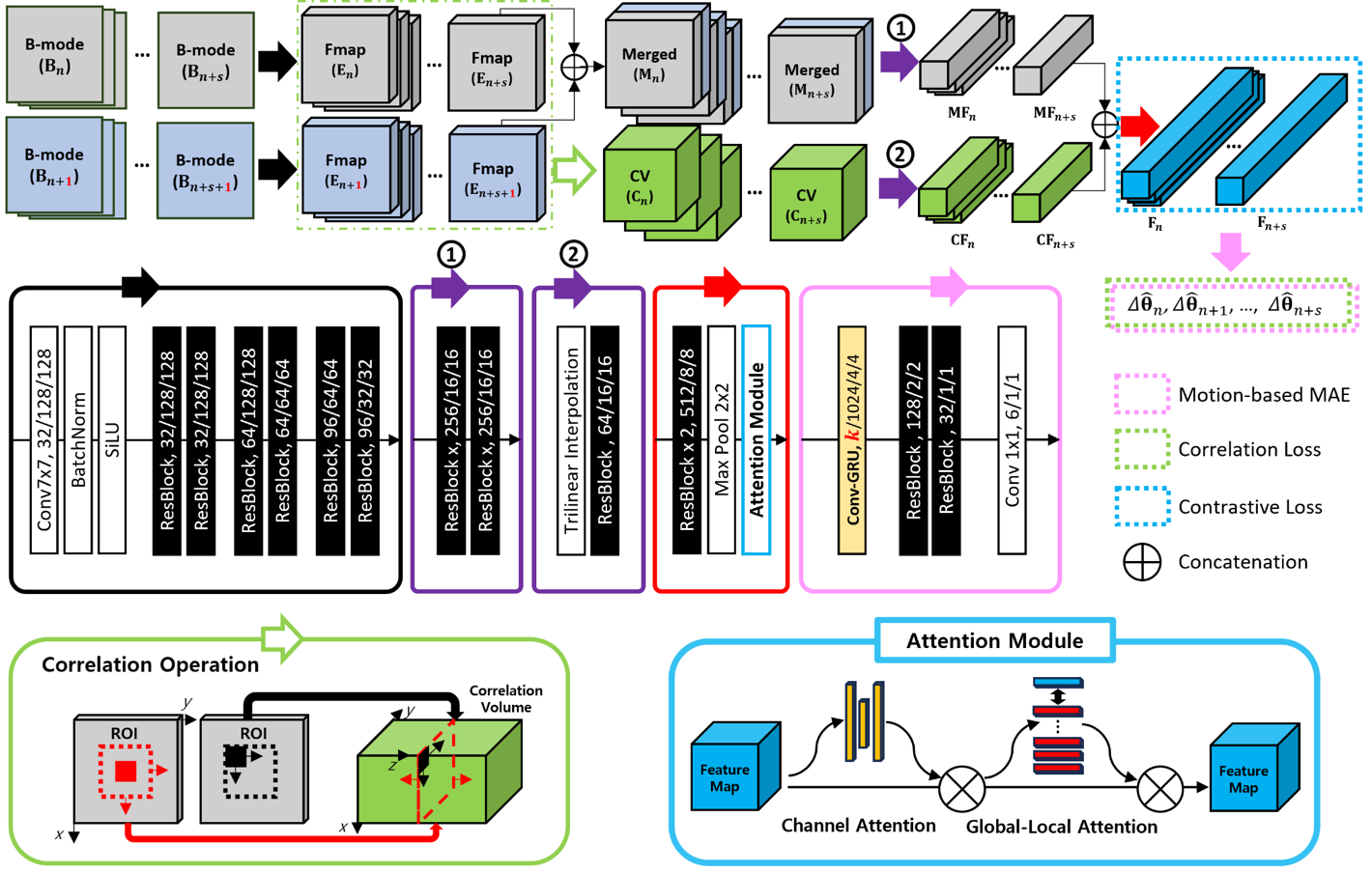}
\caption{Overview of MoGlo-Net.} \label{MoGlo-Net}
\end{figure}

\makeatletter
\renewcommand\paragraph{\@startsection{paragraph}{4}{\z@}
  {-1.25ex\@plus -1ex \@minus -.2ex}
  {0.75ex \@plus 0.1ex}
  {\normalfont\normalsize\itshape}}
\makeatother
\paragraph{MoGLo-Net\protect\footnote{\url{https://github.com/guhong3648/MoGLo}}}\label{MoGLo-Net}
Input images are cropped to square regions to remove background artifacts and normalised to $[-1, 1]$. Fig.~\ref{MoGlo-Net} shows overview of MoGLo-Net, a motion-based learning network with global-local attention. Two ultrasound sequences, each consisting of $s+1$ frames, are processed in parallel through a ResNet-based encoder for consistent feature refinement.

The correlation volumes $C$ are computed from encoded features of the two sequences. A \textit{patch-wise correlation operation}, inspired by \cite{chen1997determination, tuthill_1998_SD_2, gee_2006_SD_3}, models local relationships between successive frames by (1) defining a common region of interest (ROI) on adjacent feature maps, (2) computing cosine similarity between local patches by sliding one patch over the entire ROI, and (3) aggregating results from multiple ROIs to form the correlation volume.
The correlation volume encodes motion cues, enhancing the model's motion estimation accuracy.

The encoded features of the two input sequences are merged into $M$ and refined via encoder blocks (purple arrows) together with $C$. The resulting features are concatenated to form the final feature map $F$, where the \textit{global-local attention module} is applied. This module is designed as a self-attention mechanism~\cite{huang_2021_gloria,li_2017_GLA_1,le_2022_GLA_2}: (1) global (GF) and local features (LF) are downsampled by factors of 4 and 2, respectively; (2) LFs are extracted from early encoder layers as patch-wise feature blocks; (3) both GFs and LFs are refined using a conventional attention mechanism; (4) cosine similarity between GFs and LFs serves as attention scores to weight LFs; and (5) weighted LFs are projected to aggregate local information.
The final recalibrated feature, formed by concatenating GFs and LFs, is fed into the RNN-based estimator (Conv-GRU) to predict 6-DoF transformations $\Delta \theta$. 

Three loss functions are used: \textit{Correlation Loss} \cite{guo2022ultrasound} ensures motion consistency; \textit{Triplet Loss} \cite{guo2022ultrasound} contrasts the final feature maps; and \textit{Motion-based Mean Absolute Error (MMAE)} emphasises errors in fast-motion regions:
\begin{align}
    \mathcal{L}_{\text{MMAE}} = 
    \frac{1}{6(s+1)}
    \sum_{i=n}^{n+s} \sum_{k=1}^{6} 
    w_i \left| \Delta{\theta}_{i,k}^{gt} - \Delta{\theta}_{i,k}\right|
\end{align}
where $s+1$ is the length of input sequence and $k$ denotes dimensions of 6-DoF prediction. $w_i = \left| \Delta{\theta}_i \right| + \varepsilon$ is a motion-based weighting term. Fast motion errors are penalised more heavily, as $w_i$ increases with larger motion vectors. The smoothing term $\varepsilon$ prevents over-amplification.
This model predicts 6-DoF for the input sequences, but only the final frame's estimation is used during inference. Therefore, predicting the motion for the entire scan requires $N$ sequences, where $N$ is the number of frames in the whole scan. To do this, we add padding at the beginning of the scan frames. The global and local transformations are derived based on the TUS-REC2024 baseline code.

\makeatletter
\renewcommand\paragraph{\@startsection{paragraph}{4}{\z@}
  {-1.25ex\@plus -1ex \@minus -.2ex}
  {0.75ex \@plus 0.1ex}
  {\normalfont\normalsize\itshape}}
\makeatother
\paragraph{\texorpdfstring{PLPPI\protect\footnote{\url{https://github.com/Alphafrey946/PLPPI}}\protect\footnote{This work is summarised in \cite{10684746}, which provides further details on the method design.}}{PLPPI}}\label{PLPPI}

To address the complexities in trackerless freehand ultrasound reconstruction, particularly out-of-plane motion \cite{10684746}, a lightweight, physics-informed deep learning model is proposed. The dual-stream network decouples spatial and temporal learning, incorporating learnable operators to capture data priors for modeling temporal relationships and integrating a physical model to simplify learning, offering flexibility for various scanning paths.

The PLPPI model consists of spatial and temporal branches, followed by a fusion module and prediction head. The spatial branch uses 2D convolutions to aggregate intra-frame spatial context, while the temporal branch extracts inter-frame motion cues via speckle decorrelation patterns. This involves constructing a correlation volume to quantify the underlying motion information. The outputs from both branches are fused to represent the input sequence in the feature space, with speckle decorrelation serving as a key physics-based prior.
During training, input image stacks are split into two sub-volumes and passed through 2D convolutions to obtain $c_{21}$ and $c_{22}$. The temporal branch computes a correlation volume $cv$ by measuring patch-wise similarity between $c_{21}$ and $c_{22}$, where $c \in \mathbb{R}^{h \times w \times d}$. $cv$ is defined as: $cv(x_1, x_2) = \sum_{s \in [-p,p] \times [-p,p]} c_{21}(x_1+s)^T \cdot  c_{22}(x_2+s)$, with $x_1$, $x_2$ denoting the patch locations centered at $c_{21}$ and $c_{22}$, respectively, and $p$ the maximum displacement between $x_1$ and $x_2$. The squared patch size is $K=2p+1=21$. Temporal features are then bilinearly upsampled and fused with spatial features for joint representation. 

Compared to their preceding work \cite{10684746}, two key modifications are introduced: (1) replacing the ResNet backbone with the pretrained foundation model Biomedical CLIP \cite{zhang2023biomedclip}, and (2) redesigning the loss function to better leverage the capabilities of the foundation model. The new loss is defined as:
\begin{equation}
\mathcal{L} = \alpha \parallel\theta^{gt}-\theta\parallel^2 + \beta \parallel C( {I^{gt}})-C(I^{recon})\parallel^2 + \gamma \parallel {\theta^{gt}} \cdot p_{lmk} -\theta \cdot p_{lmk}  \parallel^2 
\end{equation}

The loss has three terms: (1) MSE between predicted and ground truth pose, (2) embedding consistency using Biomedical CLIP \cite{zhang2023biomedclip} on a ``reconstructed'' image $I^{recon}$, obtained by taking pixel-wise average of the two closest images to predicted ${\theta}$ and (3) projection loss as Euclidean distance between projected and true 3D landmarks, projected from 2D landmarks $p_{lmk}$. $\alpha$, $\beta$, $\gamma$ are hyperparameters. The Biomedical CLIP is finetuned with provided training data. 

During inference, the model outputs $n-1$ local transformations estimated from $n$ input images. Sliding window averaging is applied to obtain the final local transfromations: ${\theta}_{local}^{(i)} = \frac{1}{W} \sum^{i}_{j=i-W+1} \theta^{(j)}$, with window size $W$. Global transformation is then computed as $T_{global}^{(i)} = \prod^{i}_{j=1} T_{local}^{(j)}$ where $T_{local}^{(j)}$ is converted from ${\theta}_{local}^{(j)}$.

\subsubsection{Methodology Analysis Among Teams}
\label{Comparative_analysis}

Most of the proposed approaches leverage both spatial and temporal learning to capture long-term dependencies within ultrasound sequences. Examples include the use of Mamba modules in FiMoNet and LSTM networks in RecuVol. ResNet and EfficientNet are selected as backbone architectures across several methods. All models predict 6-DoF transformations. While most methods estimate frame-to-frame transformations, FlowNet predicts transformations between non-adjacent (interval) frames.
Regarding loss functions, the primary objective across methods is to minimise the discrepancy between the predicted and ground truth transformation parameters, commonly using $L1$, MSE, or Pearson correlation-based losses (e.g., FiMoNet, MoGLo-Net). Additional loss formulations are also utilised: MoGLo-Net uses a triplet loss; FlowNet uses a point-based loss on transformed coordinates and an MSE loss between original and warped ultrasound images; and PLPPI integrates embedding consistency loss and MSE loss on landmark coordinates.
Pre-training is adopted by three approaches: FiMoNet and RecuVol use ImageNet-pretrained weights, while PLPPI uses foundation model Biomedical CLIP. Ensemble learning is another common strategy. FiMoNet combines two distinct models; RecuVol aggregates eight models derived from two rounds of 4-fold cross-validation; and FlowNet selects three models from different training epochs and also combines predictions from sequences with different offsets.
\renewcommand{\arraystretch}{2}  
\setlength{\tabcolsep}{6pt} 
{\small
\begin{longtable}{>{\raggedright\arraybackslash}p{1.8cm}>{\raggedright\arraybackslash}p{1.8cm}>{\raggedright\arraybackslash}p{1.95cm}>{\raggedright\arraybackslash}p{1.95cm}>{\raggedright\arraybackslash}p{1.95cm}>{\raggedright\arraybackslash}p{1.95cm}>{\raggedright\arraybackslash}p{1.95cm}}

\caption{Implementation details of the baseline and top five participating methods, including model architectures, training setups, loss functions, and data processing strategies.}\label{imp}

\\\toprule
\multicolumn{1}{p{1.8cm}}{\raggedright Model abbreviation} & Baseline & FiMoNet & RecuVol & FlowNet & MoGLo-Net & PLPPI \\ 
\midrule
\endfirsthead

\toprule
\multicolumn{1}{p{1.8cm}}{\raggedright Model abbreviation} & Baseline & FiMoNet & RecuVol & FlowNet & MoGLo-Net & PLPPI  \\
\midrule
\endhead

\bottomrule
\endfoot

\bottomrule
\endlastfoot

Architecture & 2D CNN & \multicolumn{1}{p{1.95cm}}{\raggedright 2D CNN; State Space Model} & \multicolumn{1}{p{1.95cm}}{\raggedright 2D CNN (extracts features) followed by LSTM}  & 2D CNN& \multicolumn{1}{p{1.95cm}}{\raggedright 2D ResNet; Conv-GRU}& 2D CNN    \\ 
\hline

Backbone & \multicolumn{1}{p{1.8cm}}{\raggedright EfficientNet-B1} &  \multicolumn{1}{p{1.95cm}}{\raggedright ResNet18; Mamba} & ResNet & \multicolumn{1}{p{1.95cm}}{\raggedright EfficientNet-B6} & ResNet & \multicolumn{1}{p{1.95cm}}{\raggedright ResNet-50 from~\cite{lin2023pmc}} \\ 
\hline

\multicolumn{1}{p{1.8cm}}{\raggedright Input sequence length} & 2 & \multicolumn{1}{p{1.95cm}}{\raggedright Not fixed, depends on scan length} & $\sim$16 & 10  &5& 6  \\ 
\hline

Output & \multicolumn{1}{p{1.8cm}}{\raggedright Rigid; 6-DoF of adjacent frames} & \multicolumn{1}{p{1.95cm}}{\raggedright Rigid; 6-DoF of adjacent frames} & 6-DoF& 6-DoF &6-DoF& \multicolumn{1}{p{1.95cm}}{\raggedright Rigid; 6 DoF (utilising the representation in~\cite{zhou2019continuity})}  \\ 
\hline

\multicolumn{1}{p{1.8cm}}{\raggedright Model size (number of parameters)} & $\sim$6.5e6 & 1.8e7 & $\sim$1e7 & 4.1e7&3.3e7& 4.6e7  \\ 
\hline

\multicolumn{1}{p{1.8cm}}{\raggedright Model weights initialisation} & \multicolumn{1}{p{1.8cm}}{\raggedright Random initialisation}  & \multicolumn{1}{p{1.95cm}}{\raggedright ResNet (ImageNet-1K initialisation); Mamba (random initialisation)} & \multicolumn{1}{p{1.95cm}}{\raggedright ImageNet initialisation for CNN}& \multicolumn{1}{p{1.95cm}}{\raggedright Kaiming normal distribution}&\multicolumn{1}{p{1.95cm}}{\raggedright Random initialisation}& \multicolumn{1}{p{1.95cm}}{\raggedright Kaiming normal distribution}  \\ 
\hline

Pretraining & N/A & N/A & \multicolumn{1}{p{1.95cm}}{\raggedright ImageNet pretrained CNN backbone} & N/A&N/A& Biomedical CLIP   \\ 
\hline

\multicolumn{1}{p{1.8cm}}{\raggedright Train/Val/Test splits} & 3:1:1 & 5:1:4& \multicolumn{1}{p{1.95cm}}{\raggedright 5 fold cross validation} & 3:1:1 &45:5:3& 8:1:1  \\ 
\hline

Pre-processing & N/A & \multicolumn{1}{p{1.8cm}}{\raggedright Resize image to 50\% width and height}& \multicolumn{1}{p{1.95cm}}{\raggedright Normalising, downsampling  by 1.25 (for half of the final ensemble models)} & Normalising to $[0,1]$&\multicolumn{1}{p{1.95cm}}{\raggedright Cropping; scaling to $[-1,1]$}  & Fine-tuned Biomedical CLIP on the training dataset \\ 

\multicolumn{1}{p{1.8cm}}{\raggedright Data augmentation} & N/A & \multicolumn{1}{p{1.95cm}}{\raggedright Randomly sampling scans at different intervals; randomly flipping scans} & \multicolumn{1}{p{1.95cm}}{\raggedright PyTorch TrivialAugment} & \multicolumn{1}{p{1.95cm}}{\raggedright Flip the order of consecutive frames} &N/A& \multicolumn{1}{p{1.95cm}}{\raggedright Adding Gaussian noise, random cropping}   \\ 
\hline

\multicolumn{1}{p{1.8cm}}{\raggedright Data sampling}&N/A& \multicolumn{1}{p{1.95cm}}{\raggedright Randomly sampling scans of different lengths, ranging from 60 to 180}&\multicolumn{1}{p{1.95cm}}{\raggedright Sequences of 16 consecutive frames of the same scan}&N/A& \multicolumn{1}{p{1.95cm}}{\raggedright Randomly sample ultrasound sequence with 5 frames} &N/A\\ 
\hline

 External data  &  N/A  &  N/A  & N/A  &  N/A  &  N/A &  N/A  \\  
 \hline

Loss & \multicolumn{1}{p{1.8cm}}{\raggedright MSE loss on transformed points coordinates} & \multicolumn{1}{p{1.95cm}}{\raggedright  $L1$ loss and Pearson correlation loss on transformation parameters} & \multicolumn{1}{p{1.95cm}}{\raggedright  MSE loss on transformation  parameters} & MSE loss on transformed point coordinates, transform parameters and warped images& \multicolumn{1}{p{1.95cm}}{\raggedright MMAE loss, Correlation loss and Triplet loss} & \multicolumn{1}{p{1.95cm}}{\raggedright MSE loss, Consistency loss, Projection loss ($\alpha=1$, $\beta=0.69$, $\gamma=0.67$)} \\ 
\hline

 Optimiser &  Adam  & Adam  &  Adam  &  Adam  &  AdamW  &  AdamW  \\ 
 \hline

\multicolumn{1}{p{1.8cm}}{\raggedright Other details (e.g., any specific technique used)} & N/A & \multicolumn{1}{p{1.95cm}}{\raggedright Multi-directional state space model~\cite{yan2024fine}} & \multicolumn{1}{p{1.95cm}}{\raggedright Out of 5 folds, one was withheld due to unstable training} & N/A& \multicolumn{1}{p{1.95cm}}{\raggedright Motion-based MAE; correlation operation; global-local attention}
 & \multicolumn{1}{p{1.95cm}}{\raggedright Self-attention and shift-invariance~\cite{zhang2019making}; Bayesian search~\cite{dewancker2016bayesian} for hyperparameter tuning} \\ 

\end{longtable}
}

Illustrated in Table~\ref{imp}, all methods are trained in an end-to-end manner and utilise offline inference. Most teams adtopt the Adam optimizer, with two opting for AdamW. A uniform base learning rate of 1e-4 is used across all submissions, though learning rate scheduling varies, including approaches such as StepLR, ReduceLROnPlateau, and cosine annealing with warmup. Training epochs span from under 100 to 13,400. Batch size varies between 1 and 32. The teams utilise a variety of GPU configurations for model training, including single-GPU setups with NVIDIA Quadro GV100, RTX 3090, 4090, and A6000, as well as a dual-GPU setup with A40s. Training times range from 1.2 to 9.7 GPU days, depending on resources and setup.
None of the teams report the use of external data during training.
Standard preprocessing steps, including scaling, cropping, and normalisation, along with data augmentation techniques such as temporal sampling and flipping, are commonly applied.

\subsection{Results Analysis}
\label{Results}

\subsubsection{Overall Performance}
\label{Overall}

\begin{figure}[htbp]
    \centering
    \begin{subfigure}[b]{0.57\textwidth}
        \includegraphics[width=\textwidth]{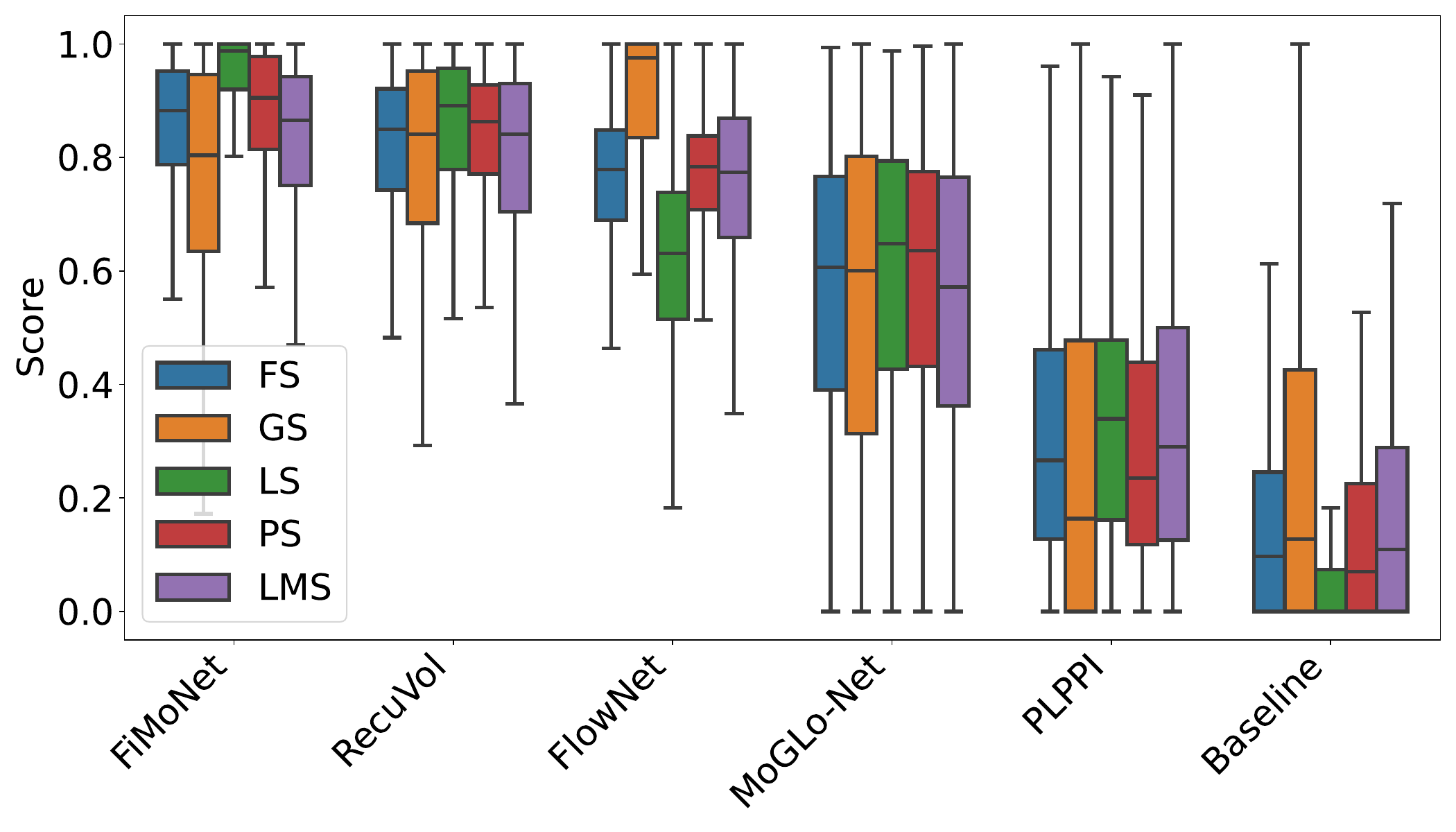}
        \caption{}\label{scores}
    \end{subfigure}
    \hfill
    \begin{subfigure}[b]{0.42\textwidth}
        \includegraphics[width=\textwidth]{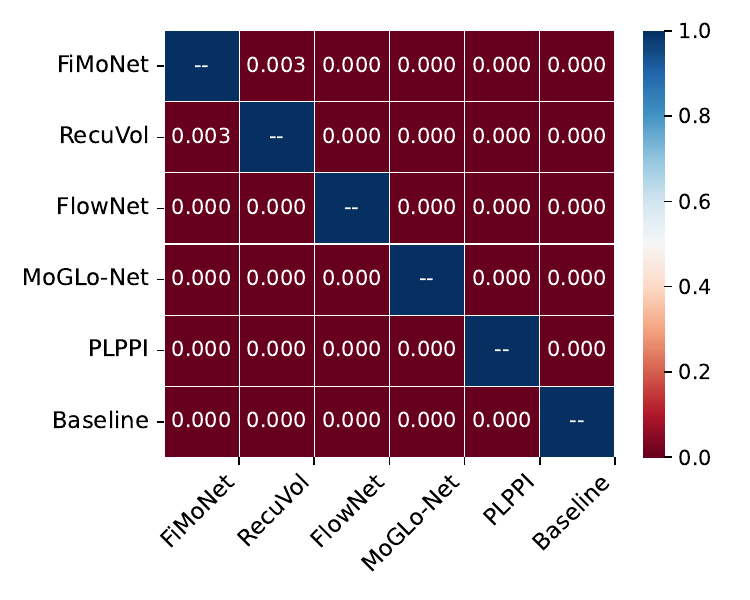}
        \caption{}\label{p-values}
    \end{subfigure}
    \caption{(a) Box plots illustrating the distribution of performance scores (FS, GS, LS, PS, LMS) across all test cases for each evaluated method (FiMoNet, RecuVol, FlowNet, MoGLO-Net, PLPPI, and Baseline). The central line in each box represents the median score, with box edges indicating the first and third quartiles, and whiskers extending to 1.5 times the interquartile range (IQR). (b) $p$-values for pairwise, subject-level statistical comparison of final scores across methods. Values close to zero indicate statistically significant differences in performance.
}
    \label{scores_p-values}
\end{figure}

\begin{figure}[htbp]
    \centering
    \begin{subfigure}[b]{0.49\textwidth}
        \includegraphics[width=\textwidth]{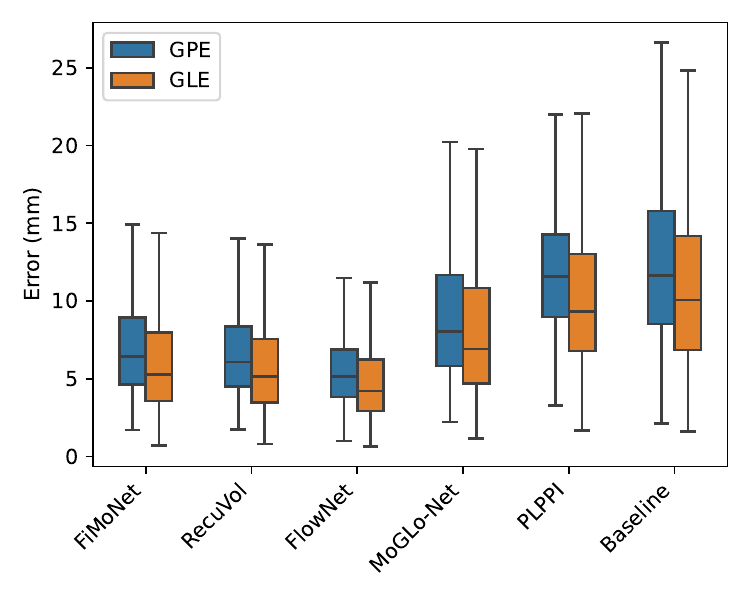}
        \caption{}
        \label{GPE_GLE_all_scans}
    \end{subfigure}
    \hfill
    \begin{subfigure}[b]{0.49\textwidth}
        \includegraphics[width=\textwidth]{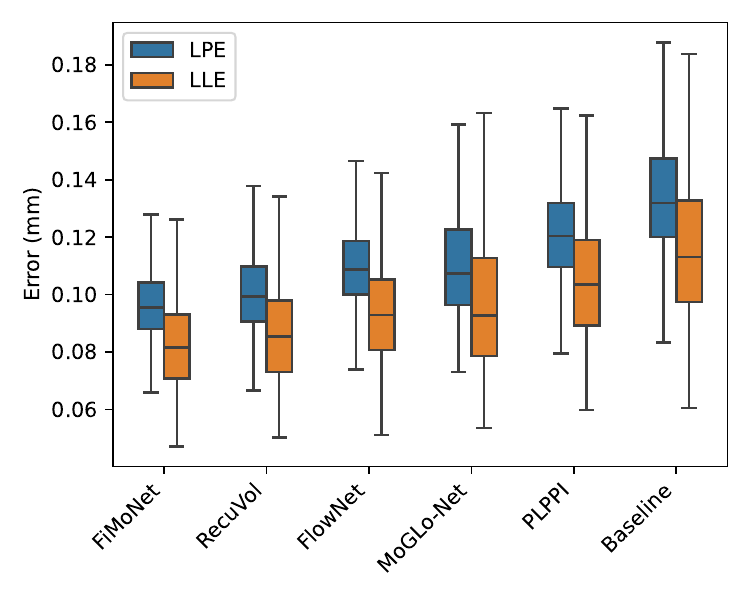}
        \caption{}
        \label{LPE_LLE_all_scans}
    \end{subfigure}
    \caption{(a): Box plots of GPE and GLE for each team. (b) Box plots of LPE and LLE for each team. In both subfigures, lower error values indicate better performance. The central line in each box represents the median value, with box edges indicating the first and third quartiles, and whiskers extending to 1.5 times the interquartile range (IQR).}
    \label{raw_metric_values_all_scans}
\end{figure}

Tables~\ref{results_score} and~\ref{results_metric} present the performance of each team, assessed using four evaluation metrics along with their corresponding normalised scores, as defined in Section~\ref{Evaluation_metrics} and \ref{Ranking_scheme}. Figs.~\ref{scores} and~\ref{raw_metric_values_all_scans} provide a graphical representation with more detailed distributions of the evaluation metrics and scores. The abbreviations FS, GS, LS, PS, and LMS refer to the final score, global score, local score, pixel score, and landmark score, respectively. 
The evaluation results demonstrate that composite metrics effectively capture the strengths and limitations of participating methods across multiple spatial levels. 

FiMoNet leads in 3 out of 4 normalised scores, particularly in local scores (LS: $0.951\pm0.074$), and in two unnormalised metrics about frame-to-frame accuracy (LPE: $0.097\pm0.014$, LLE: $0.084\pm0.019$), reflecting the advantage of its use of Mamba for temporal modeling, Pearson correlation-based loss, and dual-model ensembling. Its relatively low runtime also highlights a favorable balance between accuracy and efficiency. Close behind, 
RecuVol achieves better global metrics (GPE: $6.858\pm3.526$, GLE: $5.978\pm3.719$), but slightly lower local precision than FiMoNet. This indicates that while its LSTM-based temporal modeling and extensive ensemble setup (eight models) improve robustness, it may not capture local spatial structures as effectively. FlowNet, although achieving the lowest global errors (GPE: $5.970\pm3.523$, GLE: $5.167\pm3.682$), ranks lower in local-related metrics. This suggests that its interval-based frame prediction strategy and point/image-based loss functions capture coarse alignment well, but are less suited for precise local alignment. Its high inference time also presents a practical limitation.
MoGLo-Net and PLPPI show lower performance across all scores, with notably low landmark ($0.551\pm0.270$ and $0.322\pm0.240$) and global scores ($0.548\pm0.322$ and $0.272\pm0.302$). This suggests that their strategies, such as triplet loss in MoGLo-Net and embedding/projection losses in PLPPI, may not compensate for the lack of strong temporal modeling or ensemble learning. The Baseline model shows the lowest performance across all metrics, particularly in local alignment (LS: $0.056\pm0.106$), but requires the shortest run time due to its simplicity.

The final score, which normalises performance based on global and local transformations, on all pixel and landmark level errors, ranks FiMoNet ($0.852$) and RecuVol ($0.817$) highest, indicating superior overall accuracy. These methods employ temporal modeling (Mamba and LSTM, respectively) and ensemble strategies, suggesting that integrating spatial-temporal features and model ensembling contributes to consistent performance across spatial scales.
Disaggregated metrics reveal further insights. The global score, based on GPE and GLE, highlights models that excel in aligning entire ultrasound scan. FlowNet, despite ranking third overall, achieves the best GPE ($5.970$ mm), reflecting strong global transformation learning. However, its local score is substantially lower ($0.622$), indicating that precise local alignment is not guaranteed by low global error alone. In contrast, FiMoNet achieves the highest local score (LS: 0.951), suggesting its fine-grained feature extraction at multiple scales is particularly effective at capturing anatomical detail.

\begin{table}[H]
\begin{center}
\caption{Performance of participating teams expressed as normalised scores based on evaluation metrics. An upward arrow ($\uparrow$) denotes that higher values indicate better performance, while a downward arrow ($\downarrow$) indicates that lower values correspond to better performance. Values highlighted in bold represent the best-performing results for each score.}\label{results_score}
\small
\begin{tabular*}{\textwidth}{@{\extracolsep{\fill}}cccccccc@{\extracolsep{\fill}}}
\hline
Rank & \makecell{Model\\Abbreviation} & \makecell{FS ($\uparrow$)} & \makecell{GS ($\uparrow$)} & \makecell{LS ($\uparrow$)}  & \makecell{PS ($\uparrow$)} & \makecell{LMS ($\uparrow$)}  & Run Time (s) ($\downarrow$)     \\
\hline
1 & FiMoNet & $\mathbf{0.852\pm0.130}$   & $0.753\pm0.230$  &  $\mathbf{0.951\pm0.074}$   &  $\mathbf{0.875\pm0.122}$  &  $\mathbf{0.829\pm0.148}$   & $9.213\pm1.153$  \\
2 & RecuVol  & $0.817\pm0.140$   & $0.790\pm0.205$  &  $0.844\pm0.153$   &  $0.835\pm0.131$  &  $0.799\pm0.169$    & $17.173\pm1.800$    \\ 
3 & FlowNet   & $0.754\pm0.145$   & $\mathbf{0.886\pm0.182}$  & $0.622\pm0.169$   &  $0.757\pm0.135$  &  $0.751\pm0.175$   & $46.956\pm5.617$    \\ 
4 & MoGLo-Net   & $0.573\pm0.240$   & $0.548\pm0.322$  &  $0.598\pm0.246$  &  $0.595\pm0.233$  &  $0.551\pm0.270$   & $16.964\pm2.015$   \\ 
5 & PLPPI  & $0.303\pm0.215$   & $0.272\pm0.302$  &  $0.334\pm0.200$   & $0.285\pm0.209$  &  $0.322\pm0.240$ & $15.112\pm1.656$   \\ 
6 & Baseline  &  $0.146\pm0.159$   & $0.236\pm0.273$ &  $0.056\pm0.106$   &  $0.125\pm0.148$  &  $0.167\pm0.186$  & $\mathbf{8.135\pm0.996}$    \\ 

\hline
\end{tabular*}
\end{center}
\end{table}

\begin{table}[H]
\begin{center}
\caption{Performance of participating teams measured by evaluation metrics. An upward arrow ($\uparrow$) denotes that higher values indicate better performance, while a downward arrow ($\downarrow$) indicates that lower values correspond to better performance. Values highlighted in bold represent the best-performing results for each metric.}\label{results_metric}
\small
\begin{tabular*}{\textwidth}{@{\extracolsep{\fill}}ccccccc@{\extracolsep{\fill}}}
\hline
Rank & \makecell{Model\\Abbreviation}  & GPE (mm) ($\downarrow$) & GLE (mm) ($\downarrow$) & LPE (mm) ($\downarrow$) &  LLE (mm) ($\downarrow$) &  \\
\hline
1 & FiMoNet &  $7.191\pm3.687$  &  $6.281\pm3.812$  & $\mathbf{0.097\pm0.014}$ & $\mathbf{0.084\pm0.019}$  \\
2 & RecuVol  &  $6.858\pm3.526$  &  $5.978\pm3.719$  & $0.101\pm0.016$ & $0.088\pm0.021$     \\ 
3 & FlowNet    &  $\mathbf{5.970\pm3.523}$  &  $\mathbf{5.167\pm3.682}$  & $0.111\pm0.016$ & $0.096\pm0.022$     \\ 
4 & MoGLo-Net   &  $9.388\pm5.358$  & $8.459\pm5.699$  & $0.112\pm0.024$ & $0.100\pm0.033$   \\ 
5 & PLPPI  & $12.093\pm4.460$ &  $10.366\pm5.006$ & $0.122\pm0.019$ & $0.107\pm0.025$    \\ 
6 & Baseline  & $12.490\pm5.462$ &  $11.129\pm5.838$ & $0.135\pm0.024$ & $0.118\pm0.031$   \\ 

\hline
\end{tabular*}
\end{center}
\end{table}

The statistical testing results below demonstrate the validity and effectiveness of the evaluation metrics:
 \begin{itemize}[itemsep=-4pt, topsep=1pt]
\setlength{\labelsep}{5pt} 
  \renewcommand{\labelitemi}{\scalebox{0.6}{\textbullet}} 
     \item For the five normalised scores, at scan level, all pairwise team comparisons yield $p$-values below 0.001, except for the comparison between PLPPI and baseline ($p$-value $= 0.035$) in global score, and FlowNet vs MoGLo-Net ($p$-value $= 0.033$) in local score.
At subject level, for the five normalised scores, all pairwise team comparisons yield $p$-values below 0.001, except for the comparison between FiMoNet and RecuVol ($p$-value $= 0.003$) in final score, FiMoNet and RecuVol ($p$-value $= 0.021$) in global score, PLPPI and baseline ($p$-value $= 0.358$) in global score, FlowNet and MoGLo-Net ($p$-value $= 0.087$) in local score, FiMoNet and RecuVol ($p$-value $= 0.021$) in landmark score.

\item For the four error metrics, at scan level, all pairwise team comparisons result in $p$-values less than 0.001, except for the comparison between PLPPI and the baseline method ($p$-value $= 0.037$) in GPE metric, and FlowNet vs MoGLo-Net ($p$-value $= 0.011$) in LPE metric.
At subject level, all pairwise team comparisons result in $p$-values less than 0.001, except for the comparison between FiMoNet and RecuVol ($p$-value $= 0.008$) in GPE metric, PLPPI and baseline ($p$-value $= 0.348$) in GPE metric, FiMoNet and RecuVol ($p$-value $= 0.021$) in GLE metric, PLPPI and baseline ($p$-value $= 0.081$) in GLE metric, and FlowNet vs MoGLo-Net ($p$-value $= 0.043$) in LPE metric.
 \end{itemize}

\begin{figure}[htbp]
    \centering
    \begin{subfigure}[b]{0.365\textwidth}
        \includegraphics[width=\textwidth]{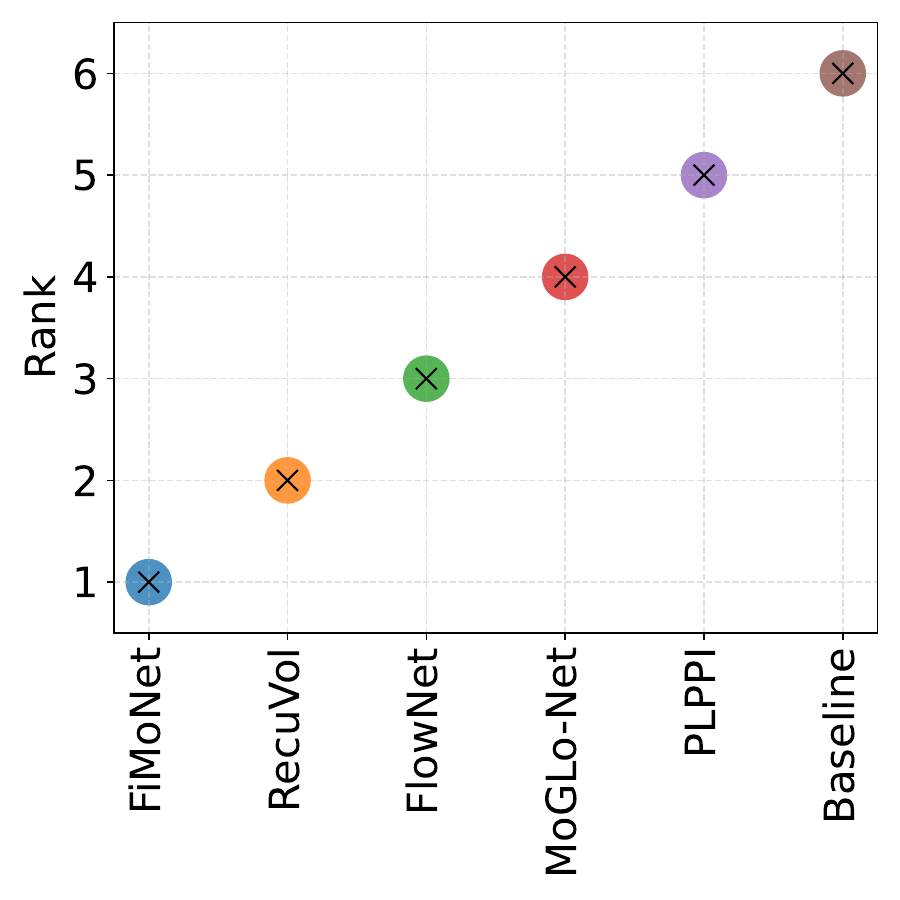}
        \caption{}
        \label{blob_ranking}
    \end{subfigure}
    \hfill
    \begin{subfigure}[b]{0.62\textwidth}
        \includegraphics[width=\textwidth]{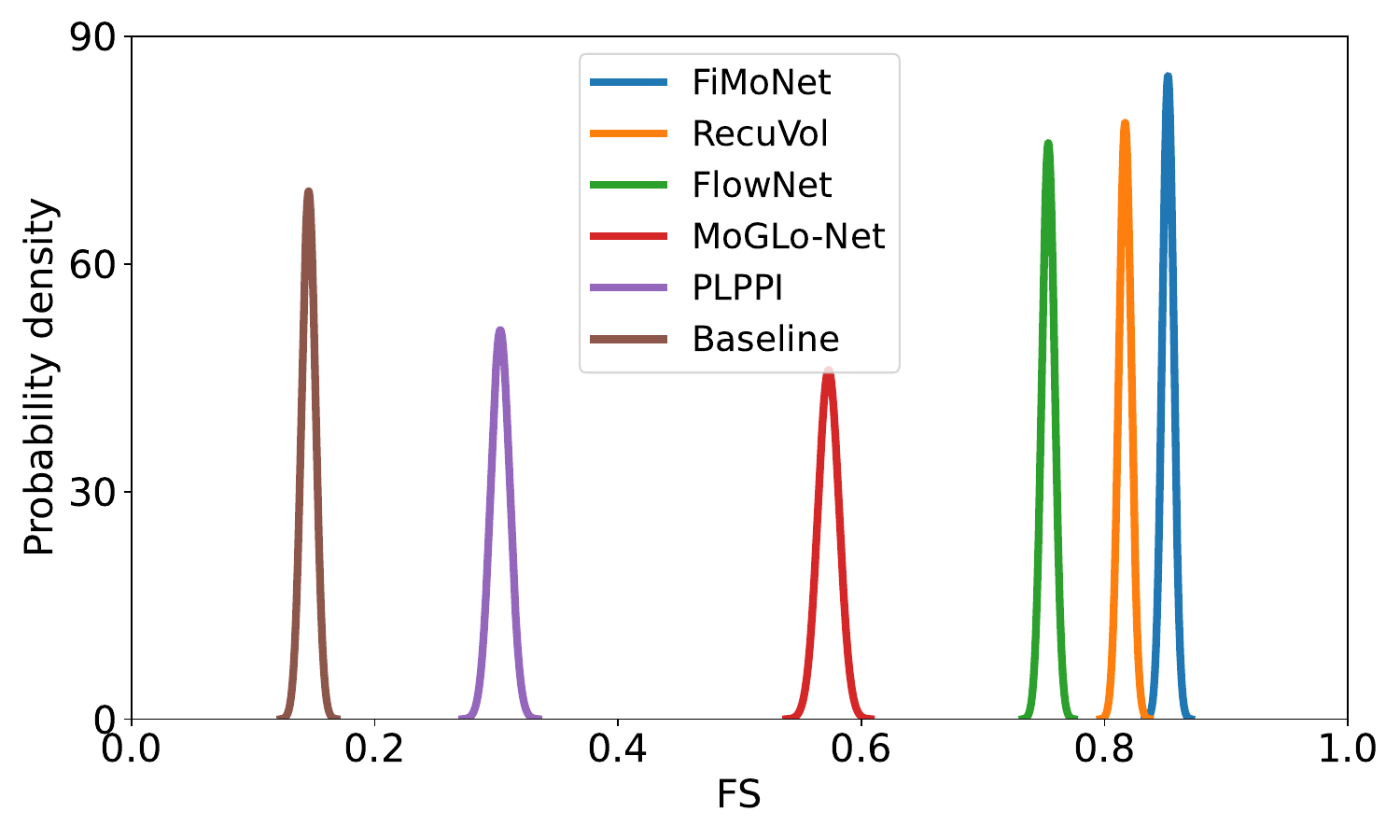}
        \caption{}
        \label{clt_distributions}
    \end{subfigure}
    \caption{(a) Ranking stability of participating teams using bootstrap sampling (2,000 bootstrap
samples, each equal to test set size). The area of each bubble corresponds to the relative frequency of the respective ranks observed across bootstrap samples. The median rank for each algorithm is represented by a black cross. (b) Sampling distributions of the mean final score for each algorithm, approximated using the Central Limit Theorem. The curves show the probability density of the sample mean, highlighting spread and separation between algorithm performances.}
    \label{blob_ranking_pdf}
\end{figure}

To assess the robustness of algorithm rankings, we conducted a bootstrap analysis using 2,000 resampled test sets. Specifically, each bootstrap sample was created by resampling the test cases with replacement, keeping the sample size unchanged. Fig.~\ref{blob_ranking} visualises the resulting rankings using a blob plot. The size of each bubble is proportional to the relative frequency of the corresponding ranks obtained across bootstrap samples. The median rank for each algorithm is denoted by a black cross. Notably, all algorithms exhibit perfect ranking consistency: each algorithm achieves the same rank in all 2,000 bootstrap samples, resulting in a single bubble per algorithm with 100\% frequency. This suggests that performance differences among the algorithms are highly stable under resampling. Fig.~\ref{clt_distributions} shows the estimated sampling distributions of the mean final score for each algorithm~\cite{dudley2014uniform}. Each distribution is modeled as a Gaussian (normal) distribution, where the center of the curve corresponds to the empirical mean of the final score for that algorithm, calculated across all bootstrap samples. The standard deviation is determined by the standard error of the mean, computed as the sample standard deviation divided by the square root of the number of bootstrap samples. The tightness and separation of these curves provide insight into the consistency and distinguishability of algorithm performance rankings.

\subsubsection{Team-wise Performance Comparison Across Scan Patterns}\label{analysis_sub_type_per_team}
\begin{figure}[H]
\centering
\includegraphics[width=\textwidth]{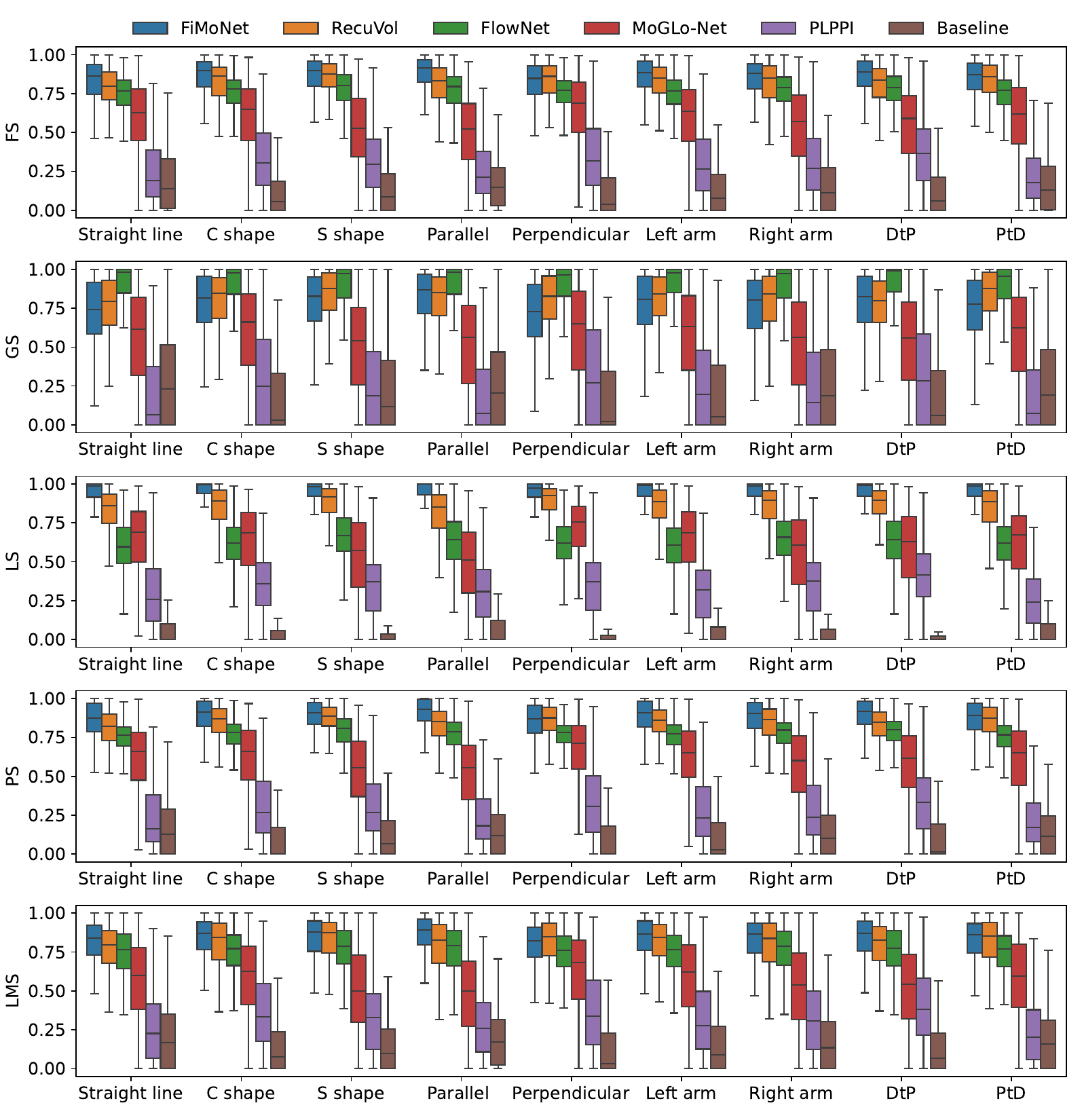}
\caption{Normalised scores for each team across various scan patterns: \textit{Straight line shape}, \textit{C shape} and \textit{S shape}; \textit{parallel} and \textit{perpendicular}; \textit{left arm} and \textit{right arm}; \textit{distal-to-proximal (DtP)} and \textit{proximal-to-distal (PtD)}.}\label{scores_sub_type_per_team}
\end{figure}

\begin{figure}[H]
\centering
\includegraphics[width=\textwidth]{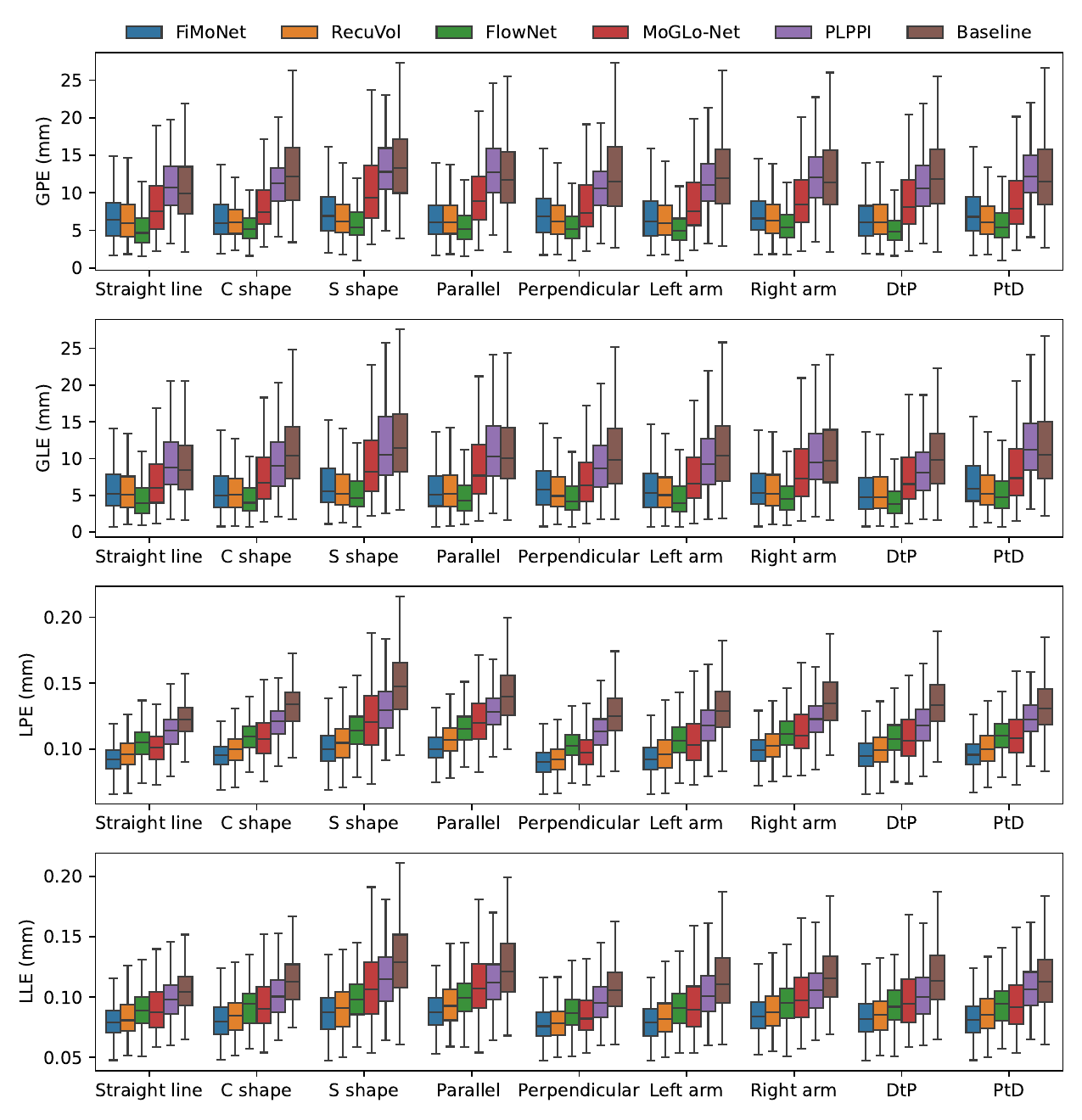}
\caption{Raw error metrics for each team across various scan patterns: \textit{Straight line shape}, \textit{C shape} and \textit{S shape}; \textit{parallel} and \textit{perpendicular}; \textit{left arm} and \textit{right arm}; \textit{distal-to-proximal (DtP)} and \textit{proximal-to-distal (PtD)}.}\label{metrics_sub_type_per_team}
\end{figure}

Figs.~\ref{scores_sub_type_per_team} and~\ref{metrics_sub_type_per_team} present the normalised scores and raw error metrics, respectively, for each team across various scan patterns. Consistently for all teams, \textit{Straight line shape} scans have the lowest local metrics (LPE, LLE) while \textit{S shape} scans have the highest local metrics. For global metrics (GPE, GLE), different scan shapes show more consistent performance, although \textit{S shape} scans still show slightly higher global errors, especially for methods like MoGLo-Net and PLPPI. This indicates that \textit{straight line shape} scans are more tractable for the evaluated methods, while \textit{S shape} scans are prone to greater drift and lower global consistency due to their complex trajectories. Overall, linear trajectories such as those in \textit{straight line shape} scans are handled more robustly by current methods, while more complex paths, especially \textit{S shape} scans, lead to significant degradation in both global and local performance. These findings highlight the importance of evaluating reconstruction robustness under diverse motion patterns and reinforce the need for algorithms that generalise effectively across varied scanning conditions.

Probe orientation also influences performance: all methods show higher LPE and LLE in \textit{parallel} scans than in \textit{perpendicular} scans, while GPE and GLE are comparable across orientations for all methods but MoGLo-Net and PLPPI. This indicates that the parallel orientation may induce more local misalignment, possibly due to less frame-to-frame overlap.

For most methods, \textit{proximal-to-distal} scans result in slightly higher error metrics compared to \textit{distal-to-proximal} scans.
In contrast, the influence of arm side (\textit{left} vs. \textit{right}) appears minimal, with comparable performance observed across both groups. 
Overall, these trends reveal the impact of scan shape, orientation, and direction on method robustness, and highlight the importance of developing reconstruction algorithms that generalise well across diverse scanning conditions. As shown in Fig.~\ref{scores_sub_type_per_team}, performance scores (FS, GS, LS, PS, LMS) remain consistent across different scan patterns, as they are normalised metrics tend to be independent of absolute values.

\subsubsection{Aggregated Performance over All Teams and Influence of Scan Length}
\label{analysis_sub-dataset}
Scan length (SL) is a critical factor affecting reconstruction performance. This section analyses the average performance of all methods with respect to both scan patterns and scan length. Overall, the observed relationship between performance and scan patterns is consistent with the findings reported in Section~\ref{analysis_sub_type_per_team}.

\begin{figure}[t]
\centering
\includegraphics[width=\textwidth]{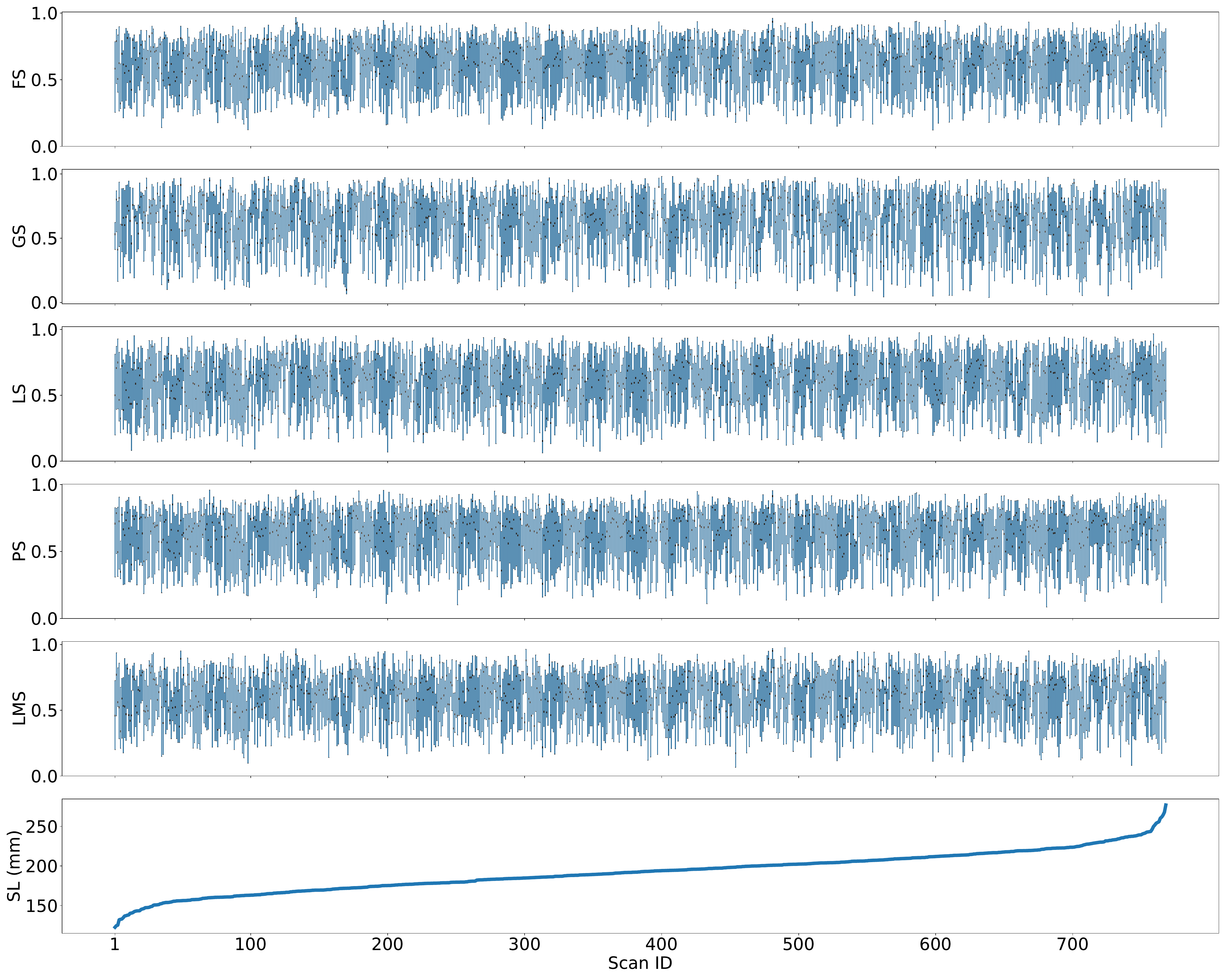}
\caption{Performance scores (FS, GS, LS, PS, LMS) for individual scans across all methods, arranged in ascending order of scan length (SL). Higher scores indicate better performance. Scan length is computed as the cumulative distance between the four corner points of consecutive frames.}\label{scores_per_scan_per_score}
\end{figure}

\begin{figure}[t]
\centering
\includegraphics[width=\textwidth]{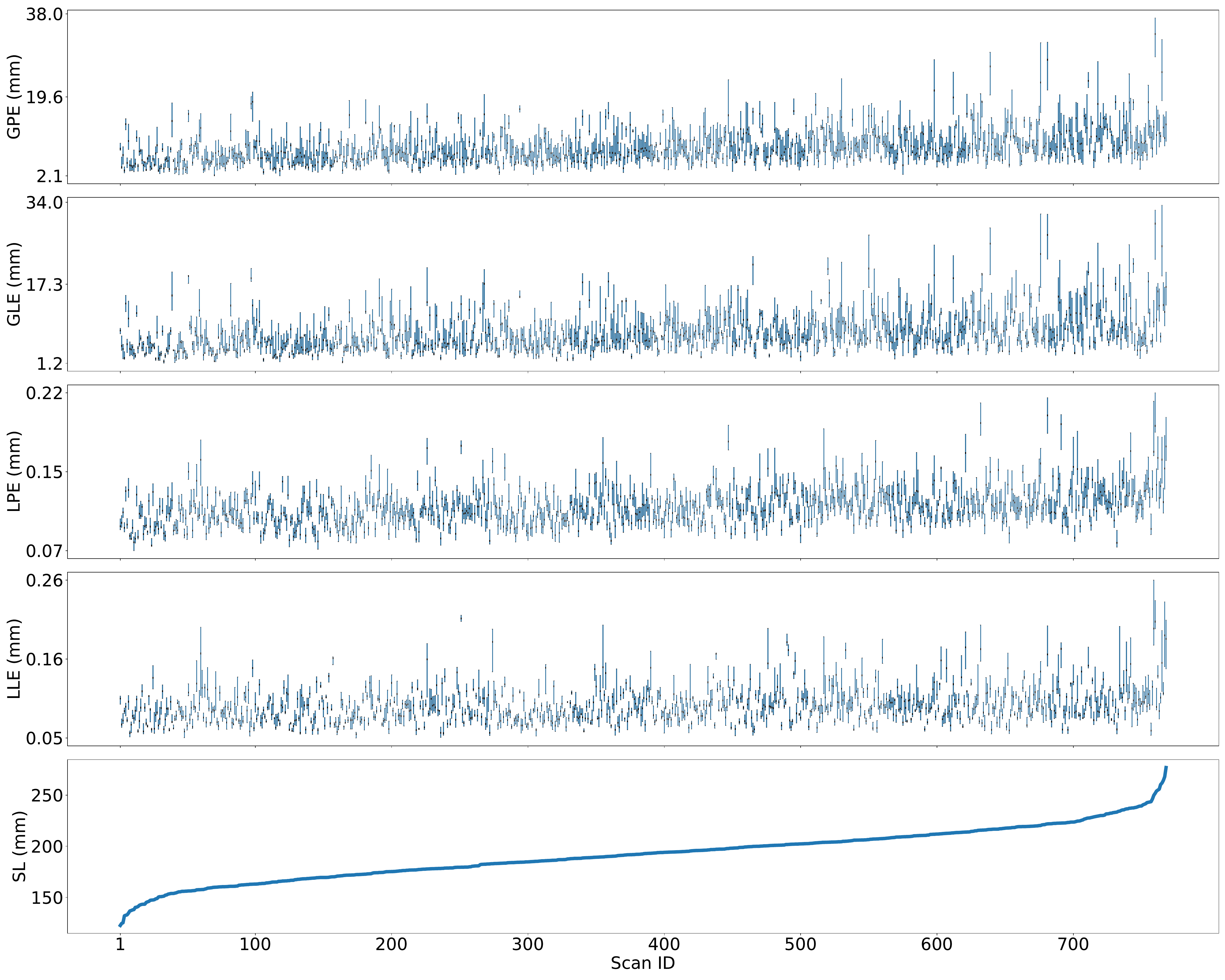}
\caption{Error metrics (GPE, GLE, LPE, LLE) for individual scans across all methods. All metrics are in millimeters (mm). Scans are sorted in ascending order of scan length (SL).}\label{metrics_per_scan_per_metric}
\end{figure}

Figs.~\ref{scores_per_scan_per_score} and \ref{metrics_per_scan_per_metric} show the distribution of performance scores and error metrics across all methods for each scan, respectively. In the horizontal axis, the scans are sorted by increasing scan length. The trend suggests that, in general, all error metrics tend to increase with scan length, indicating a potential deterioration in both global trajectory accuracy and local alignment. However, this pattern is not strictly consistent across all cases, and some scans show minimal or no significant degradation. This overall tendency indicates that longer sequences may lead to greater drift, which affects performance at both global and local spatial scales.
Performance scores (FS, GS, LS, PS, LMS) remain relatively stable across scan lengths, as they are normalised metrics designed to reduce sensitivity to overall error magnitude.

\begin{figure}[htbp]
    \centering
    \begin{subfigure}[b]{0.49\textwidth}
        \includegraphics[width=\textwidth]{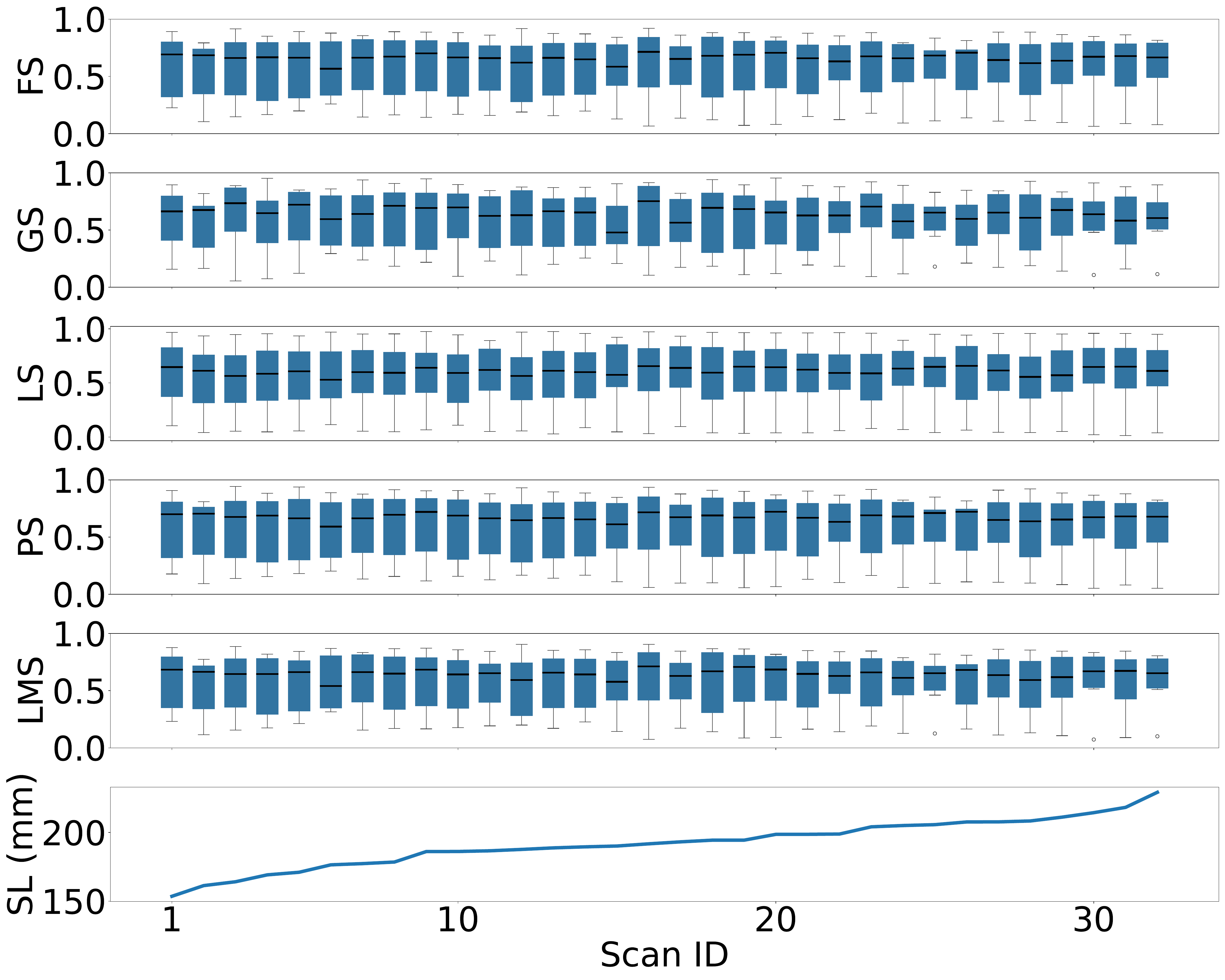}
        \caption{}
        \label{scores_per_subject_per_score}
    \end{subfigure}
    \hfill
    \begin{subfigure}[b]{0.49\textwidth}
        \includegraphics[width=\textwidth]{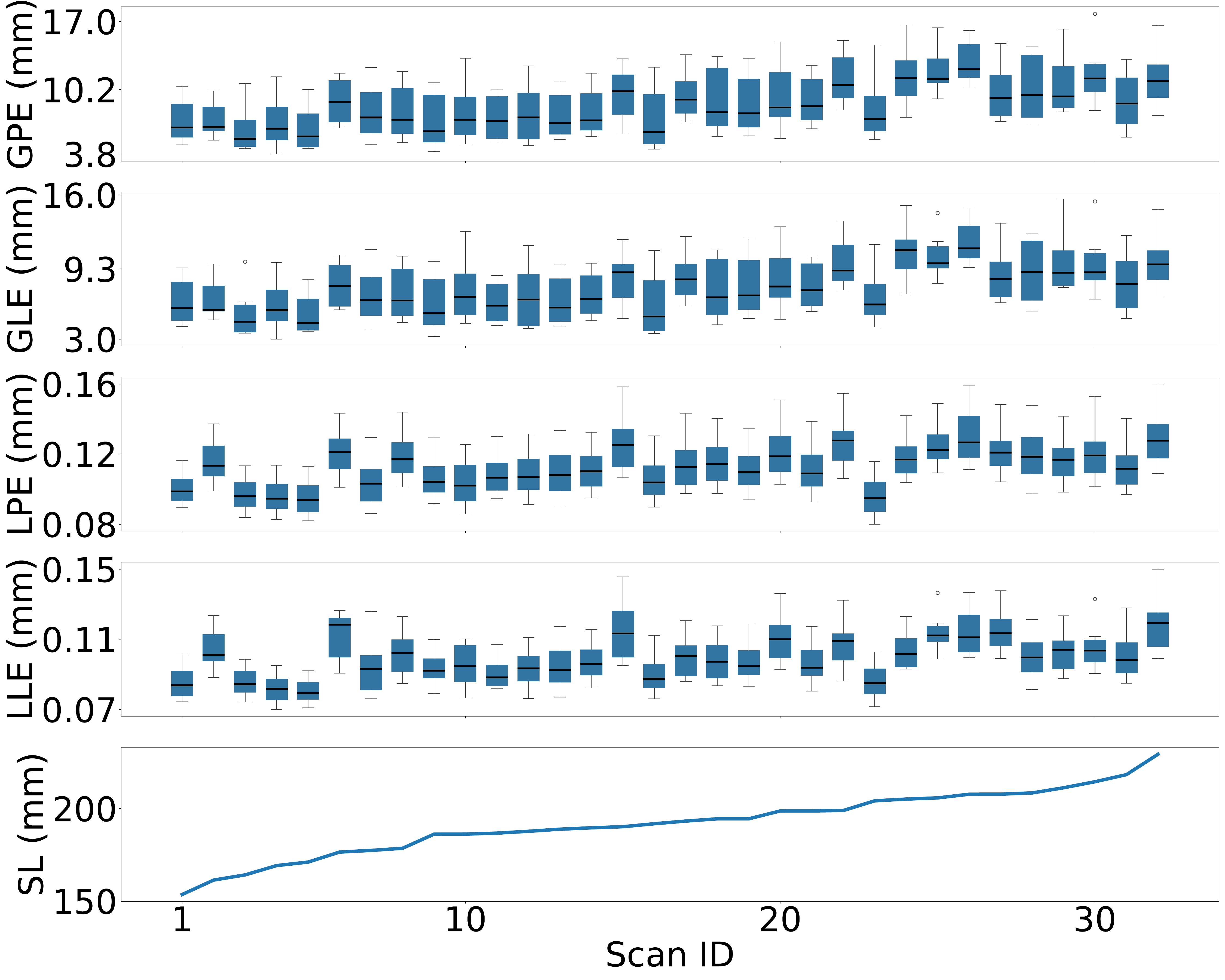}
        \caption{}
        \label{metrics_per_subject_per_metric}
    \end{subfigure}
    \caption{(a) Performance scores for individual subjects across all methods, arranged in ascending order of scan length. Higher scores indicate better performance. (b) Error metrics for individual subjects across all methods. Lower error indicates better performance. All metrics are in millimeters (mm). Scans are sorted in ascending order of scan length.}
    \label{score_metric_values_all_subjects}
\end{figure}

Fig.~\ref{score_metric_values_all_subjects} presents the distribution of performance scores and error metrics across all methods for each subject, with subjects sorted by increasing scan length. The performance scores remain relatively stable across subjects. 
The error metrics demonstrate a positive correlation with scan length, increasing in subjects with longer scans. 

\begin{figure}[t]
\centering
\includegraphics[width=\textwidth]{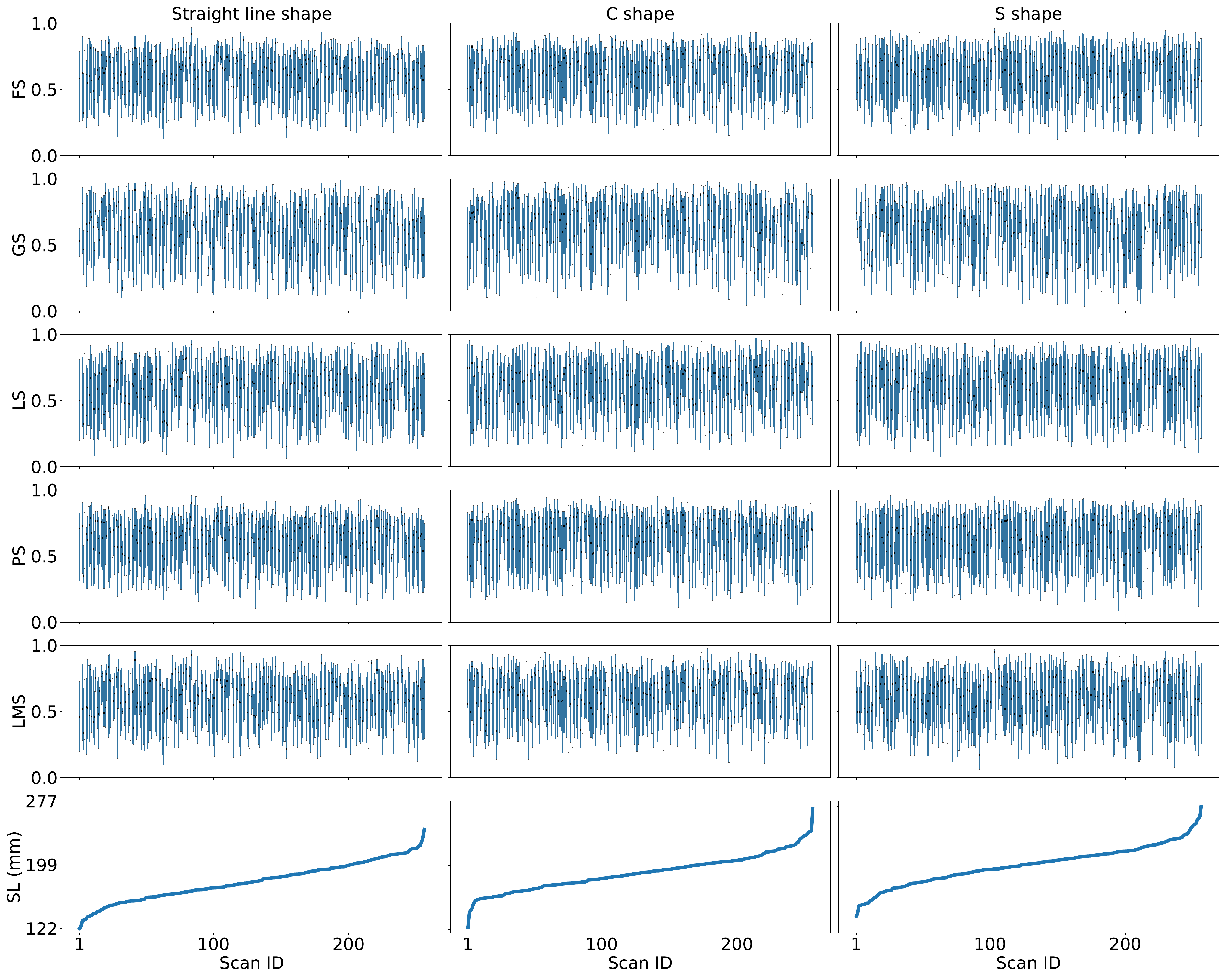}
\caption{Distribution of normalised performance scores for individual scans grouped by scanning protocol: \textit{Straight line shape}, \textit{C shape}, and \textit{S shape}. Scans are ordered by increasing scan length within each group.}
\label{scores_per_type_per_score}
\end{figure}

\begin{figure}[t]
\centering
\includegraphics[width=\textwidth]{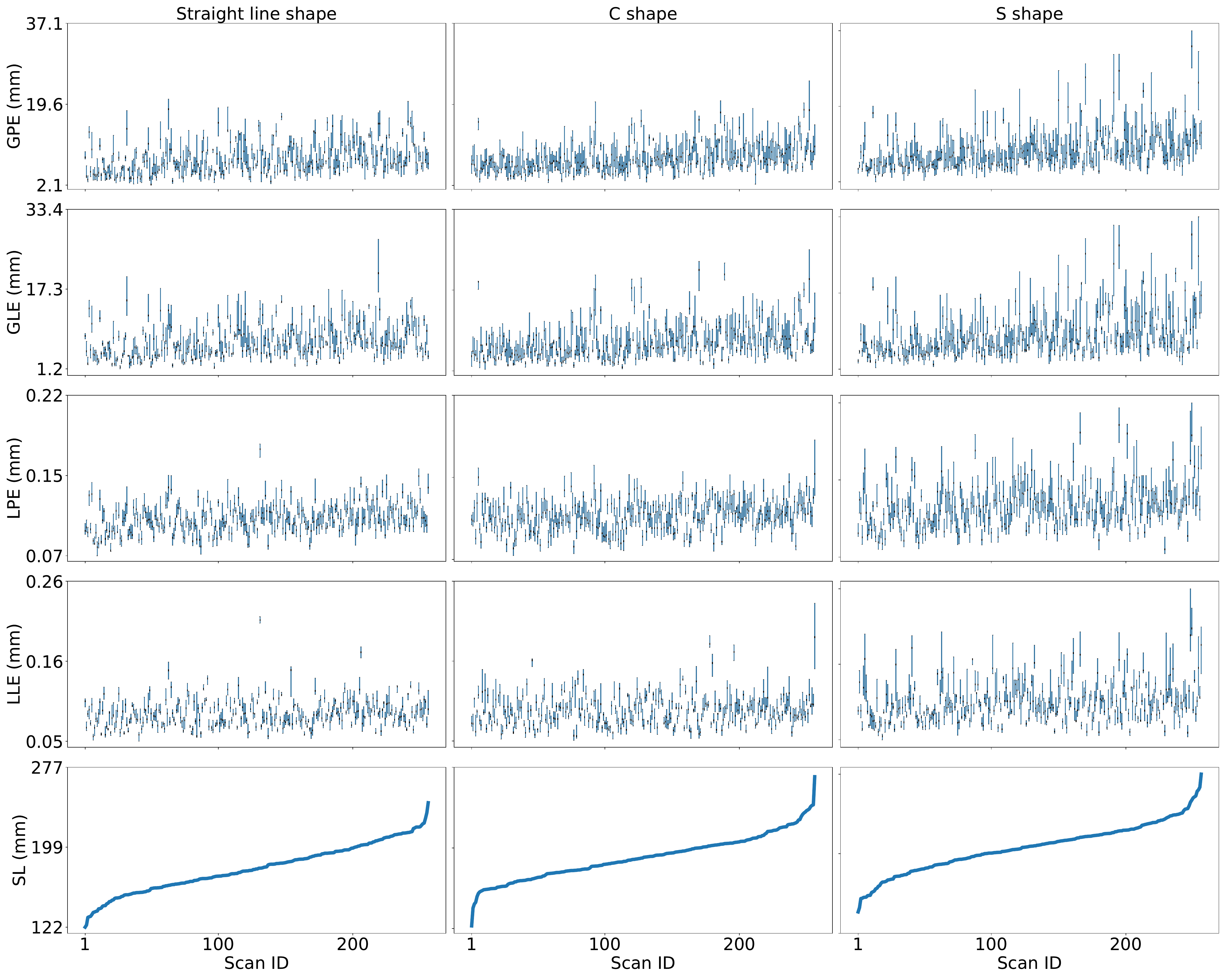}
\caption{Distribution of error metrics for individual scans grouped by scanning protocol: \textit{Straight line shape}, \textit{C shape}, and \textit{S shape}. Scans are ordered by increasing scan length within each group.}
\label{metrics_per_type_per_metric}
\end{figure}

Figs.~\ref{scores_per_type_per_score} and~\ref{metrics_per_type_per_metric} illustrate the distribution of performance scores and error metrics across individual scans, grouped by three scanning protocols: \textit{Straight line shape}, \textit{C shape}, and \textit{S shape}. Within each protocol, scans are ordered by increasing scan length, enabling assessment of both protocol-specific and length-dependent trends.
Across all three protocols, the normalised performance scores (FS, GS, LS, PS, LMS) generally remain within a consistent range, as expected. For the error metrics, global metrics (GPE, GLE) increase progressively with scan length in all scanning protocols.
\noindent Local errors (LPE and LLE) show stable trend, suggesting that local alignment is comparatively less sensitive than global consistency.
Comparing across scanning protocols, the \textit{straight line shape} scans yield the lowest and most stable error values across most metrics. In contrast, the \textit{S shape} scans are associated with the largest error and highest variability. \textit{C shape} scans fall between these extremes, showing moderate errors and variability. Notably, for scans of comparable length, \textit{S shape} sequences still perform worse, suggesting again that both scan length and path complexity jointly negatively impact method effectiveness.

\begin{figure}[t]
\centering
\includegraphics[width=\textwidth]{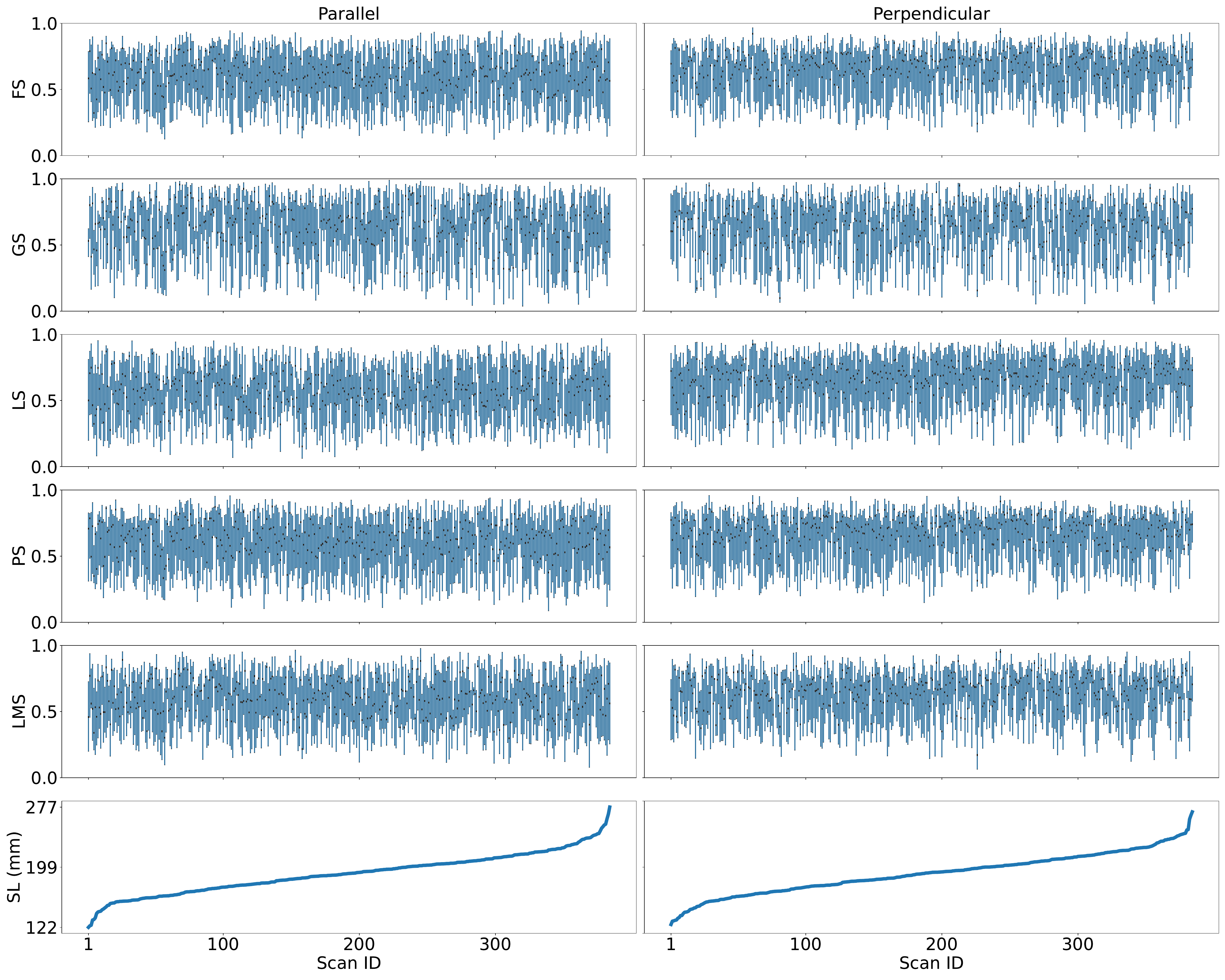}
\caption{Distribution of performance scores (FS, GS, LS, PS, LMS) for individual scans across all methods, categorised by probe orientation (\textit{parallel} and \textit{perpendicular}) and ordered by increasing scan length within each group.}\label{scores_par_per_per_score}
\end{figure}

\begin{figure}[t]
\centering
\includegraphics[width=\textwidth]{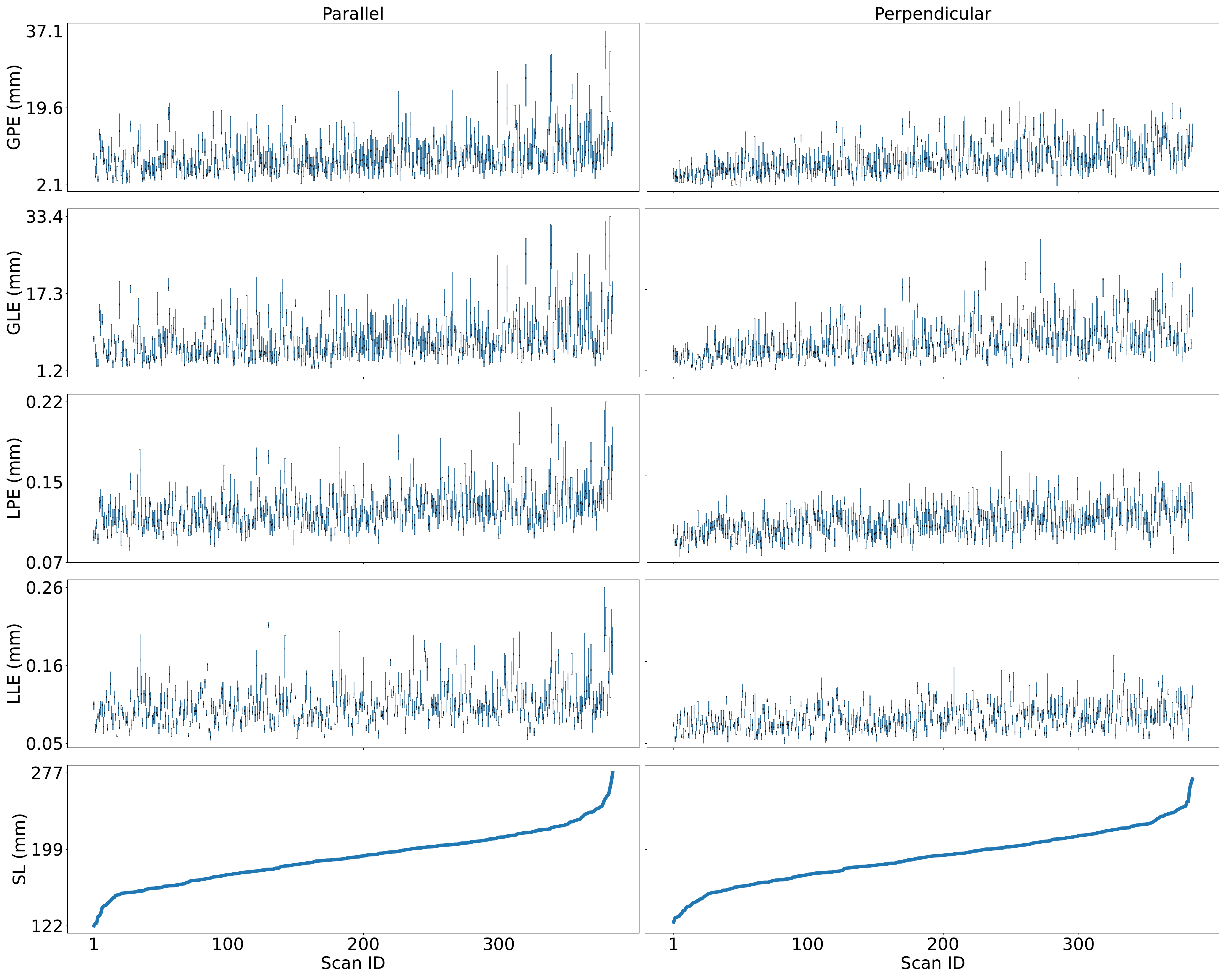}
\caption{Distribution of error metrics (GPE, GLE, LPE, LLE) for individual scans across all methods, categorised by probe orientation (\textit{parallel} and \textit{perpendicular}) and ordered by increasing scan length within each group.}\label{metrics_par_per_per_metric}
\end{figure}

Figs.~\ref{scores_par_per_per_score} and~\ref{metrics_par_per_per_metric} present the distribution of performance scores and error metrics across all evaluated methods for scans categorised by probe orientation, \textit{parallel} and \textit{perpendicular}, respectively. Scans are sorted by their length within each group. This configuration allows analysis of both orientation-dependent effects and scan length sensitivity.
Normalised performance scores are stable.
Both global errors and local errors increase with scan length across both probe orientations. However, the increase is more noticeable in the \textit{parallel} group, where both the magnitude and variability of errors are significantly higher. 
They demonstrate that both scan length and probe orientation significantly affect reconstruction performance. 

\begin{figure}[t]
\centering
\includegraphics[width=\textwidth]{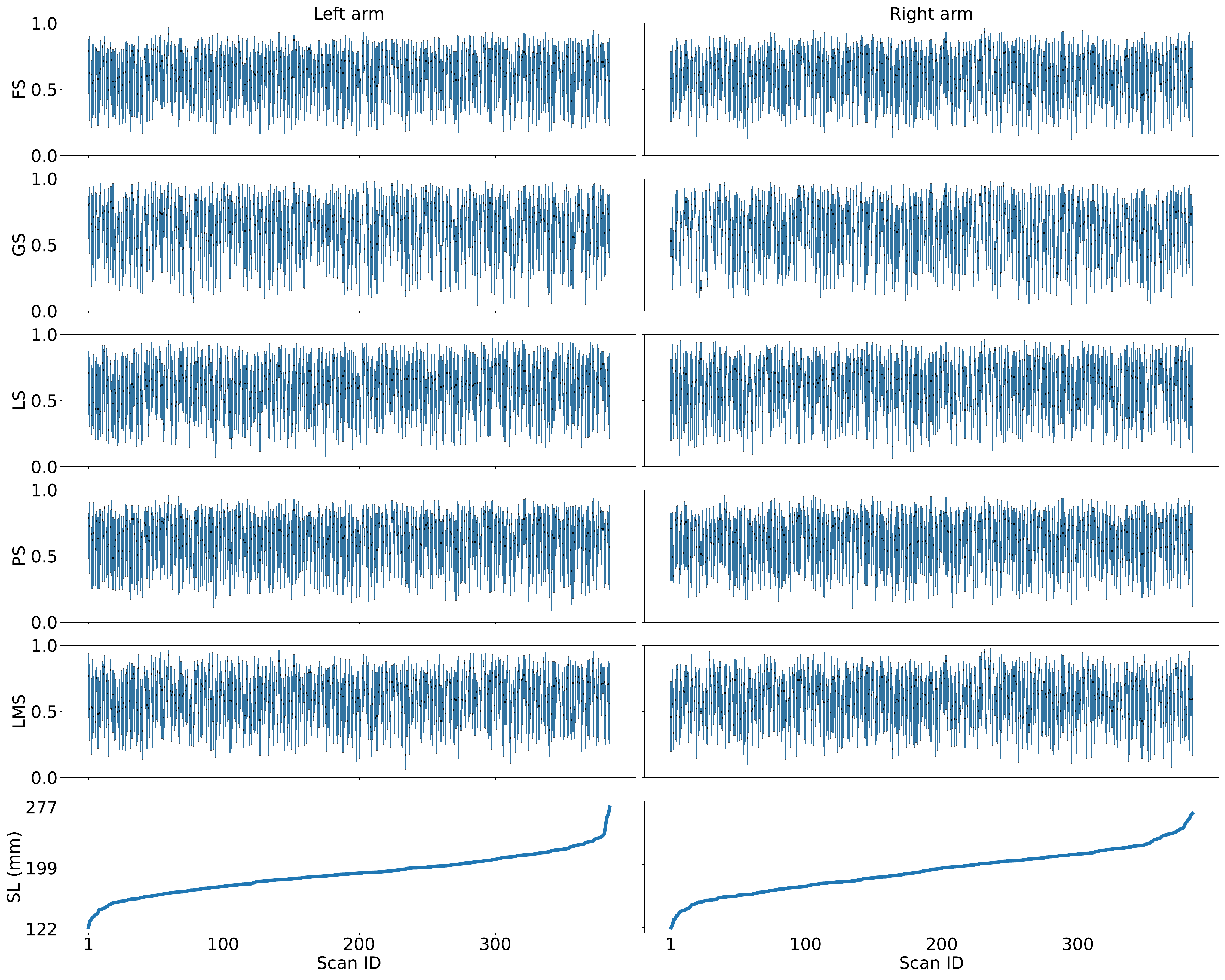}
\caption{Distribution of individual scores across all methods for each scan, categorised by scanned arm (\textit{left arm} and \textit{right arm}), and presented in ascending order of scan length.}\label{scores_LH_RH_per_score}
\end{figure}

\begin{figure}[t]
\centering
\includegraphics[width=\textwidth]{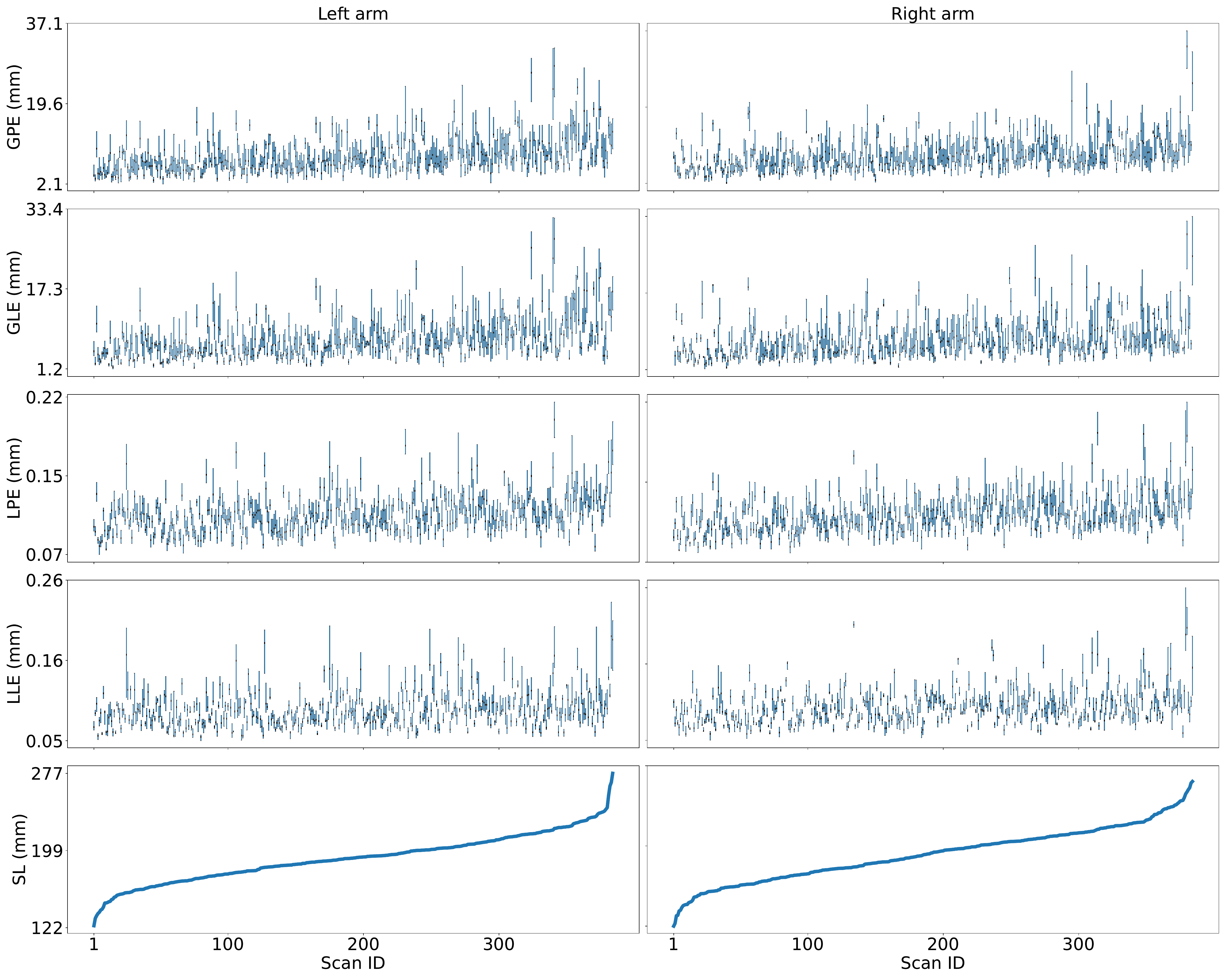}
\caption{Distribution of individual raw metric values across all methods for each scan, categorised by scanned arm (\textit{left arm} and \textit{right arm}), and presented in ascending order of scan length.}\label{metrics_LH_RH_per_metric}
\end{figure}

Figs.~\ref{scores_LH_RH_per_score} and~\ref{metrics_LH_RH_per_metric} show the distribution of performance scores and error metrics across all methods for individual scans, categorised by the scanned arm (\textit{left} or \textit{right}) and arranged in ascending order of scan length. 
Across both groups, normalised scores remain within a consistent range overall.
Error metrics consistently increase with scan length across both left and right arm scans. In addition, the error metrics are generally consistent between left and right arm scans, suggesting that the arm being scanned does not systematically influence the metric values across methods compared to scan length.

\begin{figure}[t]
\centering
\includegraphics[width=\textwidth]{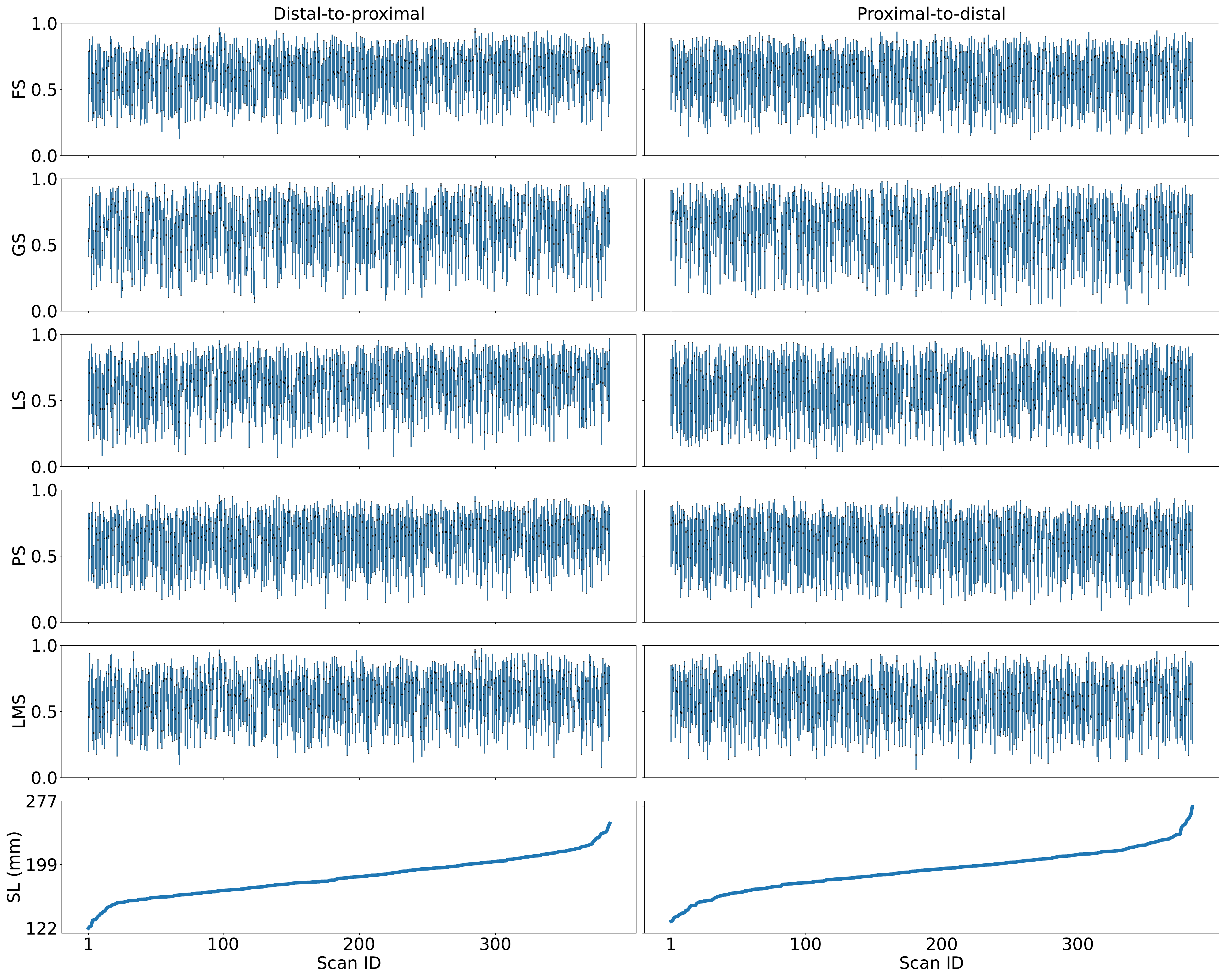}
\caption{Distribution of individual scores across all methods for each scan, categorised by scanning direction (\textit{distal-to-proximal} and \textit{proximal-to-distal}), and presented in ascending order of scan length.}\label{scores_DtP_PtD_per_score}
\end{figure}

\begin{figure}[t]
\centering
\includegraphics[width=\textwidth]{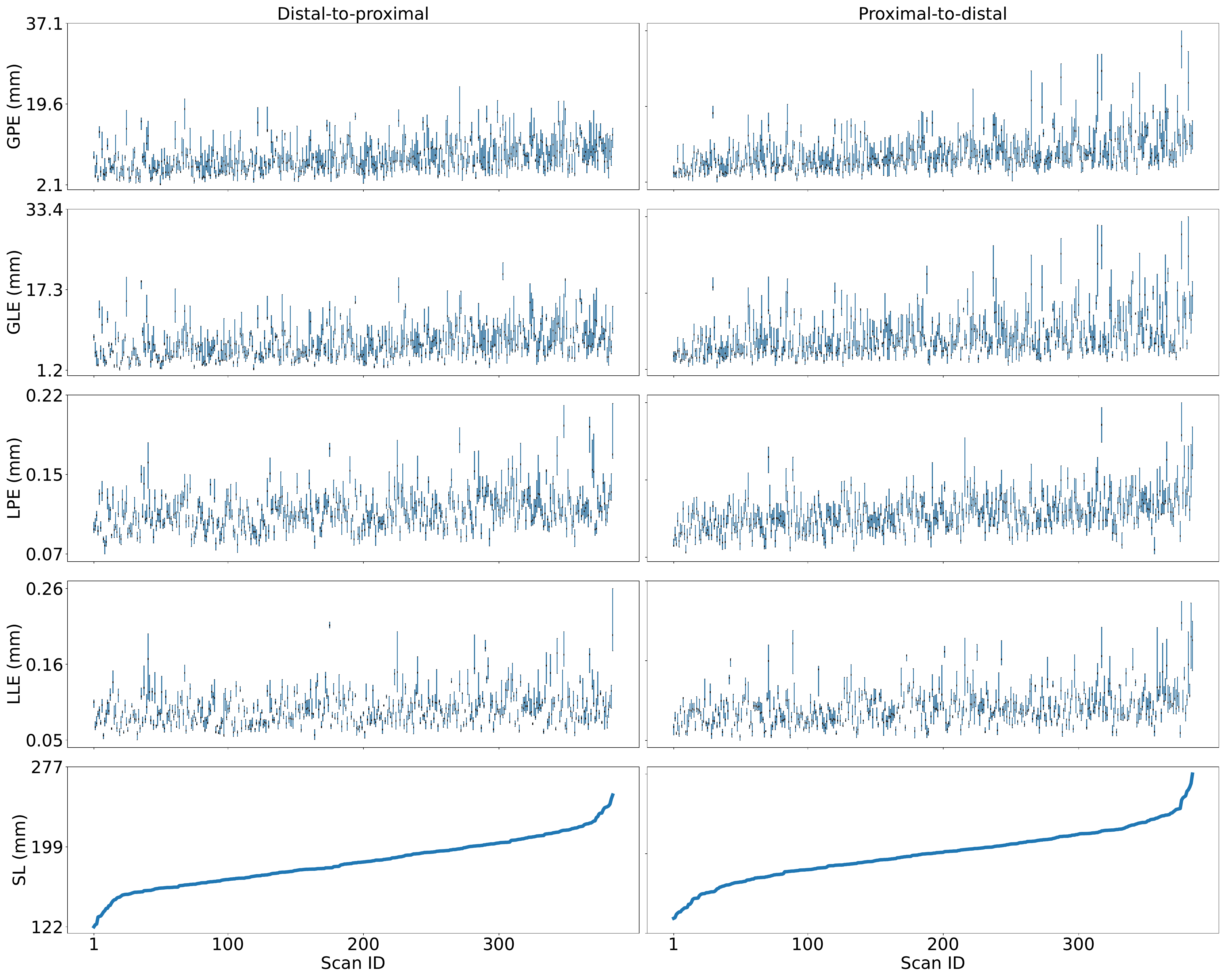}
\caption{Distribution of individual raw metric values across all methods for each scan, categorised by scanning direction (\textit{distal-to-proximal} and \textit{proximal-to-distal}), and presented in ascending order of scan length.}\label{metrics_DtP_PtD_per_metric}
\end{figure}

Figs.~\ref{scores_DtP_PtD_per_score} and~\ref{metrics_DtP_PtD_per_metric} explore the influence of scanning direction on performance, comparing scans acquired in a \textit{distal-to-proximal} (DtP) versus \textit{proximal-to-distal} (PtD) direction. 
Both global and local errors increase progressively with scan length in both directions, but with higher magnitudes and greater variability observed in the PtD group. In particular, GPE and GLE values are noticeably elevated in longer PtD scans, indicating more noticeable drift and loss of global alignment. Local metrics (LPE and LLE) also follow this trend but remain comparatively bounded, supporting the idea that global performance is more sensitive to scanning strategy and sequence length.
These findings suggest that scanning direction influences the spatial alignment and reconstruction performance of sequential frames. 
The amplified degradation in PtD scans indicates that the choice of scanning direction is a non-negligible factor in performance, especially for longer or more complex trajectories.
Overall, these results highlight the importance of considering scanning direction as a variable in both evaluation and design of spatial tracking reconstruction systems. Future research may investigate direction-sensitive strategies for drift correction to improve robustness across scan patterns.

The trends observed across figures above consistently demonstrate that both global and local error metrics increase with scan length, regardless of scan patterns (e.g., protocol, orientation or direction). This suggests a strong dependence of metric magnitude on scan length. To validate this observation quantitatively, the correlation between scan length and each metric is assessed using the Pearson correlation coefficient ($r$).
For most metrics, $r$ values range from 0.3 to 0.5, indicating moderate positive correlations between scan length and error magnitude. Notably, the four subject-level metrics exhibit stronger correlations, with $r$ values between 0.58 and 0.78, suggesting the association between scan length and performance degradation at the subject level. In all cases, the correlations are statistically significant ($p < 0.05$).

\subsubsection{Qualitative Results}
\label{qualitative_results}
\begin{figure}[H]
\centering
\includegraphics[width=\textwidth]{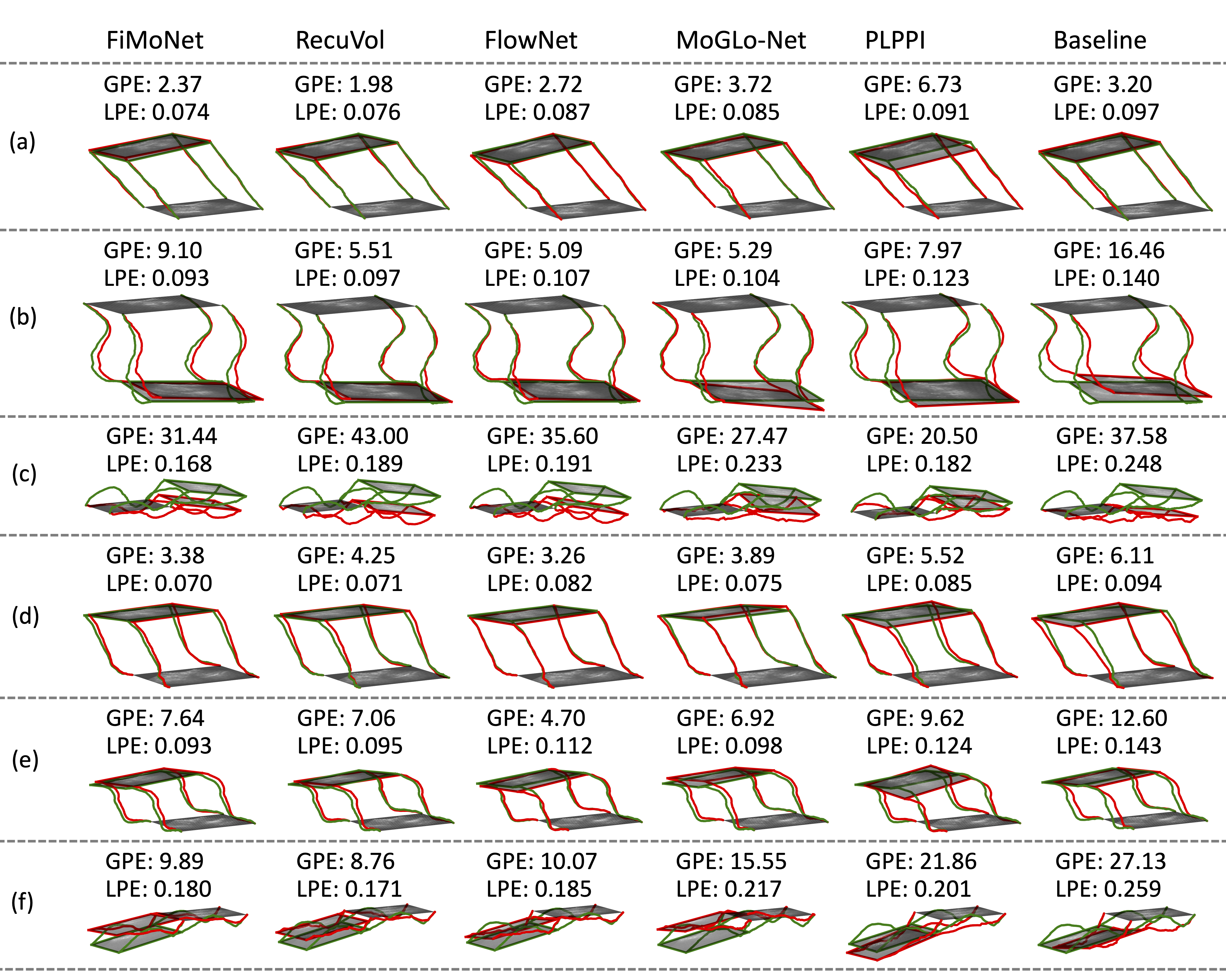}
\caption{Trajectories of the four corner points from the ground truth (green lines) and model predictions (red lines) on selected scans. (a): straight line shape scan on the left arm along a perpendicular scanning path, from the distal to the proximal direction; (b) S shape scan on the right arm along a perpendicular scanning path, from the proximal to the distal direction; (c): S shape scan on the right arm along a parallel scanning path, from the proximal to the distal direction; (d): C shape scan on the left arm along a perpendicular scanning path, from the distal to the proximal direction; (e): S shape scan on the left arm along a perpendicular scanning path, from the distal to the proximal direction; (f): S shape scan on the right arm along a parallel scanning path, from the distal to the proximal direction.}\label{traj_plots}
\end{figure}

Fig.~\ref{traj_plots} plots the predicted trajectories for each participating teams, selecting the scans with the best, worst, and median performance based on average GPE (subfigures a-c) and average LPE (subfigures d-f) of all teams, respectively. To include as diverse scan patterns as possible so that representativeness can be guaranteed, the scan with the next-lowest (or next-highest / next-median) error was selected until the three cases selected based on each metric have different types of scan shapes. To show the quantitative difference, the corresponding GPE and LPE values are also provided. Fig.~\ref{traj_plots} illustrates that accurate local predictions do not necessarily ensure accurate global reconstructions. In addition, the error magnitude varies across different scans.

\subsubsection{Further Analysis}
\label{Further_analysis}

Across all evaluated methods, a primary limitation is the sensitivity to scan length, as reflected by the correlation between scan length and error magnitude. This suggests that most approaches struggle to maintain lower global error over long sequences, likely due to lack of long range correlation. Furthermore, performance degradation is particularly prominent in geometrically complex trajectories (e.g., \textit{S shape} scans), indicating an overall lack of robustness to scanning path variability. 

While all teams share these global limitations, individual methods demonstrate distinct advantages. Some teams excel in local consistency, achieving low LPE and LLE across a wide range of scans, indicating effective frame-to-frame alignment strategies. Others achieve stronger performance on long scans, suggesting better global modeling. However, no team consistently outperform others on all metrics. This inconsistency across metrics highlights the current trade-offs between local precision and global robustness. 

Notably, the performance gain brought by ensemble learning and pretraining highlights the value of integrating uncertainty calibration and prior visual knowledge, which could be considered when clinically deployed. Taken together, these results suggest that future development should prioritise hybrid approaches that combine local precision, temporal consistency, and robust drift correction. Additionally, performance difference across scan patterns emphasises the need for improving adaptability in clinical scanning environments.

\section{Discussions}
\label{Discussions}

While the current Challenge setup offers a comprehensive evaluation of trackerless freehand 3D ultrasound reconstruction methods, several limitations remain, which also highlight opportunities for improvement in future iterations.

\textit{Reducing barriers to participation.}
A key priority for future editions of the TUS-REC Challenge is to lower the technical entry barrier, thereby enabling broader participation from research groups across computer vision, robotics, and medical imaging domains, including those without prior experience in freehand ultrasound reconstruction. Currently, the prerequisite knowledge of ultrasound imaging principles, spatial calibration procedures, and coordinate systems may discourage otherwise capable participants. By providing more detailed documentations, the Challenge can broaden accessibility and encourage participation from a more diverse range of research communities. Notably, with existing effort provided by this Challenge paper, this barrier has already been significantly lowered, laying a strong foundation for continued growth and engagement.

\textit{Wider anatomical areas and clinical applications.}
Another limitation of the present Challenge lies in the anatomical scope of the dataset, which is restricted to forearm scans. While the forearm offers a tractable and clinically relevant use case, it presents relatively constrained geometry and motion characteristics. Therefore, it may not fully capture the broader spectrum of challenges encountered in other anatomical regions. This may limit the generalisability of method performance observed in the current setting. 

\textit{Score weighting and ranking methodologies.}
In TUS-REC2024 Challenge, a min-max normalisation strategy is applied at the scan level, scaling performance scores into a fixed range of $[0, 1]$. While this approach helps reduce the influence of extreme values, it introduces several limitations. For example, it can over-amplify marginal differences in performance when overall variability is low, leading to exaggerated score separation between similarly performing methods. Additionally, the presence of a poorly performing team can artificially boost the normalised scores of the other teams. These effects may distort the overall fairness and interpretability of rankings.
It is important to acknowledge that all normalisation strategies inherently involve trade-offs, and no single method is universally optimal across all evaluation contexts. Min-max normalisation can suppress the impact of outliers but may exaggerate differences when overall performance is similar. In contrast, z-score normalisation avoids exaggerated differences but cannot suppress the impact of outliers as min-max normalisation does.     
Ultimately, each method emphasises different aspects of relative performance, and the choice of normalisation inevitably shapes the interpretation of results. While careful selection and justification of normalisation methods can improve transparency, there is no definitive solution that fully eliminates bias or distortion in score scaling. Therefore, normalisation should be viewed as an empirical approach with acknowledged limitations.

\textit{Risk of data leakage.}
As the Challenge allows multiple submission attempts and uses a fixed test set, there is a risk that participating methods may become inadvertently overfitted to the test data over time. This is especially relevant when teams refine their models based on feedback from repeated evaluations, potentially optimising for the specific test distribution rather than generalisable performance. Such overfitting can undermine the fairness and validity of the ranking results.
To mitigate this issue, future editions of the TUS-REC Challenge should consider introducing additional unseen test data in later evaluation phases. This could include a hold-out set only revealed after the final submission deadline or a progressive test set release strategy. Incorporating fresh data would better evaluate the generalisation ability of submitted methods and reduce the likelihood of overfitting to a static benchmark. Moreover, it would more closely reflect real-world deployment conditions, where models must perform reliably on previously unseen patients and scanning conditions.

\section{Conclusion}
\label{Conclusion}

Trackerless 3D freehand ultrasound reconstruction represents a critical advancement in enabling cost-effective, portable, and workflow-friendly 3D imaging solutions for diverse clinical settings. By eliminating the need for external tracking hardware, these methods promise improved accessibility in point-of-care diagnostics and interventional guidance. However, it also introduces new algorithmic challenges in accurate motion estimation under unconstrained probe motion. 

TUS-REC2024 Challenge represents a major step forward in benchmarking the current state-of-the-art for trackerless 3D freehand ultrasound reconstruction. With the large publicly available dataset for this task and participation from leading international research teams, it has successfully enabled a comparative evaluation of modern methods under a standardised test framework. The dataset, containing over two thousand scans across multiple subjects, provides an invaluable resource for the community and will continue to support method development and reproducibility beyond the Challenge.

The submitted methods reflect a rich diversity of algorithmic strategies, including spatial and temporal modeling, data augmentation, and ensemble learning. These contributions offer a strong foundation for future research. Despite notable advancements, this task is not yet solved to a clinically satisfactory degree. Sensitivity to scan length and scan pattern reveals that generalisation remains a challenge, and further work is needed to bridge the gap between experimental performance and clinical deployment.

The Challenge website and infrastructure will continue to be available beyond the official competition period, welcoming post-deadline submissions from the research community. It is intended to serve as a long-term benchmark for trackerless freehand ultrasound reconstruction, enabling ongoing method development, reproducibility studies, and performance comparisons as the field advances.

In summary, TUS-REC2024 Challenge provides not only a rigorous benchmark for current methods but also a catalyst for methodological advancement and clinical translation. Its biomedical and technical impact lies in establishing a shared framework to accelerate the development of practical, high-performance trackerless freehand ultrasound systems.

\section*{CRediT authorship contribution statement}
\label{CRediT}

\textbf{Qi Li}: Conceptualization, Data curation, Formal analysis, Investigation, Methodology, Resources, Software, Validation, Visualization, Writing – original draft, Writing – review \& editing, Challenge organisation.
\textbf{Shaheer U. Saeed}: Conceptualization, Resources, Validation, Writing – review \& editing, Challenge organisation.
\textbf{Yuliang Huang}: Conceptualization, Investigation, Methodology, Resources, Software, Writing – review \& editing, Challenge organisation.
\textbf{Mingyuan Luo}: Investigation, Methodology, Software, Writing – review \& editing.
\textbf{Zhongnuo Yan}: Investigation, Methodology, Software, Writing – review \& editing.
\textbf{Jiongquan Chen}: Investigation, Methodology, Software, Writing – review \& editing.
\textbf{Xin Yang}: Investigation, Methodology, Software, Writing – review \& editing.
\textbf{Dong Ni}: Investigation, Methodology, Software, Writing – review \& editing.
\textbf{Nektarios Winter}: Investigation, Methodology, Software, Writing – review \& editing.
\textbf{Phuc Nguyen}: Investigation, Methodology, Software, Writing – review \& editing.
\textbf{Lucas Steinberger}: Investigation, Methodology, Software, Writing – review \& editing.
\textbf{Caelan Haney}: Investigation, Methodology, Software, Writing – review \& editing.
\textbf{Yuan Zhao}: Investigation, Methodology, Software, Writing – review \& editing.
\textbf{Mingjie Jiang}: Investigation, Methodology, Software, Writing – review \& editing.
\textbf{Bowen Ren}: Investigation, Methodology, Software, Writing – review \& editing.
\textbf{SiYeoul Lee}: Investigation, Methodology, Software, Visualization, Writing – review \& editing.
\textbf{Seonho Kim}: Investigation, Methodology, Software, Visualization, Writing – review \& editing.
\textbf{MinKyung Seo}: Investigation, Methodology, Software, Visualization, Writing – review \& editing.
\textbf{MinWoo Kim}: Investigation, Methodology, Software, Visualization, Writing – review \& editing.
\textbf{Yimeng Dou}: Investigation, Methodology, Software, Writing – review \& editing.
\textbf{Zhiwei Zhang}: Investigation, Methodology, Software, Writing – review \& editing.
\textbf{Yin Li}: Investigation, Methodology, Software, Writing – review \& editing.
\textbf{Tomy Varghese}: Investigation, Methodology, Software, Writing – review \& editing.
\textbf{Dean C. Barratt}: Funding acquisition, Project administration, Writing – review \& editing, Challenge organisation.
\textbf{Matthew J. Clarkson}: Funding acquisition, Project administration, Writing – review \& editing, Challenge organisation.
\textbf{Tom Vercauteren}: Project administration, Supervision, Writing – review \& editing, Challenge organisation.
\textbf{Yipeng Hu}: Conceptualization, Funding acquisition, Project administration, Resources, Supervision, Writing – review \& editing, Challenge organisation.

\section*{Declaration of competing interest}

The authors declare that they have no known competing financial interests or personal relationships that could have appeared to influence the work reported in this paper.

\section*{Data availability}
\begin{sloppypar}
The training and validation datasets used in TUS-REC2024 Challenge are publicly available at the following repositories: \url{https://doi.org/10.5281/zenodo.11178508}, \url{https://doi.org/10.5281/zenodo.11180794}, \url{https://doi.org/10.5281/zenodo.11355499}, and \url{https://doi.org/10.5281/zenodo.12752245}. The baseline model code has been released and can be accessed via GitHub at \url{https://github.com/QiLi111/tus-rec-challenge\_baseline}. Participants are encouraged to release their code voluntarily. All publicly released codes related to the Challenge are listed on the official Challenge website: \url{https://github-pages.ucl.ac.uk/tus-rec-challenge/TUS-REC2024/leaderboard.html}.
\end{sloppypar}

\section*{Acknowledgement}
\label{Acknowledgement}

This work was supported by the EPSRC [EP/T029404/1], a Royal Academy of Engineering/Medtronic Research Chair [RCSRF1819\textbackslash7\textbackslash734] (TV), Wellcome/EPSRC  Centre  for Interventional  and  Surgical  Sciences [203145Z/16/Z], and the International Alliance for Cancer Early Detection, an alliance between Cancer Research UK [C28070/A30912; C73666/A31378; EDDAPA-2024/100014], Canary Center at Stanford University, the University of Cambridge, OHSU Knight Cancer Institute, University College London and the University of Manchester. TV is co-founder and shareholder of Hypervision Surgical. Qi Li was supported by the University College London Overseas and Graduate Research Scholarships. For the purpose of open access, the authors have applied a CC BY public copyright licence to any Author Accepted Manuscript version arising from this submission.

\begin{sloppypar}
Participating teams acknowledge support from the following funding agencies: \textbf{MUSIC Lab}: National Natural Science Foundation of China (Nos. 12326619, 62171290, 62101343), Science and Technology Planning Project of Guangdong Province (2023A0505020002), and Shenzhen-Hong Kong Joint Research Program (SGDX20201103095613036). \textbf{ISRU@DKFZ}: DKFZ (German Cancer Research Center) Heidelberg; DAAD. \textbf{AMI-Lab}: National Research Foundation of Korea (NRF) grant funded by the Korea government (MSIT) (No. RS2021-NR059679); Institute of Information \& communications Technology Planning \& Evaluation (IITP) under the Artificial Intelligence Convergence Innovation Human Resources Development (IITP-2025-RS-2023-00254177) grant funded by the Korea government (MSIT). \textbf{UW-Madison Elastography Lab}: National Heart, Lung, and Blood Institute, grant number 1R01HL147866; National Science Foundation, grant number CNS 2333491.
\end{sloppypar}

\section*{Declaration of generative AI and AI-assisted technologies in the writing process}
\label{AI}
During the preparation of this work the author(s) used ChatGPT in order to improve language and readability. After using this tool/service, the author(s) reviewed and edited the content as needed and take(s) full responsibility for the content of the published article.

\clearpage
\appendix
\section{Comparative summary of the literature on trackerless freehand ultrasound reconstruction}
\label{related_work_app}

\renewcommand{\arraystretch}{2}  
\setlength{\tabcolsep}{6pt} 
{
\scriptsize
\begin{longtable}{>{\raggedright\arraybackslash}p{0.6cm}>{\raggedright\arraybackslash}p{2.1cm}>{\raggedright\arraybackslash}p{2cm}>{\raggedright\arraybackslash}p{1.4cm}>{\raggedright\arraybackslash}p{1.4cm}>{\raggedright\arraybackslash}p{2cm}>
{\raggedright\arraybackslash}p{2.4cm}>{\raggedright\arraybackslash}p{1.5cm}}
\caption{Summary of the literature analysed with respect to input, output, network architecture, and loss function. Electronic version is attached as the supplementary material.}\label{related_work_table1}

\\\toprule
Paper  & Algorithm input & Network input & Secondary information extracted from input (e.g., optical flow) & Network architecture & Network output & Loss function 

(fidelity term) & Loss function (regularisation term) \\ 
\midrule
\endfirsthead

\toprule
Paper &  Algorithm input & Network input & Secondary information extracted from input (e.g., optical flow) & Network architecture & Network output & Loss function (fidelity term) & Loss function (regularisation term) \\ 
\midrule
\endhead

\bottomrule
\endfoot

\bottomrule
\endlastfoot

Prevost et al. \cite{prevost2017deep} & Two successive images, sampled from 20 scans (phantom dataset), or from 88 scans (forearm dataset), or from 12 scans (low legs dataset), using 2-fold patient cross validation for each dataset separately. & 1) Two successive images; 

2) Optical flow (two channels) & Optical flow & CNN & One transformation between two successive images, consisting of 6 degree of freedoms (3 for translation, 3 for rotation) & Euclidean distance between output (6-DoF) and ground truth & No \\
\hline

Prevost et al. \cite{prevost20183d} & 1) Two successive images, sampled from 20 scans (phantom dataset), or from 88 scans (forearm dataset), or from 12 scans (low legs dataset), or from 600 scans (forearm dataset), or from 100 scans (carotid dataset) using 2-fold cross validation (additional pairs were generated by considering image pairs that are not consecutive); 

2) IMU. & 1) Two successive images; 

2) Optical flow (two channels); 

3) IMU (concatenate the three Euler rotation angles from the IMU with the 512-dimensional vector from the penultimate layer of the network) &	Optical flow &	CNN & One transformation between two successive images, consisting of 6 degree of freedoms (3 for translation, 3 for rotation) &	$L2$ loss, the squared difference between the network output and the ground truth. & No \\
\hline

Hou et al. \cite{hou2018computing} & One image & One image & No & Any CNN architecture, e.g., GoogLeNet & The Riemannian Logarithm of canonical left-invariant metric on SE(3) & Geodesic distance between ground truth and prediction & No \\
\hline

Guo et al. \cite{guo2020sensorless} & Sequences of 5 frames, sampled from 500 ultrasound scans for training and 70 scans for validation. & A sequence of 5 frames & No & 3D ResNext model & One transformation (mean motion vector), consisting of 6 degree of freedoms (3 for translation, 3 for rotation) & 1) Mean squared error (MSE) loss (ground truth labels are defined as the mean parameters of the relative transform parameters between the neighboring frames); 

2) Case-wise correlation loss: the Pearson correlation coefficient between the estimated motion and the ground truth mean, for every degree-of-freedom. & No \\

Wein et al. \cite{wein2020three} & CNN: two successive ultrasound frames, sampled from 100 sweeps during training;

Registration (optimision): transversal and sagittal sweeps, reconstructed from CNN. & CNN: two successive ultrasound frames; two channels of 2D optical flow; Registration (optimision): transversal and sagittal sweeps, reconstructed from CNN. & CNN: two channels of 2D optical flow & CNN & CNN: one transformation between two successive images, consisting of 6 degree of freedoms (3 for translation, 3 for rotation) & 1) CNN: Euclidean distance between output (6-DoF) and ground truth;

2) Registration: image similarity on the B-Mode intensities (optimising a number of trajectory parameters and the rigid transformation between the sweep centers). & No \\
\hline

Miura et al. \cite{miura2020localizing} & Two successive ultrasound frames, sampled from 180 scans (27,948 frames) for training and 30 scans (5,176 frames) for validation. & Two successive ultrasound frames & Extracted optical flow using FlowNetS & Localization network: ResNet34; Optical flow estimation network: FlowNetS (pre-trained by the Flying Chairs dataset) & One transformation between two successive images, consisting of 6 degree of freedoms (3 for translation, 3 for rotation) & 1) Euclidean distance between ground truth and the estimated 6 parameters; 

2) Reprojection error (left-right consistency loss): Euclidean distance between a point on the image and a point reprojected from the consecutive image frame. & No \\
\hline

Miura et al. \cite{miura2021probe} & Two successive ultrasound images, sampled from 370 ultrasound scans for training and 40 scans for validation. & 1) Two successive ultrasound images; 

2) Optical flow & Optical flow & AlexNet & One transformation between two successive images, consisting of 6 degree of freedoms (3 for translation, 3 for rotation) & 1) Euclidean distance between prediction and ground truth; 

2) Forward consistency loss: Euclidean distance between points projected forward through successive frame-to-frame transformations and the ground truth points;

3) Backward consistency loss: Euclidean distance between the original points and those reprojected forward and then back through successive frame-to-frame transformations. & No \\
\hline

Guo et al. \cite{guo2021transducer} & Sequences of 5 consecutive frames, sampled from 500 source domain scans for training, 70 source domain scans for validation, and 9 target domain scans for training. & A sequence of 5 consecutive frames & No & CNN & One transformation (mean motion vector), consisting of 6 degree of freedoms (3 for translation, 3 for rotation) & Mean squared error (MSE) loss between network output and ground truth DOF labels. & Discrepancy loss: the $L2$ norm between the two feature extractors’ output feature vectors, from source domain data and target domain data. \\
\hline

Yeung et al. \cite{yeung2021learning} & Sets of an arbitrary number of 2D images, sampled from 3D ultrasound fetal brain volumes. & A set of an arbitrary number of 2D images & No & CNN + attention mechanism & The set of corresponding predicted locations (3 Cartesian coordinates (i.e. $x,y,z$) of the centre, the bottom right and left corners of a plane) & Mean squared error loss between the ground truth and predicted locations & No \\

Luo et al. \cite{luo2021self} & $N$-length scans, sampled from 8500 sequences generated from 85 3D DDH scans for training, or 6500 sequences generated from 65 3D fetus ultrasound scans for training. & $N$-length scan & No & Convolutional LSTM (ConvLSTM) & N-1 transformations between adjacent frames; each transformation has 6 degree of freedoms (3 for translation, 3 for rotation) & Training phase: 

1) Mean absolute error (MAE) loss between prediction and ground truth; 

2) Case-wise correlation loss; 

3) A cross-entropy loss that trains the classifier to distinguish reconstructed volumes from real volumes.

Test phase: 

Self-supervised loss between the original frames and the corresponding slices extracted from the reconstructed volume at the estimated positions. & Test phase: 

1) Adversarial loss; 

2) Quadratic potential divergence. \\
\hline

Miura et al. \cite{miura2021pose} & Sequences of 11 frames, sampled from 300 scans for training and 30 scans for validation. & 1) CNN (feature extraction): two consecutive frames and two channels of optical flow; 

2) ConvLSTM: one feature map from CNN. & Two channels of optical flow (estimated by FlowNetS, trained on Flying Chairs dataset) & CNN (AlexNet; trained on ImageNet) + ConvLSTM + spatial-temporal attention module & 10 transformations between adjacent frames and 11 transformations (absolute pose); each transformation has 6 degree of freedoms (3 for translation, 3 for rotation) & For relative pose estimation: 

1) Euclidean distance between prediction and ground truth; 

2) Forward consistency loss: Euclidean distance between points projected forward through successive frame-to-frame transformations and the ground truth points; 

3) Backward consistency loss: Euclidean distance between the original points and those reprojected forward and then back through successive frame-to-frame transformations.

For absolute pose estimation: 

MSE between ground truth pose and estimated absolute pose. & No\\
\hline

Yeung et al. \cite{yeung2021implicitvol} & 3D locations of $N$ images, computed from estimated transformations (3 for Euler angle, 3 for translation); 2D images are sampled from fifteen 3D volumes around the central axis of the brain non-uniformly, or sampled from two videos of real 2D freehand fetal brain ultrasound; $N = 128$ or $256$ & 3D locations of $N$ images, computed from estimated transformations (3 for Euler angle, 3 for translation); $N = 128$ or $256$ & Positional encoding & MLP & Intensities of pixels of $N$ input frames & Structural similarity (SSIM) loss between the actual and network-predicted intensities & No \\

Xie et al. \cite{xie2021image} & Two adjacent ultrasound images, sampled from 100 ultrasound sweeps using 10-fold cross validation. & 1) Two adjacent ultrasound images; 2) Optical flow. & Optical flow with 4 different scales, predicted from FlowNetCSS module & ResNet + channel attention + FlowNetCSS + spatial attention & One transformation between two adjacent images, consisting of 6 degree of freedoms (3 for translation, 3 for rotation) & 1) Mean absolute error (MAE) loss: distance between predicted 6-DoF and ground truth; 

2) Transformation matrix loss: the Frobenius norm between $4\times4$ transformation matrixes converted from 6-DoF of output and ground truth. & No \\
\hline

Ning et al. \cite{ning2022spatial} & 1) Sequences of $N$ frames, sampled from 60 cases for training and 20 cases for validation (forearm dataset), or from 96 cases for training and 32 cases for validation (clinical dataset); 

2) Quaternions of $N$ frames from IMU data. & 1) A sequence of $N$ frames; 

2) Quaternions of $N$ frames from IMU data. & 1) Position embedding; 

2) Difference map. & CNN + Transformer & $N-1$ transformations between adjacent frames; each transformation has 6 degree of freedoms (3 for translation, 3 for rotation) & MSE between the ground truth of relative poses and predictions & No \\
\hline

Leblanc et al. \cite{leblanc2022stretched} & Two consecutives frames, sampled from $13,087\sim15,687$ consecutives pairs of original frames for training and $4323\sim4916$ consecutives pairs of original frames for validation. & Two consecutives frames & No & Mask R-CNN with inception backbone (for segmenting the femoral artery) + CNN & Relative longitudinal motion (in mm) between the input registered pair of frames & $L1$ loss between ground truth and prediction & No \\
\hline

Di Vece et al. \cite{di2022deep} & Ultrasound images, sampled from 75088 images for training (phantom data) or 22029 images for training (real fetal data). & An ultrasound image & No & ResNet-18 & A vector of nine parameters relative to the centre of the fetal brain ultrasound volume (3 for translation and 6 for rotation); the last six parameters are used internally by the CNN to reconstruct the rotation matrix using Gram-Schmidt process. & Mean squared error (MSE) between prediction and ground truth, on translation and rotation matrices separately. & No \\

Guo et al. \cite{guo2022ultrasound} & Sequences of 5 frames, sampled from 488 scans for training and 66 scans for validation, or 80 scans for training and 10 scans for validation using 10-fold cross-validation. & A sequence of 5 frames & No & 3D ResNeXt + self-attention block & One transformation (mean motion vector), consisting of 6 degree of freedoms (3 for translation, 3 for rotation) & 1) Mean squared error (MSE) loss between prediction and ground truth (Ground-truth labels are defined as the mean parameters of the relative transform parameters between the neighboring frames); 

2) Case-wise correlation loss: the Pearson correlation coefficient between the estimated motion and the ground truth mean, for every degrees-of-freedom. & Margin ranking loss: correlates two random samples' feature vectors distance with their transformation parameters discrepancy. \\
\hline

Luo et al. \cite{luo2022deep} & 1) $N$-length scans, sampled from 196 scans for training (arm dataset) or from 136 scans for training (carotid dataset) (each scan was randomly augmented to 40 sequences at each epoch); 

2) Acceleration from IMU; 

3) Euler angle calculated using orientation data from IMU. & 1) A $N$-length scan; 

2) Acceleration from IMU; 

3) Euler angle calculated using orientation data from IMU. & No & ResNet + LSTM & $N$-1 transformations between adjacent frames; each transformation has 6 degree of freedoms (3 for translation, 3 for rotation) & Training phase: 

1) Mean absolute error (MAE) between the estimated transformation parameters and ground truth; 

2) Pearson correlation loss;

Test phase: 

1) Pearson correlation loss between the estimated acceleration and acceleration from IMU; 

2) MAE loss between the estimated Euler angle and IMU Euler angle. & No \\
\hline

Mikaeili et al. \cite{mikaeili2022trajectory} & Sequences of 3 or 5 frames, sampled from 2220 frames during training. & A sequence of 3 or 5 frames & No & DenseNet + FlowNet & Transformation between two adjacent images, consisting of 6 degree of freedoms (3 for translation, 3 for rotation) & Not clear & $L2$ regularization (the squared magnitude of all weights in the network) \\
\hline

Yeung et al. \cite{yeung2022adaptive} & A set of $n$ 2D ultrasound slices, randomly sampled from 50 3D ultrasound volumes which are co-aligned with the predefined 3D anatomical atlas. Fine-tuned on $m$ volume-sampled 2D images or native 2D freehand images. & A set of $n$ sampled 2D slices from 3D volumes for training; a new set of $m$ 2D ultrasound images for fine-tuning. & No & CNN & 1) Locations ($x$, $y$ and $z$ coordinates of the top right, top left and bottom right corners) of input slices, in a predefined 3D anatomical atlas; 

2) Displacement between each pair of slices. & 1) Mean least-squared error (MSE) between the predicted and ground truth locations; 

2) MSE between the predicted and ground truth displacements; 

3) Cycle consistency loss: MSE between the overall displacement of a sequence of images and the displacement from the first image to the last of that sequence. & No \\

Chen et al. \cite{chen2023freehand} & Three original images sampled from one scan, and eight interpolated images. & Three original images and eight interpolated images & Bezier interpolation among input ultrasound frames, to make the FPS is consistent with the tracking device. & 3DCNN + LSTM & 1) Position of ROI middle pixel; 

2) Six offset parameters. & Mean square error loss & No \\
\hline

Wysocki et al. \cite{wysocki2024ultra} & Positions of points in a Cartesian coordinate space, sampled along a ray from four tilted scans for training (synthetic data), or four tilted scan pairs for training (phantom data). & Positions of points in a Cartesian coordinate space, sampled along a ray & No & MLP + ultrasound volume rendering model & Parameter vector (attenuation, reflectance, border probability, scattering density, and scattering intensity) from MLP; pixel intensity from rendering. & 1) Structural Similarity Index Measure (SSIM) between the generated B-mode images with their corresponding target images; 

2) Squared Error Loss ($L2$) & No \\
\hline

Luo et al. \cite{luo2023recon} & $N$-length 2D ultrasound scans, sampled from 13500 generated scans from 135 3D DDH volumes for training, 16 volumes for validation, or 9800 generated scans from 98 3D fetus ultrasound volumes, 15 volumes for validation, or 45 2D spine ultrasound scans for training, 12 scans for validation. & $N$-length 2D ultrasound scan (features for each pair of adjacent frames from ResNet are fed into LSTM) & 1) Canny edge map for each frame; 

2) Optical flow from the adjacent frames. & ResNet + LSTM & $N-1$ transformations between adjacent frames; each transformation has 6 degree of freedoms (3 for translation, 3 for rotation) & Training phase: 

1) Mean absolute error (MAE) between predicted 6-DoF and ground truth of adjacent frames; 

2) Pearson correlation; 

3) Cross-entropy loss that trains the classifier to distinguish reconstructed volumes from real volumes.

Test phase: 

1) Frame-level contextual consistency: MAE between the original frames and the corresponding slices extracted from the reconstructed volume at the estimated positions; 

2) Path-level similarity constraint that selects the prior training scan whose ground truth transformations are most similar to those of the current test scan and optimises the network by minimising the MAE and Pearson correlation loss between the estimated transformations of the selected training scan and its ground truth. & Training phase: 

Motion-weighted regularization that builds a positive correlation between motion and appearance variance.

Test phase: 

Global adversarial shape prior identifies the reconstructed volume from numerous real volumes in the training dataset. \\

Li et al. \cite{li2023trackerless} & Sequences of 20 frames, sampled from 137 scans for training and 46 scans for validation. & A sequence of 20 frames for EfficientNet (b1) or one frame for LSTM & No & EfficientNet (b1) or EfficientNet + LSTM & 79 selected transformations from $C_{20}^2$ transformations; each transformation has 6 degree of freedoms (3 for translation, 3 for rotation) & 1) Mean-square-errors (MSEs) between four corner points, transformed from predictions and ground truth, average over all predictions;

2) Accumulated loss: MSE between four corner points obtained from the ground truth transformation and from an indirect prediction composed of two direct predictions. & Consistency loss: MSE between four corner points obtained from a direct prediction (e.g., transformation from frame 1 to frame 10) and from an indirect prediction composed of two direct predictions (e.g., frame 1 to frame 5, then frame 5 to frame 10) \\
\hline

Li et al. \cite{li2023privileged} & Sequences of $M$ frames ($M=100$ or $140$), sampled from 137 scans for training and 46 scans for validation. & A sequence of $M$ frames ($M=100$ or $140$) & No & EfficientNet (b1) & A number of selected transformations from $C_M^2$ transformations; each transformation has 6 degree of freedoms (3 for translation, 3 for rotation) & 1) MSE between four corner points, transformed from predictions and ground truth, respectively;

2) Two cross-entropy losses between predicted class probability vectors and ground truth targets, for classifying anatomical and protocol. & No \\

Luo et al. \cite{luo2023multi} & 1) $N$-length scanning sequences, sampled from 200 scans for training and 40 scans for validation (arm dataset), or from 150 scans for training and 30 scans for validation (carotid dataset); 

2) $M$ independent IMU data. & 1) $N$-length scanning sequence; 2) $M$ independent IMU data. & No & ResNet + LSTM & $M\times(N-1)$ transformations between adjacent frames; each transformation has 6 degree of freedoms (3 for translation, 3 for rotation) & Training phase: 

1) Mean absolute error (MAE) between prediction and ground truth; 

2) Pearson correlation loss between prediction and ground truth;

The prediction is obtained by averaging the $M$ estimated parameters from repeatedly inputting the ultrasound images and $M$ different IMU data into the backbone.

Test phase: 

1) Pearson correlation loss for measuring the difference between the estimated acceleration (obtained from the estimated transformation parameters) and IMU acceleration; 

2) MAE for measuring the difference between the estimated angle (obtained from the estimated transformation parameters) and IMU angle. & Test phase: 

1) Multi-IMU consistency constraints, including i) Pearson correlation loss for measuring the difference between the estimated accelerations; and ii) MAE loss for measuring the difference between the estimated angles;

2) Self-consistency constraint: MAE between the estimated 6-DoF from a sampled sub-sequence and the sampled 6-DoF from the original estimation. \\

\hline

EI hadramy et al. \cite{el2023trackerless} & Sequences of two images with 3 channels each, 2 heatmaps and an ultrasound frame. Sampled from 9614 clips for training and 2060 clips for validation. & Two frames (the first, middle frames or the middle, last frames of the ultrasound sequence with 7 frames) and 4 Gaussian heatmaps (obtained from sparse optical flow for 2 selected points), for each of the two Siamese networks. & Sparse optical flow for 2 selected points & Sequence to Vector (Seq2Vec) network & Two transformations (one is from the first frame to the fourth frame, and the other is from the fourth frame to the last frame); each transformation has 6 degree of freedoms (3 for translation, 3 for rotation) & 1) Mean square error (MSE) between the estimated transformations (from the first frame to the fourth frame, and from the fourth frame to the last frame) at each corner point of the frames and their respective ground truth; 

2) MSE between the estimated transformation (from the first frame to the last frame, obtained by composing the two transformations above) at the corner points of the frames and the ground truth. & No \\
\hline

Li et al. \cite{li2023long} & Sequences of $M$ frames, sampled from 137 scans for training and 46 scans for validation. ($M=10,20,30,40,49,60,$ $75,100$) & A sequence of $M$ frames for EfficientNet (b1) or one frame for LSTM & No & EfficientNet (b1) or EfficientNet (b1) + LSTM & $\tau+1$ selected transformations from $C_M^2$ transformations; each transformation has 6 degree of freedoms (3 for translation, 3 for rotation); $\tau+1 = 45,80,124,157,165,$ $177,197,218$ & MSE between four corner points, transformed from predictions and ground truth, respectively. & No \\

Dou et al. \cite{dou2023sensorless} & Two adjacent ultrasound images, sampled from 1,928 scans. & Two adjacent ultrasound images & No & ResNet & Relative transformation matrix, including translation, tilt, and rotation. & Mean square error (MSE) loss & No \\
\hline

Yeung et al. \cite{yeung2024sensorless} & 1) $N$ 2D ultrasound images, sampled from a set of co-aligned 3D brain volumes; (for datasets A and B: non-uniformly sampled around the central axis of the brain; for native 2D freehand video sequences: randomly sampled from the sequences; $N=128$ or $256$); 

2) 3D locations, computed from estimated 3 Euler angles and 3 translations. & 1) $N$ 2D ultrasound images ($N=128$ or $256$); 

2) 3D locations, computed from estimated 3 Euler angles and 3 translations. & Positional encoding & MLP & Intensity values of the corresponding pixels & Structural similarity (SSIM) loss between the predicted and actual intensity values & No \\
\hline

Gaits et al. \cite{gaits2024ultrasound} & 3D spatial coordinates & 3D spatial coordinates & Positional encoding & NeRF (eight layers of 256 fully-connected perceptrons with ReLU activation) & Ultrasound slice pixel values & Mean squared error between the network output and ultrasound slice pixel value & No \\
\hline

Dou et al. \cite{10684746} & Sequences of four ultrasound frames, sampled from 45 cases for training and 15 cases for validation (phantom dataset), or from 137 cases for training and 45 cases for validation (forearm dataset). & A sequence of four ultrasound frames (randomly selected within the range of $\pm10$ frames) & No & ResNeXt & Quaternion between input frames & 1) Mean squared error (MSE) between the estimated and ground truth transformation parameters; 

2) Pearson correlation loss. & Triplet ranking loss, learning a discriminative feature embedding by penalising estimated samples based on their distance in feature space. \\
\hline

Ramesh et al. \cite{ramesh2024geometric} & $N$ 2D ultrasound images, sampled from 4 training and 2 validation 3D volumes aligned to the 3D anatomical atlas. & $N$ 2D ultrasound images & No & 2D CNN & Five sets of poses and variances for each input image: quaternions, axis-angles, euler angles, rotation matrices, translation displacement and scaling, and direct pose (3 Cartesian coordinates of 3 reference locations (the center, bottom right, and bottom left of the plane)) & 1) Mean-squared-error (MSE) loss;

2) Heteroscedastic Gaussian Negative Log-Likelihood & No \\

Yan et al. \cite{yan2024fine} & 1) $N$-length scanning sequence, sampled from 375 cases for training and 104 cases for validation (arm dataset), or from 276 cases for training and 78 cases for validation (carotid dataset); 

2) $M$ independent IMU data. & 1) $N$-length scanning sequence; 

2) $M$ independent IMU data & No & ReMamba, consisting of stem, multiple residual blocks and ReMamba blocks with multi-directional SSM. & $N-1$ transformations between adjacent frames; each transformation has 6 degree of freedoms (3 for translation, 3 for rotation) & Training phase: 

1) Mean absolute error between network prediction and ground truth; 

2) Pearson correlation loss between network prediction and ground truth;

Test phase: 

1) Alignment loss (mutual information between image feature and weighted temporal feature of IMU acceleration); 

2) Pearson correlation loss between the estimated Euler angle and weighted IMU angle. & No \\
\hline

Li et al. \cite{li2024nonrigid} & Sequences of 100 ultrasound frames, sampled from 432 scans for training and 144 scans for validation. & A sequence of 100 ultrasound frames & No & EfficientNet (b1) + Voxelmorph & From EfficientNet (b1): 99 transformations between each frame and the first frame; each transformation has 6 degree of freedoms (3 for translation, 3 for rotation); 

From Voxelmorph: dense displacement field which transformed the rigid-reconstruction-predicted volume to the final prediction. & 1) Mean squared error between predicted points coordinates and ground truth points coordinates; 

2) Intensity similarity between the ground-truth-reconstructed volume and wrapped predicted volume. & Bending energy of predicted deformation. \\
\hline

Dou et al. \cite{dou2024pitfalls} & 3D positions of ultrasound images (randomly sampled 16384 rays at each learning step) from four scans of a volume & 3D positions of query points & Positional encoding & Ultra-NeRF & A parameter vector (attenuation, reflectance, boundary probability, scattering density, scattering intensity), which can be rendered into gray scale intensity & $L2$ loss between the generated images with their corresponding target images & No \\
\hline

Dou et al. \cite{dou2024continuous} & 3D positions from 800 two-dimensional B-mode liver images (randomly sampled 2000 rays at each learning step) & 3D positions of query points, randomly sampled 2000 rays at each learning step & No & Ultra-NeRF & A parameter vector (attenuation, reflectance, boundary probability, scattering density, scattering intensity), which can be rendered into gray scale intensity & $L2$ loss between the generated images with their corresponding target images & No \\

\raisebox{-0.8\height}{\rotatebox[origin=c]{90}{\parbox{2.6cm}{\centering Großbröhmer et al.\ \mbox{\cite{grossbrohmer20253d}}}}} & 1) Two ultrasound scans (fixed and moving), sampled from 19 training, 1 validation case (20 reference, 40 moving), using a leave-one-out cross-validation; 

2) Rotational IMU data (initialise the registration using rotational IMU data) & Two point clouds (reference and moving) & 1) Automatically generated two bone surface point clouds (2048 points each);

2) Support keypoints; 

3) Gradient keypoints & Free Point Transformer (FPT) + HeatReg + instance optimisation & Displacement field ($2048\times3$) & For registration: 

1) difference between warped moving from FPT and tracked ground truth; 

2) difference between final prediction from HeatReg and tracked ground truth;

For instance optimisation: 

1) Chamfer Loss between predicted and fixed point cloud; 

2) DiVRoC loss. & For registration: $L1$ Loss with equal weighting on predicted displacements both with and without translational regularization.\\
\hline

Eid et al. \cite{eid2025rapidvol} & 3D coordinates of all pixels on a set of 2D ultrasound images (128 and 256 axial or coronal slices linearly sampled from the 3D scan) & 3D coordinates of all pixels on one 2D ultrasound image, obtained using the pose of the image. & Positional encoding & Tri-Planar Decomposition + MLP & A grayscale image at the requested pose & Negative structural similarity index measure between the rendered image and the ground truth image & No \\
\hline

Luo et al. \cite{luo2025monetv2}  & 1) $N$-length scanning sequences, sampled from 7500 generated sequences from 375 scans in each training epoch and 1040 generated sequences from 104 scans for validation (arm dataset), or from 5520 generated sequences from 276 scans in each training epoch and 780 generated sequences from 78 scans for validation (carotid dataset); or from 3680 generated sequences from 184 scans in each training epoch and 530 generated sequences from 53 scans for validation (thyroid dataset); 

2) IMU data & 1) All pairs of adjacent images from a $N$-length scanning sequence; 

2) IMU acceleration; 

3) IMU Euler angle. & No & ResNet18 + ConvLSTM & $N-1$ transformations between adjacent frames; each transformation has 6 degree of freedoms (3 for translation, 3 for rotation) & Training phase: 

1) Mean absolute error (MAE) between the estimated transformation parameters and the ground truth; 

2) Pearson correlation loss between the estimated transformation parameters and the ground truth;

Test phase:
Online multimodal self-supervised strategy: Pearson correlation loss between the estimated acceleration (obtained from the estimated translation) and IMU acceleration; MAE loss between the estimated Euler angle and IMU Euler angle. & Test phase: 

1) Scan-level velocity consistency: velocity constrains between predictions with different intervals (the MAE between long-interval transformations and the composition of the shorter-interval transformations); 

2) Path-level appearance consistency: differences between transformations of reordered subsequences and the reordered original predictions; 

3) Patch-level motion consistency: MAE between the normalised content difference and the normalised estimated distance, for two adjacent images. \\

Lee et al. \cite{lee2025enhancing} & Two ultrasound sequences with $s+1$ frames, sampled from 30 scans for training and 6 scans for validation (for in-house forearm dataset), or from 30 scans for training and 6 scans for validation (for in-house Photoacoustic dataset), or from 120 scans for training and 48 scans for validation (for public forearm dataset). & Two ultrasound sequences with $s+1$ frames (the first sequence includes frames $n$ to $n+s$; the second sequence includes frames $n+1$ to $n+s+1$) & No & CNN + RNN + self-attention mechanism & $s+1$ transformations; each transformation has 6 degree of freedoms (3 for translation, 3 for rotation) & 1) Motion-based mean absolute error: assigns higher weight to fast motion; 

2) Correlation loss between predicted 6-DoFs and ground truth. & Margin triplet loss: encourages the model to contrast the feature maps in latent space. \\
\hline

Sun et al. \cite{sun2025ultrasom} & Sequences of 8 frames, sampled from 160 training ultrasound scans and 20 validation ultrasound scans during training. & 1) A sequence of 8 frames; 

2) Optical flow. & 1) 7 optical flow features; 

2) spatial position encoding; 

3) temporal position encoding. & Space-Time Blocks (Mamba blocks and space attention mechanisms) + Regression head & 7 transformations; each transformation has 6 degree of freedoms (3 for translation, 3 for rotation) and represents the relative pose between each adjacent frame pair & 1) Mean squared error (MSE): the average difference between predicted and ground truth 6-DoF transformation parameters across the $n-1$ relative transformations; 

2) Correlation loss based on Pearson's correlation coefficient; 

3) Motion speed loss: the average speed error between predicted and ground truth velocities for consecutive frames, computed from translation and rotation information and averaged over the $n-2$ speed values. & No \\

\end{longtable}
}

\renewcommand{\arraystretch}{2}  
\setlength{\tabcolsep}{6pt} 
{
\scriptsize
\begin{longtable}{>{\raggedright\arraybackslash}p{0.6cm}>{\raggedright\arraybackslash}p{3cm}>{\raggedright\arraybackslash}p{2.1cm}>{\raggedright\arraybackslash}p{2.3cm}>{\raggedright\arraybackslash}p{3.2cm}>{\raggedright\arraybackslash}p{3.2cm}
}
\caption{Summary of the literature analysed with respect to dataset and evaluation metric. Electronic version is attached as the supplementary material.}\label{related_work_table2}

\\\toprule
Paper &  Dataset (anatomy, protocol, size and length) & Ground truth acquisition & If training and test data are from the same dataset & Evaluation metric(s) & Expected values and std of metric(s) \\ 
\midrule
\endfirsthead

\toprule
Paper &  Dataset (anatomy, protocol, size and length) & Ground truth acquisition & If training and test data are from the same dataset & Evaluation metric(s) & Expected values and std of metric(s)  \\ 
\midrule
\endhead

\bottomrule
\endfoot

\bottomrule
\endlastfoot

Prevost et al. \cite{prevost2017deep} & 1) BluePhantom ultrasound biopsy phantom (20 ultrasound sweeps; 7168 frames; proximal to distal direction);

2) Forearms of 12 volunteers (88 in-vivo ultrasound sweeps; 41,869 frames; proximal to distal direction); 

3) Lower legs on a subset of the volunteers (12 in-vivo ultrasound sweeps; 6647 frames; proximal to distal direction); for testing generalization capabilities. & Tracked by a surgical system (Stryker Navigation System III) & 2-fold patient cross validation for each dataset & 1) Average absolute parameter-wise error, computed per frame relative to the first frame of the sweep; 

2) Final drift: the distance between the estimated and ground truth centers of the last image. & 1) Phantom dataset: avg. absolute error: $t_x$:1.32 mm; $t_y$: 2.13 mm; $t_z$: 7.79 mm; $\theta_x$: 2.32 deg; $\theta_y$: 1.21 deg; $\theta_z$: 0.90 deg; final drift: 18.30 mm (median); 

2) Forearms dataset: avg. absolute error: $t_x$: 3.54 mm; $t_y$: 3.05 mm; $t_z$: 4.19 mm; $\theta_x$: 2.63 deg; $\theta_y$: 2.52 deg; $\theta_z$: 1.93 deg; final drift: 14.44 mm (median); 

3) Lower legs dataset: avg. absolute error: $t_x$: 4.14 mm; $t_y$: 3.91 mm; $t_z$: 17.12 mm; $\theta_x$: 1.94 deg; $\theta_y$: 2.58 deg; $\theta_z$: 2.15 deg; final drift: 40.56 mm (median). \\
\hline

Prevost et al. \cite{prevost20183d} & 1) BluePhantom ultrasound biopsy phantom (20 ultrasound sweeps; from proximal to distal direction; 7168 frames; avg. length of 131 mm);

2) Forearms of 12 volunteers (88 in-vivo ultrasound sweeps, from proximal to distal direction; 41,869 frames; avg. length of 190 mm);

3) Lower legs on a subset of the 12 volunteers (12 in-vivo ultrasound sweeps, from proximal to distal direction; 6647 frames; avg. length of 175 mm); for testing generalization capabilities;

4) Forearms on another set of 15 volunteers (600 in vivo ultrasound sweeps; 307,200 frames; avg. length of 202 mm; with IMU data);

5) Carotids from 10 volunteers (100 ultrasound sweeps; 21,945 frames; avg. length of 75 mm; with IMU data); for testing generalization capabilities.
& Stryker NAV3™  Camera (Stryker Co., Kalamazoo, MI, USA) optical tracking system & 2-fold cross validation & 1) Average absolute parameter-wise error: mean absolute difference between estimated and ground truth transformation parameters across all frames in each sweep;

2) Final drift of a sweep: the distance between the last frame center in the estimated trajectory and the corresponding center in the ground truth trajectory;

3) Maximum center error: the largest distance within a sweep between any estimated frame center and its corresponding ground truth center;

4) Length error: the difference in distance between the first and last frames of the estimated trajectory and that of the ground truth trajectory.

& 1) Dataset 1 (phantom): avg. absolute error: $t_x$: 1.32 mm; $t_y$: 2.13 mm; $t_z$: 7.79 mm; $\theta_x$: 2.32 deg; $\theta_y$: 1.21 deg; $\theta_z$: 0.90 deg; final drift: 18.30 mm (median);

2) Dataset 2 (forearms without IMU): avg. absolute error: $t_x$: 3.54 mm; $t_y$: 3.05 mm; $t_z$: 4.19 mm; $\theta_x$: 2.63 deg; $\theta_y$: 2.52 deg; $\theta_z$: 1.93 deg; final drift: 14.44 mm (median);

3) Dataset 3 (calves): avg. absolute error: $t_x$: 3.11 mm; $t_y$: 5.86 mm; $t_z$: 5.63 mm; $\theta_x$: 2.75 deg; $\theta_y$: 3.17 deg; $\theta_z$: 5.24 deg; final drift: 19.69 mm (median);

4) Dataset 4: avg. absolute error: $t_x$: 2.75 mm; $t_y$: 2.41 mm; $t_z$: 4.36 mm; $\theta_x$: 0.19 deg; $\theta_y$: 0.21 deg; $\theta_z$: 0.13 deg; final drift: 10.42 mm (median). \\
\hline

Hou et al. \cite{hou2018computing} &  1) PoseNet dataset;

2) Dataset of C-Arm X-Ray to Computed Tomography (CT) alignment problem;

3) Fetal Magnetic Resonance Imaging. & Not clear & Yes & For dataset 1) and 2): average errors of each parameter (Euler angles in degrees and translation in meters/millimetres) and geodesic distance;

For dataset 3): Cross Correlation, Mean Squared Error, Peak Signal-to-Noise Ratio, Structural SIMilarity, and geodesic distance.

& 1) Dataset 1: $\theta_x$: 1.870 deg; $\theta_y$: 3.143 deg; $\theta_z$: 3.662 deg; $t_x$: 1.759 m; $t_y$: 1.240 m; $t_z$: 0.156 m; geodesic distance: 16.370;

2) Dataset 2: $\theta_x$: 10.511 deg; $\theta_y$: 6.789 deg; $\theta_z$: 11.913 deg; $t_x$: 62.588 mm; $t_y$: 68.747 mm; $t_z$: 54.110 mm; geodesic distance: 19624.246;

3) Dataset 3: CC: 0.8732; MSE: 724.9713; PSNR: 20.7484; SSIM: 0.6470; geodesic distance: 10.0836.\\

Guo et al. \cite{guo2020sensorless} &  Prostate ultrasound: 640 TRUS videos (from 640 patients); varied lengths. & EM-tracking device & Yes. training: 500 cases; validation: 70 cases; testing: 70 cases.

During inference, sliding along the video sequence with a window size $N$. The inter-frame motion of two neighboring frames is the average motion computed in all the batches. & 1) Distance error: the mean distance between corresponding corner points across all frames in the video;

2) Final drift: the distance between the center points of the transformed end frames of a video segment using the ground truth and prediction. & Distance error: 10.33 mm;

Final drift: 17.39 mm\\
\hline

Wein et al. \cite{wein2020three} & Thyroid data: 180 ultrasound sweeps from 9 volunteers; 5 transversal and 5 sagittal sweeps for each lobe of the thyroid. & Tracked using an optical tracking system;

Manually segmented on 58 sweeps (50 from the 5 volunteer training subset and 8 from the 4 volunteer testing subset.) & Yes. Training: 100 sweeps from 5 volunteers; testing: 80 sweeps from 4 volunteers. & 1) Dice coefficient: the overlap between the 3D thyroid gland segmentations derived from the transversal and sagittal sweeps;

2) Relative trajectory error: the cumulative in-plane translation error at each frame divided by the ground truth sweep length;

3) Relative length error: the ratio of the difference between the estimated and ground truth sweep lengths to the ground truth length;

4) Volume and Norm. vol. & Results from 54-DOF refinement:

1) Dice: $0.66\pm0.12$;

2) Relative trajectory error: $0.16\pm0.09$;

3) Relative length error: $0.16\pm0.14$;

4) Volume: $1.15\pm1.45$; Norm. vol.: $0.14\pm0.16$. \\
\hline

Miura et al. \cite{miura2020localizing} & 1) Forearms from 5 subjects: 190 scans (30,801 frames);

2) Breast phantom: 60 scans (8,940 frames);

3) Hypogastric phantom: 40 scans (6,242 frames). & V120:Trio (OptiTrack) & Training data: 180 scans (27,948 frames) from forearm of 2 subjects and two phantoms; validation data: 30 scans (5,176 frames) from forearm of 1 subject; test data: 80 scans (12,859 frames) from forearm of 2 subjects. & Mean absolute error (MAE) for each parameter & MAE: $\theta_x$: 0.53 deg; $\theta_y$: 1.21 deg; $\theta_z$: 0.47 deg; $t_x$: 0.64 mm; $t_y$: 0.15 mm; $t_z$: 0.80 mm. \\
\hline

Miura et al. \cite{miura2021probe} & 1) Forearm (310 scans);

2) Breast phantom (180 scans);

3) Hypogastric phantom (40 scans). & OptiTrack V120: Trio (Acuity Inc.) & Training: 370 scans from the forearms of 3 subjects and the breast phantom; validation: 40 scans from the forearm of 1 subject; test: 120 scans from the forearms of 2 subjects and the hypogastric phantom. & 1) Mean absolute error (MAE): the average of cumulative errors from the initial frame;

2) Final drift: the error in the estimated probe position of the final frame. & 1) Forearm data: MAE: $\theta_x$: 13.6 deg; $\theta_y$:  18.9 deg; $\theta_z$: 7.5 deg; $t_x$: 8.5 mm; $t_y$: 4.3 mm; $t_z$: 15.0 mm; final drift: 75.8 mm;

2) Hypogastric phantom: MAE: $\theta_x$: 9.7 deg; $\theta_y$: 16.0 deg; $\theta_z$: 11.3 deg; $t_x$: 12.9 mm; $t_y$: 9.9 mm; $t_z$: 30.5 mm; final drift: 73.6 mm.\\
\hline

Guo et al. \cite{guo2021transducer} & 1) Source domain: 640 prostate transrectal ultrasound video sequences;

2) Target domain: 12 transabdominal ultrasound video sequences.& EM-tracking device & Source domain: training: 500 cases; validation: 70 cases; testing: 70 cases;

Target domain: training: 9 cases; testing: 3 cases;

During inference, using a sliding-window procedure to get the DOF motion vector prediction for every subsequence.& 1) Distance error;

2) Final drift.& On target domain:

Distance error: 12.67 mm; 

Final drift: 20.34 mm.\\

Yeung et al. \cite{yeung2021learning} & 3D ultrasound fetal brain volumes, obtained as part of the INTERGROWTH-21st study; 18 to 22 gestational weeks; fifty and fifteen 3D volumes acquired at 21 gestational weeks were selected for training and validation; 150,000 and 45,000 images for each training and validation epoch. & Artificially sample 2D slices from aligned 3D ultrasound volumes of the fetal brain & 1) Trained on images at 21 gestational weeks, testing on 18 to 22 gestational weeks;

2) Training and testing on 18 to 22 gestational weeks;

3) Testing on real 2D images taken at the standard TT plane;

4) Testing on 15 videos of 2D brain scans acquired from 4 subjects with gestational age between 19 to 21 weeks. & 1) Euclidean distance (ED) between all the coordinates of the predicted and ground truth planes;

2) Plane angle (PA) between the predicted and ground truth planes;

3) Normalised cross-correlation (NCC) between the input image and image sampled from the predicted plane. & Testing on real 2D images taken at the standard TT plane:

1) Comparison with the first manual annotation: 

ED: $11.36\pm3.26$ voxels; 

PA: $0.179\pm0.095$ rad; 

NCC: $0.841\pm0.096$;

2) Comparison with the second manual annotation: 

ED: $11.44\pm5.02$ voxels; 

PA: $0.180\pm0.120$ rad; 

NCC: $0.837\pm0.080$.\\
\hline

Luo et al. \cite{luo2021self} & 1) 3D developmental dysplasia of the hip (DDH) ultrasound dataset from 14 volunteers (101 US volumes);

2) 3D fetus ultrasound dataset from 78 volunteers (78 fetal ultrasound volumes); gestational age ranges from 10 to 14 weeks. & Complex ultrasound scan were simulated from real 3D volumes by combining loop, fast-and-slow scan, and sector scans. The number of generated sequences is 100 and 10 for training and testing, respectively. & Yes.

1) 3D DDH dataset: training: 85 volumes; testing: 15 volumes;

2) Fetal US dataset: training: 65 volumes; testing: 13 volumes. & 1) Final drift rate (FDR): the final drift divided by the scan length;

2) Average drift rate (ADR): the mean value of the cumulative drift of all frames divided by the length from the frame to the starting point of the sequence;

3) Maximum drift (MD): the maximum accumulated drift of all frames;

4) Sum of drift (SD): the sum of accumulated drift of all frames;

5) Bidirectional Hausdorff distance (HD): the worst distances between the predicted positions and the ground truth of all the frames in the sequence. & DDH dataset:

1) FDR: $5.44\pm3.03 \%$;

2) ADR: $13.47\pm6.47 \%$;

3) MD: $4.44\pm2.29$ mm;

4) SD: $274.14\pm136.89$ mm;

5) HD: $3.91\pm1.89$ mm;

Fetus dataset:

1) FDR: $9.94\pm4.41 \%$;

2) ADR: $27.08\pm9.31 \%$;

3) MD: $16.84\pm7.56$ mm;

4) SD: $730.11\pm303.20$ mm;

5) HD: $14.12\pm5.89$ mm.\\
\hline

Miura et al. \cite{miura2021pose} & 1) Forearms from 6 subjects (230 scans);

2) Breast phantom (140 scans);

3) Hypogastric phantom (40 scans) & OptiTrack V120: Trio (Acuity, Inc.) & Yes.

Experiment 1): Only use breast phantom data with straight line scans: training: 40 scans; validation: 5 scans; testing: 5 scans;

Experiment 2): All three datasets: training: 300 scans; validation: 30 scans; testing: 80 scans. & 1) Sum of absolute errors (AE) of the relative pose parameters per frame, computed separately for each parameter;

2) Final drift: the distance between the estimated and ground truth positions of the final frame. & 1) Breast phantom: AE: $\theta_x$: 19.52 deg; $\theta_y$: 26.38 deg; $\theta_z$: 22.04 deg; $t_x$: 23.97 mm; $t_y$: 14.02 mm; $t_z$: 45.87 mm; final drift: 27.71 mm;

2) On three datasets: AE: $\theta_x$: 125.37 deg; $\theta_y$: 252.70 deg; $\theta_z$: 120.18 deg; $t_x$: 133.24 mm; $t_y$: 40.63 mm; $t_z$: 184.08 mm; final drift: 96.46 mm.\\
\hline

Yeung et al. \cite{yeung2021implicitvol} & 1) Volume-sampled 2D images: sampling planes from fifteen 3D ultrasound fetal brain volumes; 20 weeks gestational age; obtained as part of the INTERGROWTH-21st study;

2) Two videos of native freehand 2D brain scans: 250 frames each; 20 weeks gestational age. & 1) For volume-sampled 2D images: ground truth is from the original volumes;

2) For freehand 2D brain scans: lack of ground truth 3D location information. &	Yes & 1) Normalised cross-correlation (NCC);

2) Structural similarity index measure (SSIM);

3) Absolute difference between rotation angles;

4) Absolute distance between translations.& On volume-sampled 2D images ($N$=256):

1) Slices sampled at axial directions: NCC: $0.747\pm0.096$; SSIM: $0.569\pm0.085$;

2) Slices sampled at coronal directions: NCC: $0.751\pm0.098$; SSIM: $0.570\pm0.080$;

3) Slices sampled at sagittal directions: NCC: $0.701\pm0.116$; SSIM: $0.606\pm0.072$;

4) Location: Angle: $0.137\pm0.097$ rad; Distance: $6.32\pm4.61$ pixel.  \\
\hline

Xie et al. \cite{xie2021image} & Forearms of 100 ultrasound sweeps (19502 frames) & NDI Polaris Vicra & 10-fold cross validation & 1) Mean distance error: average distance between the predicted and ground truth positions of all frames within a sweep;

2) Final drift: the distance of the final frame.& 1) Distance error: 4.73 mm;

2) Final drift: 8.55 mm.  \\

Ning et al. \cite{ning2022spatial} & 1) Forearm dataset: 100 cases (200 frames each); absolute range of probe movement: 10-20 cm;

2) Clinical dataset: 160 cases (200 frames each); absolute range of probe movement: 15-25 cm. & NDI Polaris Vicra (Northern Digital Inc., Canada) & Yes. The dataset was split into training, validation, and testing sets in a 3:1:1 ratio. & 1) Final drift: the distance between the center point of the first and last frame in the label and the distance being predicted;

2) Average distance error: the mean distance between corresponding corner points of each frame pair across the entire sequence. & 1) Forearm dataset: final drift: 15.84 mm; average distance error: 8.99 mm;

2) Clinical dataset: final drift: 35.17 mm; average distance error: 27.91 mm.\\
\hline

Leblanc et al. \cite{leblanc2022stretched} & Left or right femoral artery from 18 healthy volunteers: 111 ultrasound sweeps; 40,788 2D frames; proximal to distal. & 1) For pose: optical tracking device (NDI Polaris Spectra);

2) For segmentation: manually annotated using LabelMe; manually segmented (12 out of the 111 sequences from 6 volunteers) & For pose: 5-fold cross-validation.
Training: 10 volunteers ($13,087\sim15,687$ consecutives pairs of original frames); validation: 4 volunteers ($4323\sim4916$ pairs); testing: 4 volunteers ($4058\sim5789$ pairs).

For segmentation: training: 6 sequences (4 volunteers, 2073 frames); validation: 3 sequences (1 volunteer, 702 frames); testing: 3 sequences (1 volunteer, 1004 frames).& 1) Dice score between the manual segmentation and the output masks;

2) Relative frame-to-frame longitudinal error: the absolute difference between the prediction and ground truth;

3) Final drift: the absolute difference between the position of the last frame reconstructed using tracker-provided ground truth and the position estimated from the network outputs.& 1) Dice score: a false negative rate of 7\% and a mean dice score of 0.90 computed on the true positive cases;

2) Relative frame-to-frame longitudinal error: 0.28 mm;

3) Final drift: 17.22 mm. \\
\hline

Di Vece et al. \cite{di2022deep} & 1) Phantom fetal data: six brain volumes; 23-week GA fetus US examination phantom; 18047 planes with random orientation per volume and 725 around the TV SP;

2) Real fetal data: seven brain volumes; GA ranging from 21 to 25 and 39 weeks; 20699 images at random coordinates and 1330 around the TV SP for each volume. & The slices were generated by applying rotation and translation to a plane starting at the center of the volume, with parameters drawn from a uniform random distribution within a fixed range.

The position of the TV SP were annotated by a sonographer by directly manipulating a slicing plane within Unity and chose the translation and angle sampling intervals.& Experiment 1 (only use phantom data): 
Training: 75088 images from 4 volumes; testing: 37544 images from 2 volumes;

Experiment 2 (initialisation with weights from the phantom; training and testing on real data): training: one fetus, 22029 images, GA of 23 weeks; testing: six fetus ranging from a GA of 21 to 39 weeks. & 1) Euclidean distance between the two planes for calculating translation errors (in mm);

2) Geodesic distance for calculating rotation error (in degree). & 1) Experiment 1: 
Test planes acquired at random coordinates (Test RP): translation norm: 0.90 mm (median); rotation geodesic distance: 1.17 deg (median).

Test planes acquired around TV SP (Test SP): translation norm: 0.44 mm (median); rotation geodesic distance: 1.21 deg (median).

2) Experiment 2 (Test RP):

GA 21: translation norm: 9.94 mm (median); rotation geodesic distance: 30.58 deg (median).

GA 22: translation norm: 10.74 mm (median); rotation geodesic distance: 30.81 deg (median).

GA 23: translation norm: 10.39 mm (median); rotation geodesic distance: 21.94 deg (median).

GA 24: translation norm: 17.76 mm (median); rotation geodesic distance: 37.93 deg (median).

GA 25: translation norm: 18.80 mm (median); rotation geodesic distance: 42.23 deg (median).

GA 39: translation norm: 17.15 mm (median); rotation geodesic distance: 34.43 deg (median). \\

Guo et al. \cite{guo2022ultrasound} & 1) Dataset A (Prostate): 618 transrectal ultrasound video sequences; 618 patients;

2) Dataset B (Prostate): 100 transabdominal / transperineal ultrasound video sequences; $<100$ patients; varying lengths and resolutions. & EM tracking device & Yes.

Dataset A: training: 488 cases; validation: 66 cases; testing: 64 cases;

Dataset B: training: 80 cases; validation: 10 cases; testing: 10 cases; 10-fold cross-validation. & 1) Distance Error: the average distance between corresponding corner points of the input patches across the sweep;

2) Frame Error: the difference between the ground truth and predicted frame locations of neighboring frames;

3) Final Drift: the Euclidean distance between the last frame of the ultrasound video;

4) Drift Rate: the ratio between the final drift and the ground truth sequence length;

5) Prostate Dice: the Dice coefficient between the prostate's ground truth segmentation and the predicted segmentation;

6) Prostate Error: the absolute volume difference between the ground truth segmentation and the predicted segmentation in cubic centimeters. & Dataset A: 

Distance Error: $5.52\pm2.86$ mm; 

Frame Error: $0.90\pm0.26$ mm; 

Final Drift: $10.20\pm8.47$ mm; 

Drift Rate: $9.64\pm8.14 \%$; 

Prostate Dice: $0.89\pm0.06$; 

Prostate Error: $3.21\pm1.93$ cc;

Dataset B: 

Distance Error: $5.87\pm2.68$ mm; 

Frame Error: $1.12\pm0.26$ mm; 

Final Drift: $9.85\pm5.74$ mm; 

Drift Rate: $14.58\pm12.76 \%$;\\
\hline

Luo et al. \cite{luo2022deep} & 1) Arm dataset: from 41 volunteers (250 scans); average length of 94.83 mm; linear, curved, fast-and-slow and loop scans;

2) Carotid dataset: from 40 volunteers (160 scans); average length of 53.71 mm; linear scans. & An electromagnetic (EM) positioning transmitter / receiver & Yes. 

Arm data: taining: 196 scans; test: 54 scans; Carotid data: training: 136 scans; test: 24 cans.

Each scan in training set was randomly augmented (subsequence intercepting, interval sampling, and sequence inversion) to 40 sequences and regenerated at each epoch;
Each scan in test set was randomly augmented to 10 fixed sequences.& 1) Final drift rate (FDR);

2) Average drift rate (ADR);

3) Maximum drift (MD);

4) Sum of drift (SD);

5) Symmetric Hausdorff distance (HD);

6) Mean error of angle (EA) & 1) Arm dataset: 

FDR = $12.75\pm9.05 \%$; 

ADR = $19.05\pm11.46 \% $; 

MD = $10.24\pm7.36$ mm; 

SD = $332.29\pm316.36$ mm; 

HD = $9.40\pm7.13$ mm; 

EA = $1.55\pm1.46$ deg;

2) Carotid dataset: 

FDR = $15.67\pm8.37 \%$; 

ADR = $25.08\pm9.34 \%$; 

MD = $8.89\pm4.31$ mm; 

SD = $258.83\pm208.12$ mm; 

HD = $8.28\pm4.29$ mm; 

EA = $1.50\pm0.98$ deg.  \\
\hline

Mikaeili et al. \cite{mikaeili2022trajectory} & Model 539 Multi-purpose phantom; three sweeps (2520 frames in total) & DFRobot manufacturer’s inertial measurement unit (IMU) - Bosch BNO055 IMU & Yes. The last 100 frames from each of the three sets were held out for testing. & Average absolute error of 6 DoF parameters, between the prediction and the ground truth. & Results of input of 3 images:

Data set 1: $\theta_x$: 0.6067 deg; $\theta_y$: 0.8364 deg; $\theta_z$: 0.7604 deg; $t_x$: 2.2955 mm; $t_y$: 1.4807 mm; $t_z$: 2.3093 mm;

Data set 2: $\theta_x$: 0.4309 deg; $\theta_y$: 0.7950 deg; $\theta_z$: 0.7449 deg; $t_x$: 2.3119 mm; $t_y$: 1.5365 mm; $t_z$: 2.3676 mm;

Data set 3: $\theta_x$: 0.4193 deg; $\theta_y$: 0.8120 deg; $\theta_z$: 0.7434 deg; $t_x$: 2.3166 mm; $t_y$: 1.4754 mm; $t_z$: 2.2971 mm.\\

Yeung et al. \cite{yeung2022adaptive} & 1) Volume-sampled images: 50 training and 17 testing 3D ultrasound volumes co-aligned with the predefined 3D anatomical atlas; 3000 slices for each testing volume; 

2) Native 2D freehand images: video sequences of 2D freehand ultrasound brain scans. & 1) Volume-sampled images: sampled from aligned 3D ultrasound volumes; 2) Native 2D freehand images: the ground-truth locations were not available. & No.
Consider each ultrasound acquisition from different machines as a different domain. Training data and test data are from different domains.

Training: volume-sampled 2D images, from 50 3D volumes; fine-tuning and inference: volume-sampled 2D images from 17 aligned 3D volumes and native 2D freehand images. & 1) Euclidean distance (ED) between the coordinates of the predicted and ground truth planes in the predefined 3D anatomical atlas; 

2) Dihedral angle (DA) between the predicted and ground truth planes; 

3) Normalised standard deviation of change rate: quantifies the smoothness of consecutive ultrasound image transitions, with lower values indicating greater consistency. & For volume-sampled images:

1) Fine-tune and test on the same set of images: 

ED: $23.7\pm9.01$ voxel; 

DA: $0.198\pm0.092$ rad;

2) Fine-tune and test on different set of images: 

ED: $33.0\pm15.1$ voxel; 

DA: $0.211\pm0.097$ rad.

For native freehand images:

Normalised standard deviation of change rate: 0.553.\\
\hline

Chen et al. \cite{chen2023freehand} & 1) 1 mm metal sheet: 1000 frames; average distance of 15 mm;

2) 2 mm metal sheet: 1000 frames; average distance of 16 mm;

3) Symmetrical metal block: 2000 frames; average distance of 26 mm;

4) Asymmetric metal block: 2000 frames; average distance of 21 mm;

5) Right of human arm: 1500 frames; average distance of 100 mm;

6) Left of human arm: 1500 frames; average distance of 100 mm. &Acousto-optic positioning equipment & Yes. Each of the six datasets was split into a training set and a test set at a ratio of 7:3.
& 1) Theta X;

2) Theta Y;

3) Theta Z;

4) Lateral X;

5) Lateral Y;

6) Final drift.
& 1) Dataset 1): X: 1.55 mm; Y: 1.55 mm; Theta X: 0.61 deg; Theta Y: 0.67 deg; Theta Z: 0.83 deg; Final drift: 1.25 mm (median);

2) Dataset 3): X: 1.53 mm; Y: 1.55 mm; Theta X: 0.75 deg; Theta Y: 0.75 deg; Theta Z: 1.03 deg; Final drift: 1.33 mm (median);

3) Dataset 5): X: 1.55 mm; Y: 1.55 mm; Theta X: 1.16 deg; Theta Y: 1.05 deg; Theta Z: 1.44 deg; Final drift: 1.3 mm (median);

4) Dataset 6): X: 1.55 mm; Y: 1.53 mm; Theta X: 1.1 deg; Theta Y: 1.07 deg; Theta Z: 1.49 deg; Final drift: 1.31 mm (median).\\
\hline

Wysocki et al. \cite{wysocki2024ultra} & 1) Synthetic data: simulated B-mode images of a liver from CT images using ImFusion; seven scans; 200 2D ultrasound images for each scan;

2) Phantom data: lumbar spine, gelatine-based phantom; 13 scans; 150 frames for each scan. & 1) Not clear for synthetic data;

2) Robotic tracking for phantom data. & 1) For synthetic data: four tilted scans for training, three scans of one perpendicular and two tilted for testing.

2) For phantom data: four tilted scan pairs for training, two tilted scan pairs and one perpendicular scan for testing. The visible tissue are covered both in training and test sets. & SSIM between synthetic and reference testing data & SSIM between synthetic and reference B-mode images:

1) Liver synthetic data: tilted scan: 0.45; perpendicular scan: 0.49;

2) Spine phantom data: tilted scan: 0.51; perpendicular scan: 0.54.\\

Luo et al. \cite{luo2023recon} & 1) 3D developmental dysplasia of the hip (DDH) ultrasound dataset from 28 volunteers: 169 ultrasound volumes;

2) 3D fetus ultrasound dataset from 128 pregnant volunteers: 128 ultrasound volumes; gestational age ranges from 10 to 14 weeks;

3) 2D spine ultrasound dataset from 23 volunteers: 68 scans; mean length of 186 mm. & For datasets 1) and 2): a motion-based simulation method for generating 2D scans with complex scanning strategies from volumetric regions of interest;

For dataset 3): a robotic arm with EM positioning & Yes.

1) For 3D DDH dataset: training: 135 volumes; validation: 16 volumes; test: 18 volumes;

2) For fetus US dataset: training: 98 volumes; validation: 15 volumes; test: 15 volumes;

Generating 100 different scans for each volume in each training epoch. In the testing phase,  generating 20 fixed scans for each volume.

3) For 2D spine US dataset: training: 45 scans; validation: 12 scans; test: 11 scan. & 1) Final Drift Rate (FDR): the final drift divided by the scan length;

2) Average Drift Rate (ADR): in which drift rate refers to the cumulative drift of a frame divided by the length from this frame to the starting frame of the scan;

3) Maximum Drift (MD): the maximum drift of all frames;

4) Sum of Drift (SD): the sum of all drift;

5) Symmetric Hausdorff Distance (HD): the maximum distance between the estimated and ground truth positions across all frames;

6) Mean Error of Angle (MEA): MAE between estimated and real Euler angles. & 1) DDH: 

FDR: $10.68\pm11.21 \%$; 

ADR: $10.39\pm6.71 \%$; 

MD: $5.25\pm5.17$ mm; 

SD: $71.95\pm75.33$ mm; 

HD: $3.09\pm1.87$ mm; 

MEA: $1.33\pm2.02$ deg;

2) Fetus: 

FDR: $13.31\pm10.94 \%$; 

ADR: $18.42\pm11.78 \%$; 

MD: $18.69\pm12.78$ mm; 

SD: $236.37\pm256.35$ mm; 

HD: $9.60\pm6.45$ mm; 

MEA: $2.26\pm2.66$ deg;

3) Spine: 

FDR: $10.82\pm5.36 \%$; 

ADR: $19.72\pm12.83 \%$; 

MD: $21.05\pm10.82$ mm; 

SD: $660.85\pm909.42$ mm; 

HD: $20.38\pm11.07$ mm; 

MEA: $1.25\pm0.46$ deg.  \\
\hline

Li et al. \cite{li2023trackerless} & Forearms from 19 volunteers: 228 scans; more than 40,000 ultrasound frames; straight line, C shape and S shape; distal-to-proximal direction; $36\sim430$ frames for each scan; $100\sim200$ mm per scan. & NDI Polaris Vicra (Northern Digital Inc., Canada) & Yes. The data was split into train, validation and test sets by a ratio of 3:1:1. & 1) Frame prediction accuracy: Euclidean distance between the predicted and ground truth positions of four corner points in consecutive frames;

2) Accumulated tracking error: average Euclidean distance over all reconstructed image pixel locations;

3) Volume reconstruction overlap: Dice coefficient between the reconstructed volumes of the prediction and the ground truth;

4) Final drift: Euclidean distance, averaged over the four corners, between the final predicted frame and the ground truth frame in each scan. & For a selected transformation ($T_{10\leftarrow5}$), without accumulated and consistency loss:

CNN:

1) Frame prediction accuracy: $0.53\pm0.56$ mm;

2) Accumulated tracking error: $19.09\pm14.60$ mm;

3) Volume reconstruction overlap: $0.81\pm0.11$;

4) Final drift: $36.82\pm28.01$ mm;

LSTM:

1) Frame prediction accuracy: $0.50\pm0.41$ mm;

2) Accumulated tracking error: $18.13\pm9.49$ mm;

3) Volume reconstruction overlap: $0.78\pm0.12$;

4) Final drift: $36.48\pm20.77$ mm. \\
\hline

Li et al. \cite{li2023privileged} & Forearms from 19 volunteers: 228 scans; more than 40,000 US frames; straight line, C shape and S shape; distal-to-proximal direction; $36\sim430$ frames for each scan; $100\sim200$ mm per scan. & NDI Polaris Vicra (Northern Digital Inc., Canada) & Yes. The data was split into train, validation and test sets by a ratio of 3:1:1. & 1) Frame prediction accuracy: the Euclidean distance between the ground truth and prediction- transformed four corner points of each frame;

2) Accumulated tracking error: the averaged point distance on all pixels;

3) Volume reconstruction overlap: the overlap of all pixels between the ground truth and prediction volume;

4) Final drift: the frame prediction accuracy of the last frame in a scan.

&
$M=100$:

1) Frame prediction accuracy: $0.17\pm0.08$ mm;

2) Accumulated tracking error: $3.80\pm3.97$ mm;

3) Volume reconstruction overlap: $0.76\pm0.24$;

4) Final drift: $6.56\pm7.53$ mm;

$M=140$:

1) Frame prediction accuracy: $0.13\pm0.05$ mm;

2) Accumulated tracking error: $2.90\pm2.10$ mm;

3) Volume reconstruction overlap: $0.89\pm0.00$;

4) Final drift: $6.53\pm5.98$ mm.\\

Luo et al. \cite{luo2023multi} & 1) Arm dataset: 288 scans; linear, curved, loop, and sector scans; average length of 323.96 mm;

2) Carotid dataset: 216 scans; linear, loop, and sector scans; average length of 203.25 mm. & EM positioning system &Yes.

1) Arm dataset: training: 200 scans; validation: 40 scans; test: 48 scans;

2) Carotid dataset: training: 150 scans; validation: 30 scans; test: 36 scans.

Randomly augmented (sub-sequence intercepting, interval sampling, and sequence inversion) each training scan to 20 sequences and each test scan to 10 sequences. & 1) Final drift rate (FDR);

2) Average drift rate (ADR);

3) Maximum drift (MD);

4) Sum of drift (SD);

5) Symmetric Hausdorff distance (HD);

6) Mean error of angle (EA). & 1) Arm dataset: 

FDR = $10.01\pm5.7$ \%; 

ADR = $18.86\pm6.5$ \%; 

MD = $28.61\pm11.0$ mm; 

SD = $1064.06\pm582.5$ mm; 

HD = $27.38\pm11.4$ mm; 

EA = $2.76\pm1.3$ deg;

2) Carotid dataset: 

FDR = $10.90\pm5.3 \%$; 

ADR = $19.61\pm8.5 \%$; 

MD = $21.81\pm7.2$ mm; 

SD = $804.27\pm282.8$ mm; 

HD = $19.30\pm7.6$ mm; 

EA = $2.60\pm1.6$ deg.  \\

\hline

EI hadramy et al. \cite{el2023trackerless} & \textit{Ex vivo} swine liver; six scans & Electromagnetic tracking system (trakSTAR™, NDI, Canada) & Yes. The data with 13734 clips was split into train, validation and test sets by a ratio of 7:1.5:1.5. & 1) Final drift rate (FDR): the final drift divided by the sequence length;

2) Average drift rate (ADR): the average cumulative drift of all frames divided by the length from the frame to the starting point of the sequence.
& 1) FDR: $23.11\pm11.6 \%$;

2) ADR: $28.71\pm12.97 \%$;

3) Translation error: $0.449\pm0.189$ mm;

4) Orientation error: $1.3\pm1.5$ deg.  \\
\hline

Li et al. \cite{li2023long} & Forearms from 19 volunteers: 228 scans; straight line, C shape and S shape; distal-to-proximal direction; 36 $\sim$ 430 frames each scan; 100 $\sim$ 200 mm per scan. & Optical tracker (NDI Polaris Vicra, Northern Digital Inc., Canada) & Yes. The data was split into train, validation and test sets by a ratio of 3:1:1.

This paper also tested the influence of anatomy and protocol to reconstruction performance, by training on one type of scans (e.g., straight line shape) and test on the original test set. & 1) Frame prediction accuracy: Euclidean distance between predicted and ground truth four corner points on the $j^{th}$ frame, both transformed from the $i^{th}$ frame;

2) Accumulated tracking error: average Euclidean distance of all reconstructed image pixels between prediction and ground truth;

3) Volume reconstruction overlap: the overlap between the ground-truth- and prediction- reconstructed scan volumes;

4) Final drift: the average Euclidean distance, over the four corner points on the last frame of the scan, between ground truth and prediction.

&An example when $M=100$:

1) Frame prediction accuracy: $0.19\pm0.08$ mm;

2) Accumulated tracking error: $4.01\pm4.01$ mm;

3) Volume reconstruction overlap: $0.77\pm0.17$;

4) Final drift: $7.24\pm8.33$ mm.  \\
\hline

Dou et al. \cite{dou2023sensorless} & Prostate dataset: 3D transrectal ultrasound (1,928 scans) & The freehand scans were resampled by generating oblique slices from the existing scans. & Yes & 1) Distance error;

2) Final drift

& 1) Distance error: 4.38 mm;

2) Final drift: 7.06 mm.  \\

Yeung et al. \cite{yeung2024sensorless} & 1) Dataset A: fifteen 3D ultrasound fetal brain volumes; obtained as part of the INTERGROWTH-21st study; between 19 and 21 gestational weeks;

2) Dataset B: seventeen 3D ultrasound fetal brain volumes; between 19 and 21 gestational weeks;

3) Native 2D freehand video sequences: four videos of native freehand 2D brain scans; around 250 frames each; 20 weeks' gestational age.

& 1) For datasets A and B: ground truth is the original volumes;

2) For native 2D freehand video sequences: lack of ground truth 3D location information.

&Yes.

For native 2D freehand video sequences: 80\% of each 2D ultrasound sequence for volumetric reconstruction and the remaining 20\% for testing.

&1) Structural similarity index measure (SSIM);

2) Visual information fidelity (VIF): measuring the similarity between pair of 2D slices sampled from the 3D reconstructed volumes and the corresponding slices sampled from the native volumes;

3) Absolute difference between rotation angles ($\theta_{diff}$) from refined locations and the ground truth locations;

4) Absolute distance between translations ($T_{diff}$) from refined locations and the ground truth locations;

5) Dice similarity coefficient (DSC);

6) 95th percentile Hausdorff distance ($H_{95}$);

7) Unsigned relative volume differences.

&Boxplot; exact numbers are not provided.  \\
\hline

Gaits et al. \cite{gaits2024ultrasound} & 1) Simulated dataset: a synthetic medium consisting of a highly echogenic $1.25^3$ $cm^3$ cube in a low echogenic homogenous medium; a sectorial scan (probe tilt);

2) Simulated dataset: a pelvic T2 MRI volume; double orthogonal scan;

3) A phantom mimicking a peripheral nerve block (CAE blue phantom from CAE Healthcare) & Magnetic tracker & Yes & 1) Signal-to-noise ratio (SNR);

2) Contrast-to-noise ratio (CNR)

& 1) Cube: SNR: 31.8 dB; CNR: 30.4 dB;

2) MRI simulation: SNR: 28.3 dB; CNR: 25.8 dB;

3) Nerve phantom: SNR: 14.4 dB; CNR: 14.2 dB.  \\
\hline

Dou et al. \cite{10684746} & 1) Abdominal phantom: 75 scans; linear along the phantom contour; $30\sim94$ mm;

2) Forearms from 19 volunteers: 228 scans; straight line, C shape and S shape; distal-to-proximal direction; $36\sim430$ frames each scan; $100\sim200$ mm per scan. 

& For data 1): A 6-D EM tracking system (NDI 3D Guidance, Northern Digital Inc., Waterloo, ON, Canada);

For data 2): NDI Polaris Vicra (Northern Digital Inc., Canada)
& Yes. For both datasets, the data were split into training, validation, and test sets using a 3:1:1 ratio.  & 1) Distance error: average distance between predicted and ground truth frames over corner points in the ultrasound sequence;

2) Final drift: Euclidean distance of the final frame between the prediction and ground truth;

3) Frame error, providing the mean individual FE using the relative pose between prediction and ground truth;

4) MGU: GPU memory efficiency;

5) Average Inference Time (AIT). & Phantom dataset:

1) Distance error: $3.09\pm1.40$ mm;

2) Final drift: $4.65\pm2.25$ mm;

3) Frame error: $0.15\pm0.04$ mm;

4) MGU: 6050 MiB;

5) Average Inference Time (AIT): 0.98 s;

Forearm dataset:

1) Distance error:$12.77\pm6.84$ mm;

2) Final drift: $18.01\pm10.44$ mm;

3) Frame error: $0.48\pm0.26$ mm;

4) MGU: 6932 MiB. \\[20em]

Ramesh et al. \cite{ramesh2024geometric} & 3D ultrasound fetal brain volumes: obtained as part of the INTERGROWTH-21st study; 4 training, 2 validation, and 7 test 3D volumes selected at 19 gestational weeks. & Pre-aligned to a common reference atlas space;

Artificial 2D slice sampling (training labels derived from the nature of the data itself) & Yes & 1) Euclidean distance (ED) and plane angle (PA) between all coordinates of the predicted and ground truth planes measured in radians;

2) Mean squared error (MSE) between predicted and ground truth reference points;

3) Normalized Cross-Correlation (NCC) and Structural Similarity (SSIM) between images sampled with predicted planes and ground truth planes.

& 1) ED: $0.32\pm0.24$;

2) PA: $0.39\pm0.24$;

3) MSE: $215.10\pm241.52$;

4) NCC: $0.67\pm0.22$;

5) SSIM: $0.61\pm0.25$. \\
\hline

Yan et al. \cite{yan2024fine} & 1) Arm dataset: 583 scans; linear, curved, loop, and sector scans; average length of 386.39 mm;

2) Carotid dataset: 432 scans; linear, loop, and sector scans; average length of 241.25 mm. & EM positioning system & Yes.

Arm dataset: training: 375 scans; validation: 104 scans; test: 104 scans;

Carotid dataset: training: 276 scans; validation: 78 scans; test: 78 scans. 

Performed random augmentations on each scan (sub-sequence intercepting, interval sampling, and sequence inversion).

&1) Final drift rate (FDR);

2) Average drift rate (ADR);

3) Maximum drift (MD);

4) Sum of drift (SD);

5) Symmetric Hausdorff distance (HD);

6) Mean error of angle (MEA).

&1) Arm dataset: 

FDR = $9.72\pm7.1 \%$; 

ADR = $15.53\pm9.6 \%$; 

MD = $24.68\pm13.6$ mm; 

SD = $677.48\pm498.4$ mm; 

HD = $23.05\pm13.6$ mm; 

MEA = $3.41\pm1.8$ deg;

2) Carotid dataset: 

FDR = $8.61\pm5.9 \%$; 

ADR = $16.16\pm8.1 \%$; 

MD = $13.78\pm6.4$ mm; 

SD = $391.77\pm298.4$ mm; 

HD = $12.39\pm6.2$ mm; 

MEA = $2.09\pm1.2$ deg.  \\
\hline

Li et al. \cite{li2024nonrigid} & Forearm dataset from 60 volunteers: 720 scans; straight, c-shape and s-shape; distal-to-proximal direction; average number of frames per scan (500); $200\sim300$ mm per scan. & NDI Polaris Vicra (Northern Digital Inc., Canada) & Yes. The dataset was split into train, validation and test sets by a ratio of 3:1:1. & 1) Global pixel reconstruction error (GPE), reconstruction error on all pixels based on global displacement vectors;

2) Global landmark reconstruction error (GLE), reconstruction error on landmarks based on global displacement vectors;

3) Local pixel reconstruction error (LPE), reconstruction error on all pixels based on local displacement vectors;

4) Local landmark reconstruction error (LLE), reconstruction error on landmarks based on local displacement vectors. & 1) GPE: $16.51\pm7.76$ mm;

2) GLE: $17.91\pm7.85$ mm;

3) LPE: $3.07\pm0.99$ mm;

4) LLE: $3.38\pm1.00$ mm.  \\
\hline

Dou et al. \cite{dou2024pitfalls} & Two freehand datasets on a tissue mimicking abdominal phantom (Model 057A, CIRS Inc., USA)

1) Four scans; 621 frames;

2) Four scans; 865 frames. & A 6D electromagnetic (EM) tracking system (NDI 3D Guidance, Northern Digital Inc., Canada) & Yes & 1) Peak signal to noise ratio (PSNR);

2) Structural similarity index (SSIM). & 1) Phantom Data 1: PSNR: 27.17 dB; SSIM: 0.56 \%;

2) Phantom Data 2: PSNR: 26.11 dB; SSIM: 0.51 \%.  \\
\hline

Dou et al. \cite{dou2024continuous} & Synthetic liver data: 800 two-dimensional B-mode liver images for training. & Not clear & Yes & 1) Peak signal-to-noise ratio (PSNR);

2) Structural similarity index (SSIM).

&1) PSNR: $27.73\pm0.98$ dB;

2) SSIM: $0.54\pm0.19$  \\

\raisebox{-0.8\height}{\rotatebox[origin=c]{90}{\parbox{2.6cm}{\centering Großbröhmer et al.\ \mbox{\cite{grossbrohmer20253d}}}}} & 63 left forearm scans from 21 subjects: lateral direction; straight dorsal, C-shape dorsal and S-shape dorsal; length of a scan varies between 67 and 279 frames ($157\pm42$); a length of 11.6-17.3 cm ($14.7\pm1.6$) & NDI Hybrid Polaris Spectra (Northern Digital Inc, Canada) & Yes.

``straight dorsal” as fixed and ``C-shape” / ``S-shape” as moving sweeps.

Leave-one-out cross-validation with 19 training, 1 validation and 1 testing case. & 1) Directed Surface Distance (DSD): the average distance between each point in the transformed point cloud and its nearest neighbour in the fixed point cloud;

2) Target registration error (TRE): assessment of direct point correspondences, computed with the ground truth trajectories of moving sweeps;

3) Average symmetric surface distance (ASSD), computed with the ground truth trajectories of moving sweeps;

4) 95th percentile of Hausdorff Distances (HD95), computed with the ground truth trajectories of moving sweeps;

5) Translational pose error ($E_T$): sweep trajectory recovery, computed with the ground truth trajectories of moving sweeps;

6) Runtime & 1) DSD: $1.29\pm0.38$ mm;

2) TRE: $5.20\pm2.15$ mm;

3) ASSD: $1.13\pm0.21$ mm;

4) HD95: $2.59\pm0.80$ mm;

5) $E_T$: $6.14\pm2.66$ mm;

6) Runtime: 880 ms. \\
\hline

Eid et al. \cite{eid2025rapidvol} & 15 3D US fetal brain scans: 20 gestational weeks; collected from the INTERGROWTH-21st study & Ground truth poses are sampled from the 3D scan & Yes & Structural similarity index measure (SSIM) & SSIM results:

1) Axial input training scan sets and 128 testing slices: Coronal: $.955\pm.008$; Sagittal: $.952\pm.009$;

2) Axial input training scan sets and 256 testing slices: Coronal: $.973\pm.005$; Sagittal: $.969\pm.006$;

3) Coronal input training scan sets and 128 testing slices: Axial: $.941\pm.016$; Coronal: $.932\pm.018$; Sagittal: $.935\pm.017$;

4) Coronal input training scan sets and 256 testing slices: Axial: $.952\pm.015$; Coronal: $.950\pm.016$; Sagittal: $.950\pm.014$.  \\
\hline

Luo et al. \cite{luo2025monetv2}  & 1) Arm dataset: 583 scans;  average length of 386.39 mm; linear, curved, loop, and sector scans;

2) Carotid dataset: 432 scans; average length of 241.25 mm;  linear, loop, and sector scans;

3) Thyroid dataset: 290 scans; average length of 66.25 mm;  linear and loop scans. & EM positioning system (3-D Guidance trakSTAR, Northern Digital Inc., Canada) & Yes.

Arm dataset: training: 375 scans; validation: 104 scans; test: 104 scans;

Carotid dataset: training: 276 scans; validation: 78 scans; test: 78 scans;

Thyroid dataset: training: 184 scans; validation: 53 scans; test: 53 scans.

Augmented (subsequence intercepting, interval sampling, and sequence inversion) each training scan into 20 different sequences in each training epoch; Augmented each validation or test scan into 10 fixed sequences. & 1) Final drift rate (FDR): the ratio of the distance between the estimated and actual positions of the last frame divided by the scan length;

2) Average drift rate (ADR): the mean drift rate, where each drift rate is defined as the ratio of a frame's cumulative drift to its distance from the initial frame of the scan;

3) Maximum drift (MD): the MD among all the frames;

4) Sum of drift (SD): the total sum of all the drifts;

5) Symmetric Hausdorff distance (HD): the largest distance between estimated and actual positions across all frames;

6) Mean error of angle (MEA):  the MAE between the estimated Euler angle and the actual Euler angle. & 1) Arm dataset: 

FDR: $11.04\pm6.02 \%$; 

ADR: $18.26\pm9.10 \%$; 

MD: $28.84\pm15.62$ mm; 

SD: $901.87\pm658.43$ mm; 

HD: $26.24\pm12.47$ mm; 

MEA: $3.00\pm2.25$ deg;

2) Carotid dataset: 

FDR: $12.11\pm7.55 \%$; 

ADR: $23.78\pm17.51 \%$; 

MD: $20.75\pm8.82$ mm; 

SD: $646.00\pm455.03$ mm; 

HD: $18.50\pm8.64$ mm; 

MEA: $2.40\pm1.55$ deg;

3) Thyroid dataset: 

FDR: $17.88\pm9.10 \%$; 

ADR: $26.34\pm9.12 \%$; 

MD: $8.93\pm4.79$ mm; 

SD: $207.60\pm195.20$ mm; 

HD: $8.43\pm4.73$ mm; 

MEA: $1.36\pm0.97$ deg.  \\

Lee et al. \cite{lee2025enhancing} & 1) Forearms from 9 individuals: 54 scans;  $900\sim1000$ frames per scan; 51,300 frames in total; an arbitrary S-shaped trajectory; three scans for each forearm; travel length of $199.27\pm23.81$ mm;

2) Photoacoustic data: 60 scans; 10 volunteers; 28,445 frames in total; travel length of $153.12\pm13.20$ mm;

3) Forearms from 19 volunteers: 228 scans; 30,552 frames in total; straight line, C shape and S shape; distal-to-proximal direction; 36 $\sim$ 430 frames each scan; travel length of $144.51\pm14.26$ mm. & For data 1) and 2): Electromagnetic sensor (Polhemus);

For data 3): NDI Polaris Vicra (Northern Digital Inc., Canada) & Yes.

For dataset 1): training: 30 scans; validation: 6 scans; test: 18 scans;

For dataset 2): The dataset was split into training, validation, and testing in a 5:1:4 ratio;

For dataset 3): The dataset was divided into training, validation, and testing in a 10:4:5 ratio.

During inference, only the final prediction is used to reconstruct the 3D volume. & 1) Relative average error (rAE): measures the relative motion estimation accuracy;

2) Accumulated average error (aAE): measures the accumulated motion error over the scan trajectory;

3) Relative frame error (rFE): relative displacements between frames;

4) Accumulated frame error (aFE): cumulative displacements along the scan path;

5) Correlation (Corr):  cosine similarity, to measure the underlying trends of reconstructed trajectories in 3D Euclidean space;

6) Final drift rate (FDR): aFE of the final frame divided by the total length of the scan trajectory.

&1) Dataset 1: 

rAE: $0.1430\pm0.072$ mm, $0.0663\pm0.023$ deg; 

aAE: $7.9097\pm4.89$ mm, $5.9412\pm3.83$ deg; 

rFE: $0.2917\pm0.054$ mm; 

aFE: $16.1944\pm6.09$ mm; 

Corr: $0.9217\pm0.048$; 

FDR: $12.0577\pm6.45 \%$.

2) Dataset 2: 

rAE: $0.0597\pm0.026$ mm, $0.0221\pm0.008$ deg; 

aAE: $4.3204\pm3.31$ mm, $2.2230\pm1.88$ deg; 

rFE: $0.1225\pm0.042$ mm; 

aFE: $8.9360\pm3.99$ mm; 

Corr: $0.9165\pm0.059$; 

FDR: $10.6663\pm3.87 \%$;

3) Dataset 3: 

rAE: $0.1979\pm0.124$ mm, $0.0755\pm0.026$ deg; 

aAE: $8.0867\pm6.12$ mm, $2.3956\pm1.57$ deg; 

rFE: $0.4113\pm0.159$ mm; 

aFE: $16.4062\pm7.33$ mm; 

Corr: $0.9164\pm0.105$; 

FDR: $19.8096\pm10.30 \%$. \\
\hline

Sun et al. \cite{sun2025ultrasom} & 200 ultrasound video sequences of forearms from 20 volunteers: 58,011 frames; linear, curved, fast-slow, roll; an average scan length of 103 mm. & An optical tracking device, the NDI Polaris Vega XT (Northern Digital Inc., Canada) & Yes.

The dataset was split into training, validation, and testing sets in an 8:1:1 ratio.

Using sliding window reconstruction strategy during inference, which takes the average of these repeated predictions to obtain the final result for each frame. & 1) Distance Error (DE): the average distance between corresponding corner points across all frames in the sequence;

2) Final Drift (FD): the positional difference between the last frame of the reconstructed sequence and the ground truth;

3) Final Drift Rate (FDR): defined as FD divided by the sequence length, where the sequence length is calculated by connecting the ground truth center points of each frame;

4) Average Drift Rate (ADR): the average drift per frame;

5) Maximum Drift (MD): the largest drift within the sequence;

6) Sum of Drift (SD): total drift across all frames;

7) Symmetric Hausdorff Distance (HD): the discrepancy between the true and estimated displacement parameters;

8) Mean Angular Error (MEA): the average absolute error between the true and estimated rotational parameters.

 &1) DE: $7.34\pm3.11$ mm;

2) FD: $10.67\pm5.91$ mm;

3) FDR: $10.24\pm5.34 \%$;

4) ADR: $17.15\pm4.60 \%$;

5) MD: $13.15\pm5.50$ mm;

6) SD: $1966.59\pm784.56$ mm;

7) HD: $10.81\pm5.81$ mm;

8) MEA: $2.05\pm1.74$ deg.  \\

\end{longtable}
}

\bibliographystyle{elsarticle-num} 
\bibliography{ref}

\end{document}